\documentclass[12pt]{article}
\pdfoutput=1

\usepackage{draft}
\usepackage{putex} 
\usepackage[all,cmtip]{xy}
\usepackage{fancybox}
 
 \usepackage{extarrows}
 
\usepackage{graphicx,subfig}
\usepackage{cite}
\usepackage{mciteplus}
\usepackage{skak}
\usepackage{amsmath}
\DeclareFontFamily{OT1}{pzc}{}
\DeclareFontShape{OT1}{pzc}{m}{it}{<-> s * [1.10] pzcmi7t}{}
\DeclareMathAlphabet{\mathpzc}{OT1}{pzc}{m}{it}

\newcommand{\TY}{{\rm TY}}

\def\({\left(}
\def\){\right)}

\newcommand{\pa}{\partial}

\renewcommand{\cH}{\mathcal{H}}
\newcommand{\eeq}{\end{equation}}
\newcommand{\ea}{\end{array}}

\def\eea{\end{eqnarray}}

\def\<{\langle}
\def\>{\rangle}

\def\bZ{\mathbb{Z}}
\def\bR{\mathbb{R}}
\def\bC{\mathbb{C}}
\def\bA{\mathbb{A}}
\def\bB{\mathbb{B}}
\def\cL{\mathcal{L}}
\def\cM{\mathcal{M}}
\def\cN{\mathcal{N}}
\def\cA{\mathcal{A}}
\def\cO{\mathcal{O}}
\def\cC{\mathcal{C}}

\usepackage{amsmath}
\usepackage{comment}
\usepackage{amssymb}
\usepackage{amsthm}
\usepackage{bbm}

\newtheorem{thm}{Theorem}

\theoremstyle{definition}

\usepackage{color}
\usepackage{tikz-cd}
\usepackage{comment}

\usepackage[colorlinks=true]{hyperref}

\usepackage{adjustbox}

\begin{document}
	
	\preprint{PUPT-2603}

	\institution{Weizman}{Department of Condensed Matter Physics, Weizmann Institute of Science, Rehovot, Israel}
	\institution{PU}{Joseph Henry Laboratories, Princeton University, Princeton, NJ 08544, USA}
	\institution{CMSA}{Center of Mathematical Sciences and Applications, Harvard University, Cambridge, MA 02138, USA}
	\institution{HU}{Jefferson Physical Laboratory, Harvard University,
		Cambridge, MA 02138, USA}

	\title{
    \huge Fusion Category Symmetry I:\\ \Large Anomaly In-Flow and Gapped Phases
	}

	\authors{Ryan Thorngren\worksat{\Weizman,\CMSA}    and Yifan Wang\worksat{\PU,\CMSA,\HU}}

	\abstract{We study generalized discrete symmetries of quantum field theories in 1+1D generated by topological defect lines with no inverse. In particular, we describe 't Hooft anomalies and classify gapped phases stabilized by these symmetries, including new 1+1D topological phases. The algebra of these operators is not a group but rather is described by their fusion ring and crossing relations, captured algebraically as a fusion category. Such data defines a Turaev-Viro/Levin-Wen model in 2+1D, while a 1+1D system with this fusion category acting as a global symmetry defines a boundary condition. This is akin to gauging a discrete global symmetry at the boundary of Dijkgraaf-Witten theory. We describe how to ``ungauge" the fusion category symmetry in these boundary conditions and separate the symmetry-preserving  phases from the symmetry-breaking ones. For Tambara-Yamagami categories and their generalizations, which are associated with Kramers-Wannier-like self-dualities under orbifolding, we develop gauge theoretic techniques which simplify the analysis. We include some examples of CFTs with fusion category symmetry derived from Kramers-Wannier-like dualities as an appetizer for the Part II companion paper.
		 }
	\date{}

	\maketitle
	
	\tableofcontents
	
	\section{Introduction}
	
	A fundamental problem in quantum field theory is to determine when two theories are connected by a renormalization group flow. This has applications both to high energy physics---e.g. attempting to find the standard model along a flow from a UV theory---and to condensed matter physics---e.g. determining the nearby phase diagram of a critical point. Besides perturbatively solving the RG flow itself, we can also devise non-perturbative invariants which either cannot change along the RG flow or change in a prescribed way, such as monotonically decreasing.
	
    In these problems, it is useful to track the fate of the symmetries of our theory. In fact the symmetry group is an invariant of the former sort, which cannot ever change along the flow, although it can become spontaneously broken or enhanced at the fixed points. A finer invariant one can define, which constrains the possible fates of the symmetry, is the 't Hooft anomaly.
    
    The 't Hooft anomaly is the obstruction to coupling the theory to a background gauge field for the symmetry. More precisely, there is a minimal procedure to attempt to do so, and if gauge invariance cannot be achieved by the addition of local counterterms, then we say the 't Hooft anomaly is non-trivial. A basic fact is that a theory with a non-trivial 't Hooft anomaly must flow to a theory also with a non-trivial anomaly. In particular, the trivial theory has no 't Hooft anomalies, so a theory with a non-trivial 't Hooft anomaly can never have a (symmetric) flow to a gapped, non-degenerate phase.

	It is possible to define an even better invariant by studying anomaly in-flow. That is, while our theory cannot be consistently coupled to a background gauge field, we can usually remedy this by defining our theory to live on the boundary of a classical gauge theory with a topological term. One can show that the topological term is also an invariant of the RG flow and so must match between the UV and IR fixed points. From now on we will simply refer to this topological term as the anomaly associated with the symmetry.
	
	The obvious usefulness of anomalies have led to various generalizations of the concept beyond ordinary notions of symmetry, including higher form symmetries \cite{Kapustin_2017,Gaiotto_2015} and anomalies associated with a parameter space \cite{cordova2019anomalies}. In this paper, we explore another direction of this fruitful labor where we relax the necessity that our symmetries be defined by invertible unitary transformations of the Hilbert space. This allows us to devise invariants of RG flows based on more complex transformations such as Kramers-Wannier duality.
	
	To explain what we mean, consider that a symmetry is by definition an operator which commutes with the Hamiltonian, and hence in a spacetime correlation function can be freely moved along away from other operator insertions without affecting the result. In a relativistic theory, one expects that this operator becomes completely topological, and can be moved in any direction and even deformed into any shape, so long as it does not touch itself or other operator insertions. For this reason one can make a broad definition of the symmetry algebra of a theory as its algebra of topological operators.
	
	In the case of ordinary (group-like, 0-form) symmetries, the topological operators are codimension 1 (so that they can fill a spatial slice) and invertible, since the operator for $g \in G$ should fuse with the operator for $g^{-1} \in G$ to the trivial operator. Higher form symmetries correspond to higher codimension but still invertible topological operators, and likewise parameter-spaces provide an action of their homotopy type as a higher symmetry. However, there are also many interesting \emph{non-invertible} topological operators. Note that these operators may actually be invertible on the Hilbert space, but only in a non-local way, in the sense that one cannot define a topological operator associated with the inverse transformation.
	
	Symmetries associated with non-invertible topological operators are quite common in 1+1D, where they are typically associated with rational CFTs. For example, Kramers-Wannier duality of the Ising CFT defines such an operator. A large family of lattice Hamiltonians with such symmetries can be constructed using the so-called anyon chains \cite{Feiguin_2007,Pfeifer_2012,BuicanGromov}. For instance, the ``golden chain" has a symmetry operator whose algebra is the Fibonacci algebra $W^2 = 1 + W$, and hence cannot be realized by an invertible operator on Hilbert space, but nonetheless acts. These symmetries may also be realized by matrix-product operators (MPOs) \cite{Williamson_2016,Bultinck_2017,Cirac_2017}, and were even found in some related 2d statistical mechanical models \cite{Aasen_2016}. In these works, several local stability results were observed near specific fixed points, but none proved a global constraint on the phase diagram, i.e. an anomaly.
	
	In 1+1D the topological line defects (TDLs) are best described as a fusion category, which encodes the fusion rules of the lines as well as the ``$F$-symbol" which relates the two different resolutions of a four-way junction into two three-way junctions. In \cite{theOGs}, the authors (including one of us) argued that this fusion category is an invariant of the RG flow and moreover in the case where there is a TDL of non-integer quantum dimension that the flow cannot terminate in a trivial theory. This is the first observation we are aware of of an anomaly for such symmetries, and it is our goal to explore them further.
	
	Our approach is to use anomaly in-flow. We will argue that a 1+1D theory with a fusion category symmetry forms a boundary condition of a 2+1D topological quantum field theory known as Turaev-Viro/Levin-Wen theory, which is defined by the fusion category. In a sense, this amounts to studying the theory with gauged fusion category symmetry. However, we can still adapt the usual anomaly in-flow arguments to this theory. For instance, RG flows by symmetric perturbations can only take us between different boundary conditions of this theory. Moreover, we are able to formulate what it means for this Turaev-Viro theory to admit a symmetry-preserving boundary condition and thus argue that some RG flows must end in non-trivial theories, either gapless or where the fusion category symmetry is spontaneously broken. We explore this in some detail in the case where all quantum dimensions are integers, since these constraints are beyond those discussed in \cite{theOGs}. We believe we have found the strictest-possible anomaly-vanishing condition for these fusion category symmetries in 1+1D.
	
    The outline of the paper is as follows. In Section \ref{secgapbc} we provide a gentle review of Turaev-Viro theory and discuss its gapped boundary conditions, at times specializing to the special case of twisted Dijkgraaf-Witten theory, whose study captures the familiar theory of discrete 't Hooft anomalies in 1+1D. We formulate and prove our general anomaly-vanishing condition that the fusion category admits a fiber functor in Section 
    \ref{subsecanomvanishing}. In Section \ref{secSPTs} we discuss a converse statement, whereby boundary conditions of Turaev-Viro phases are identified with symmetric gapped phases. In particular, non-degenerate symmetric phases are classified by fiber functors, which gives a characterization of fusion category ``SPTs" in 1d.
    
    In Section \ref{secgappedphases} we explore more general gapped 1+1D phases with fusion category symmetry and propose a classification in the case of an iterated group extension category which we explore in detail for some Tambara-Yamagami categories. In Section \ref{secstrategy}, we outline our strategy to solve the anomaly-vanishing condition from Section \ref{subsecanomvanishing}, either yielding an invertible symmetric phase or demonstrating the existence of an anomaly. We apply this strategy to the classification of gapped phases for Tambara-Yamagami categories in Section \ref{secTYphases}. We describe some applications to finite gauge theories in Section \ref{secfinitegauge}.
    
    In Section \ref{secisingdualities} we discuss fusion category symmetry arising from the well-known Kramers-Wannier duality of the critical Ising CFT. Besides Kramers-Wannier on each factor, the Ising$^2$ theory realizes two $\bZ_2 \times \bZ_2$ Tambara-Yamagami categories ${\rm Rep}(D_8)$ and ${\rm Rep}(H_8)$ symmetries (anomaly-free), as well as two $\bZ_4$ Tambara-Yamagami symmetries (anomalous).
    
    In a follow-up work \cite{toappear}, we will describe many more examples of CFTs with fusion category symmetry.

    We would like to thank Zohar Komargodski for collaborating on this work in its early stages. RT would also like to acknowledge Tsuf Lichtman, Erez Berg, Ady Stern, and Netanel Lindner for collaboration on a related project as well as Dave Aasen and Dominic Williamson for many useful discussions, Ehud Meir for patiently explaining the mathematical classification of module categories for $G$-extension categories in \cite{Meir_2012}, and especially Pavel Etingof for directing us to the notion of a fiber functor, which is central to this work. The work of YW is supported in part by the US NSF under Grant No. PHY-1620059 and by the Simons Foundation Grant No. 488653. YW would like to thank Ofer Aharony and Xi Yin for useful discussions. YW is also grateful to the Weizmann Institute of Science for hospitality where the project was initiated during his visit.
 
  		\section{Anomaly In-flow for Fusion Category Symmetry}\label{secgapbc}
  		
  		In this section, we discuss gapped boundary conditions of Turaev-Viro theory associated with a fusion category $\cA$. In a sense we will discuss, the boundary conditions of this theory have a \emph{gauged} fusion category symmetry described by $\cA$. Since our focus is on theories with global fusion category symmetry, we must spend some effort identifying symmetry-preserving vs. symmetry-breaking gapped phases in the gauge theory. We will see these correspond to free/Neumann and fixed/Dirichlet boundary conditions, respectively. Some Turaev-Viro theories do not admit a free boundary condition, and these correspond to fusion category symmetries which do not admit a symmetric deformation to a gapped, non-degenerate theory. Our identification of these TQFTs will lead us to our general anomaly-vanishing condition for these symmetries. We make contact with the usual group cohomology theory of SPT phases throughout. We will also describe a method to ``ungauge" the boundary, which reconstructs the symmetric phase from the Turaev-Viro boundary condition.
  		
  		\subsection{The Free and Fixed Boundary Conditions of Turaev-Viro Theory}
  		
  		As is well known, the (bulk) topological lines of a 2+1D TQFT are described by a modular tensor category $\cC$. This category encodes the fusion algebra of the lines as well as their braidings. Up to some mild ambiguity involving a choice of framing for the spacetime, $\cC$ determines the partition function of the TQFT \cite{reshetikhin1990}.
  		
  		Suppose our TQFT admits a gapped boundary condition $\bA$. On this boundary, there are also topological lines, which form a fusion category $\cA$.\footnote{Importantly, since these lines are constrained to live in 1+1D, $\cA$ does \emph{not} come with a braiding, and there are even examples of fusion categories which admit no braiding, such as ${\rm Vec}_G$, the category of $G$-graded vector spaces, where $G$ is a finite nonabelian group, or the $\frac{1}{2}E_6$ fusion category \cite{aasen2017fermion}.} These lines have well-defined fusion rules and an $F$-symbol (aka $6j$ symbol) which defines the crossing relation:

 	 \begin{equation}\label{eqnFmove}
 	\adjincludegraphics[width=4cm,valign=c]{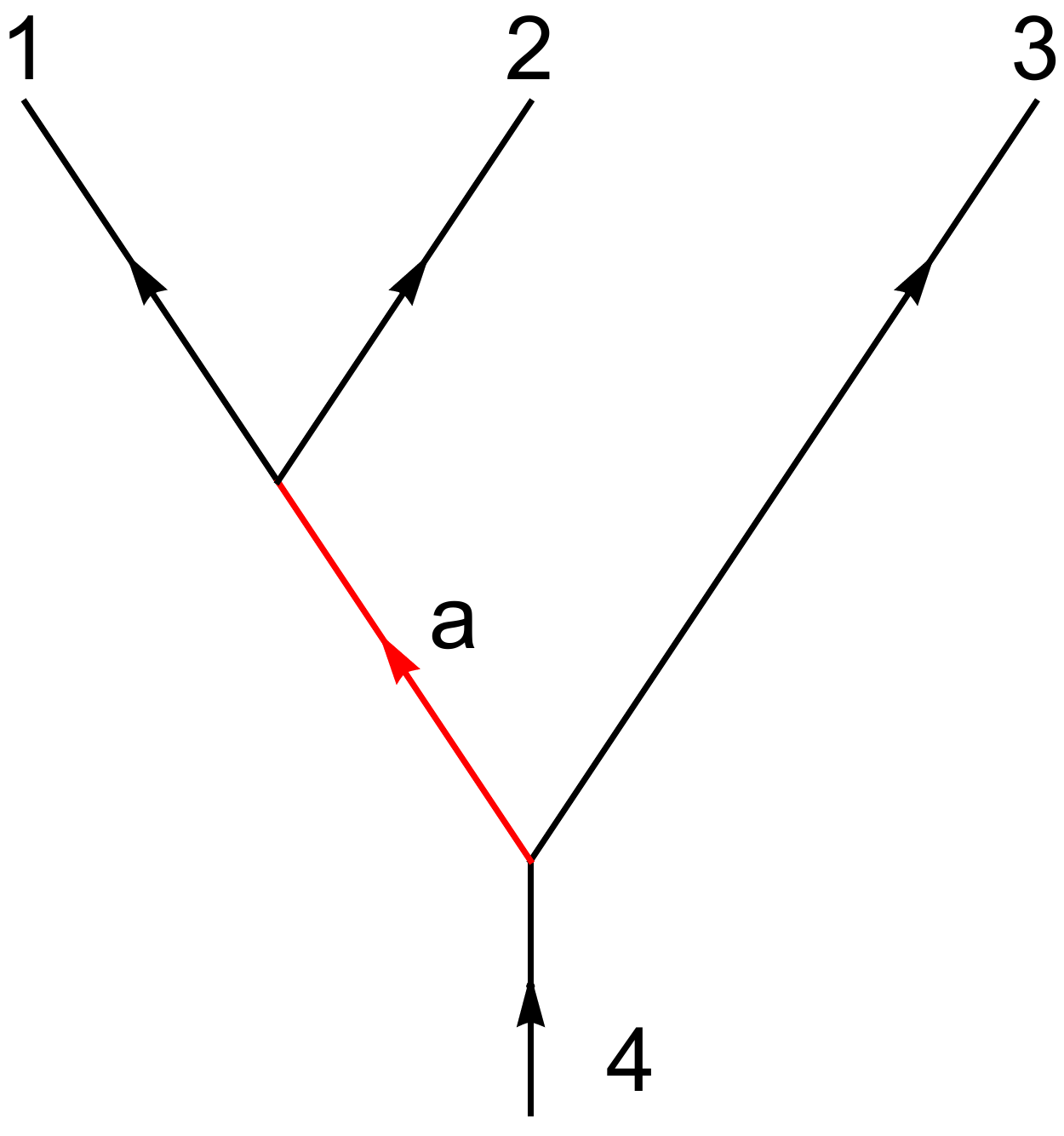} = (F^{123}_4)_{ab}\adjincludegraphics[width=4cm,valign=c]{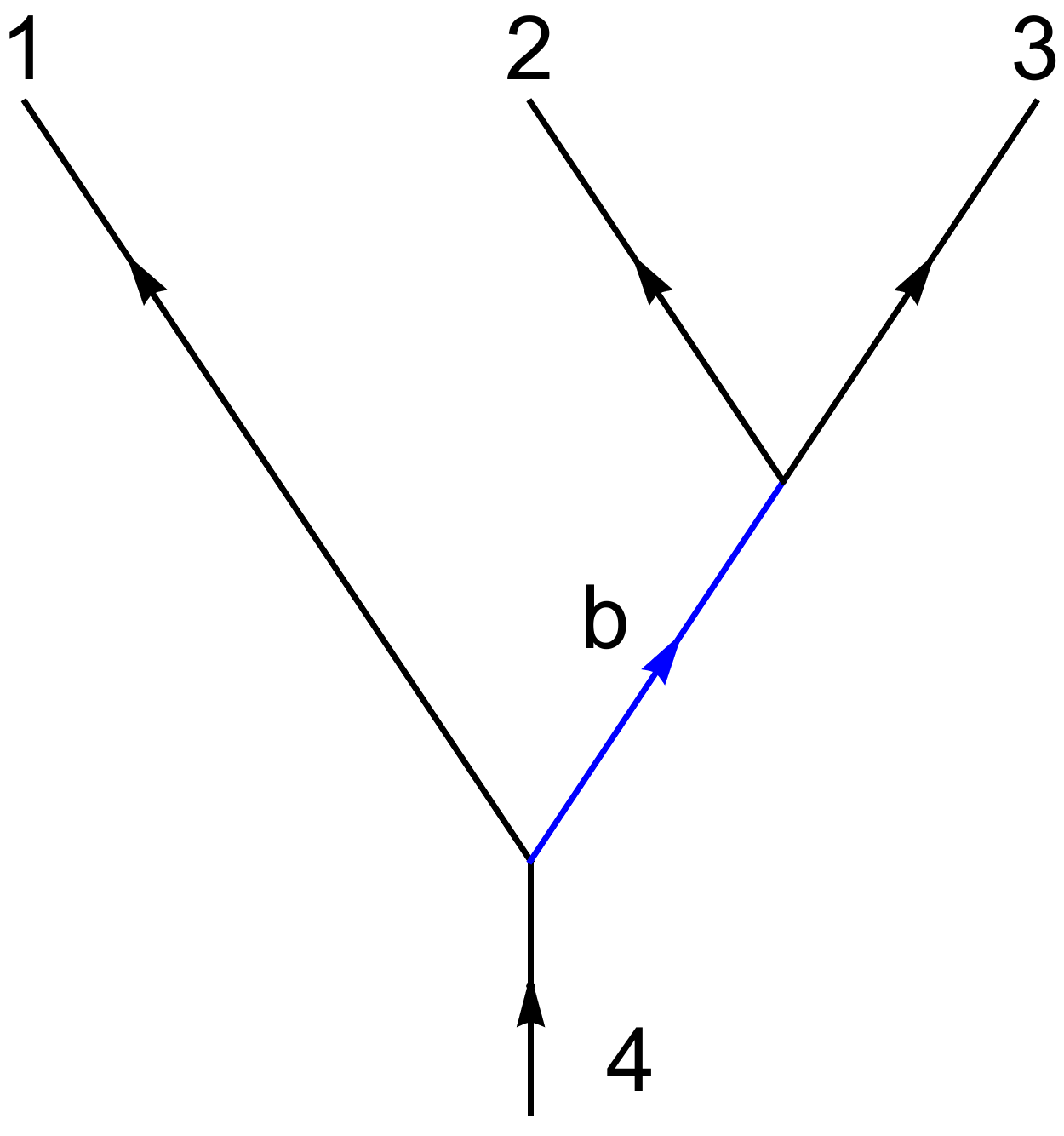}
 	 	\end{equation}
 
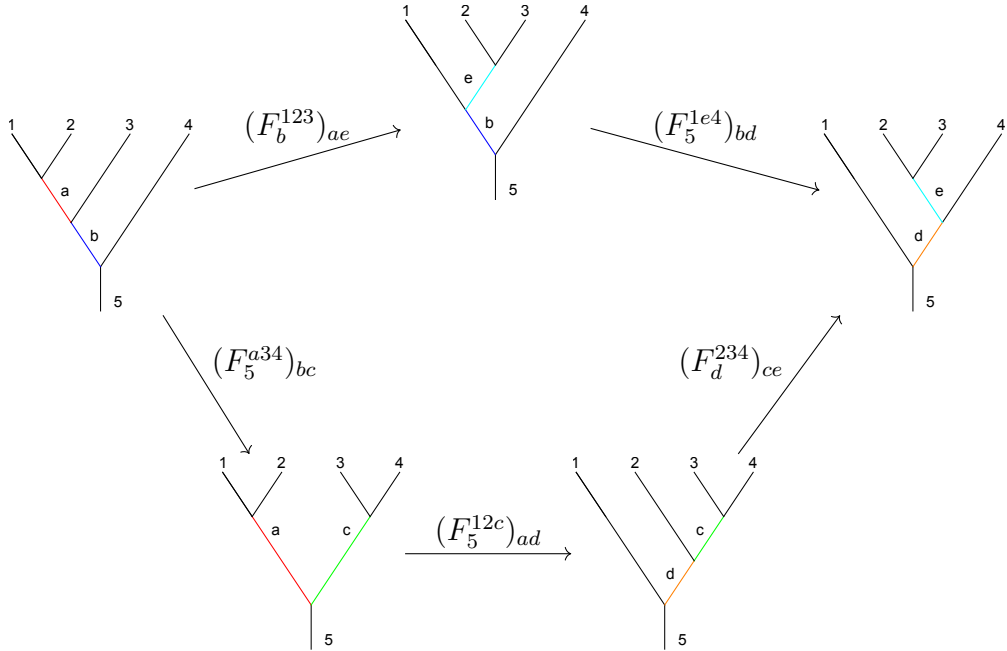
\begin{figure}[htb]
 	\centering
\begin{tikzpicture}
\node[inner sep=0pt] (A) at (0,1.5)
{\includegraphics[width=.15\textwidth]{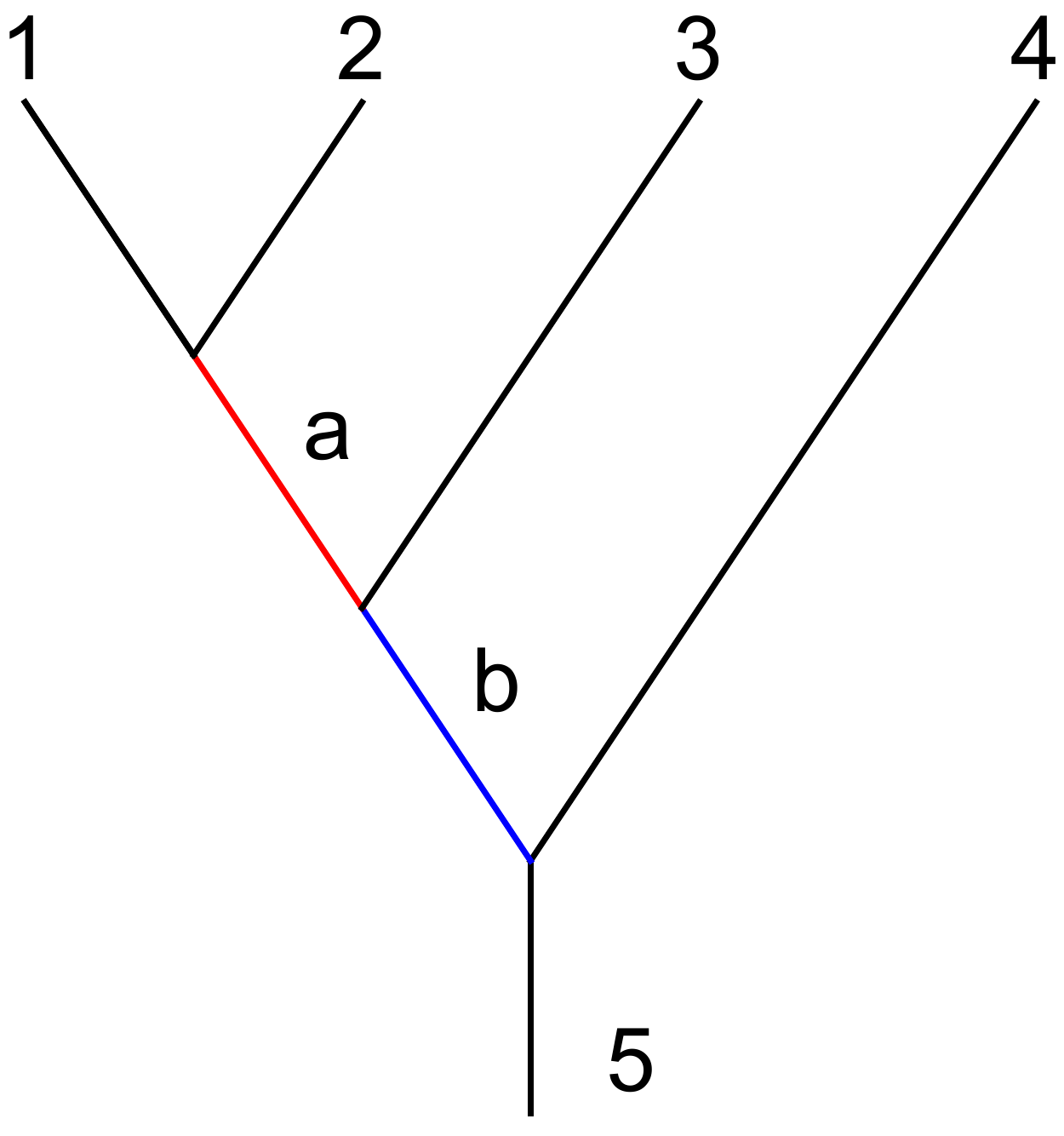}};
\node[inner sep=0pt] (B) at (2.8,-3)
{\includegraphics[width=.15\textwidth]{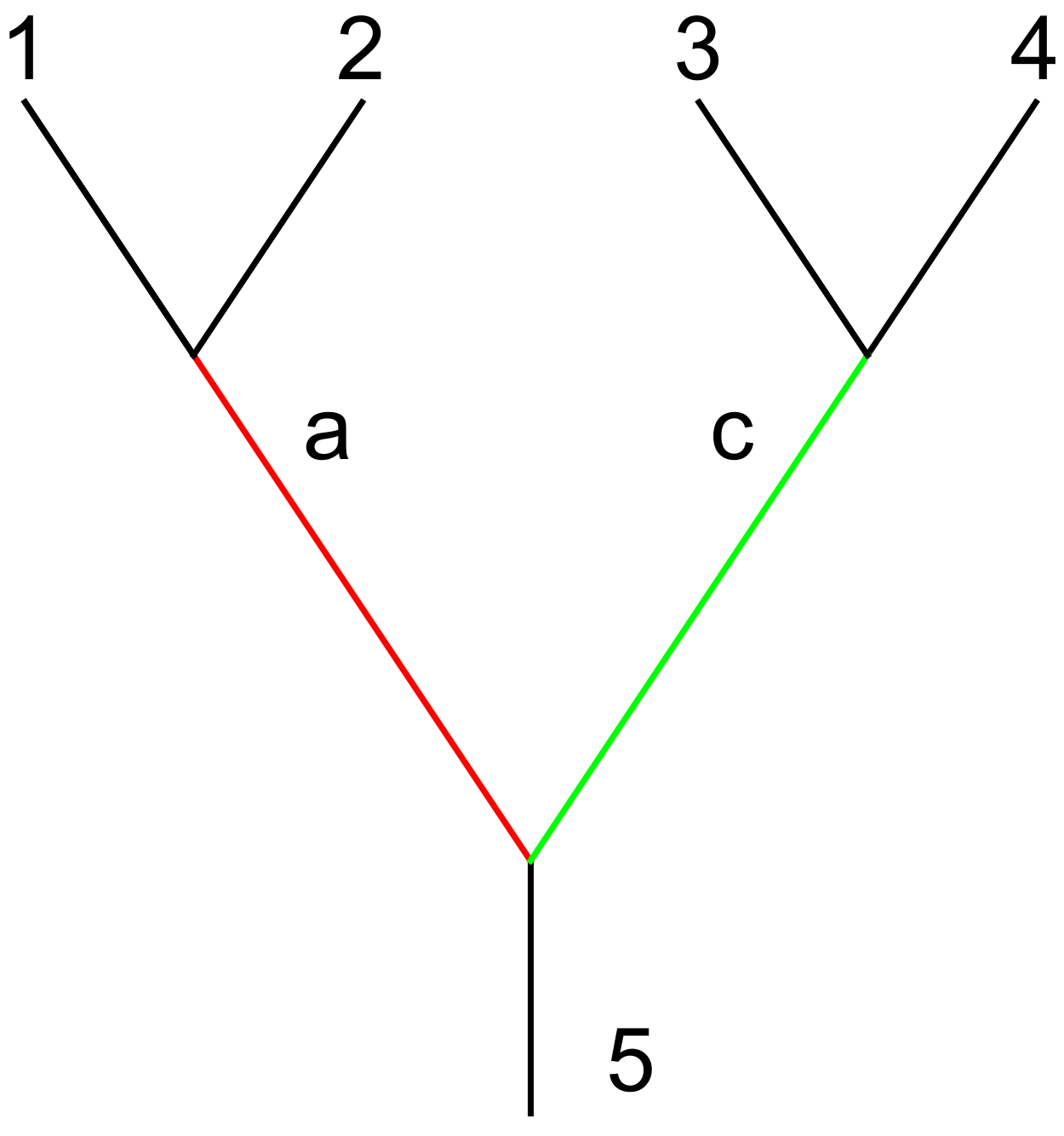}};
\node[inner sep=0pt] (C) at (7.5,-3)
{\includegraphics[width=.15\textwidth]{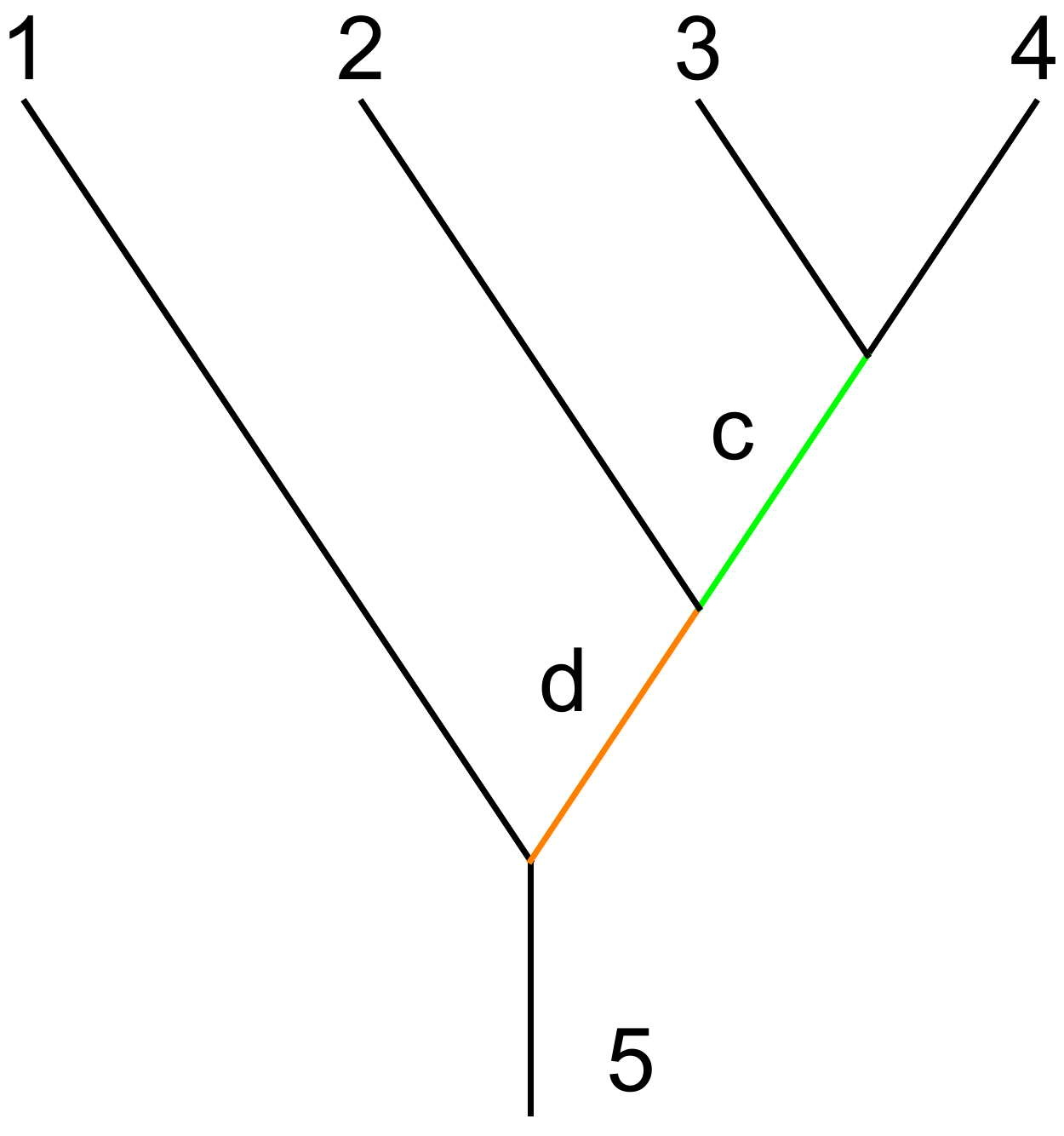}};
\node[inner sep=0pt] (D) at (10.8,1.5)
{\includegraphics[width=.15\textwidth]{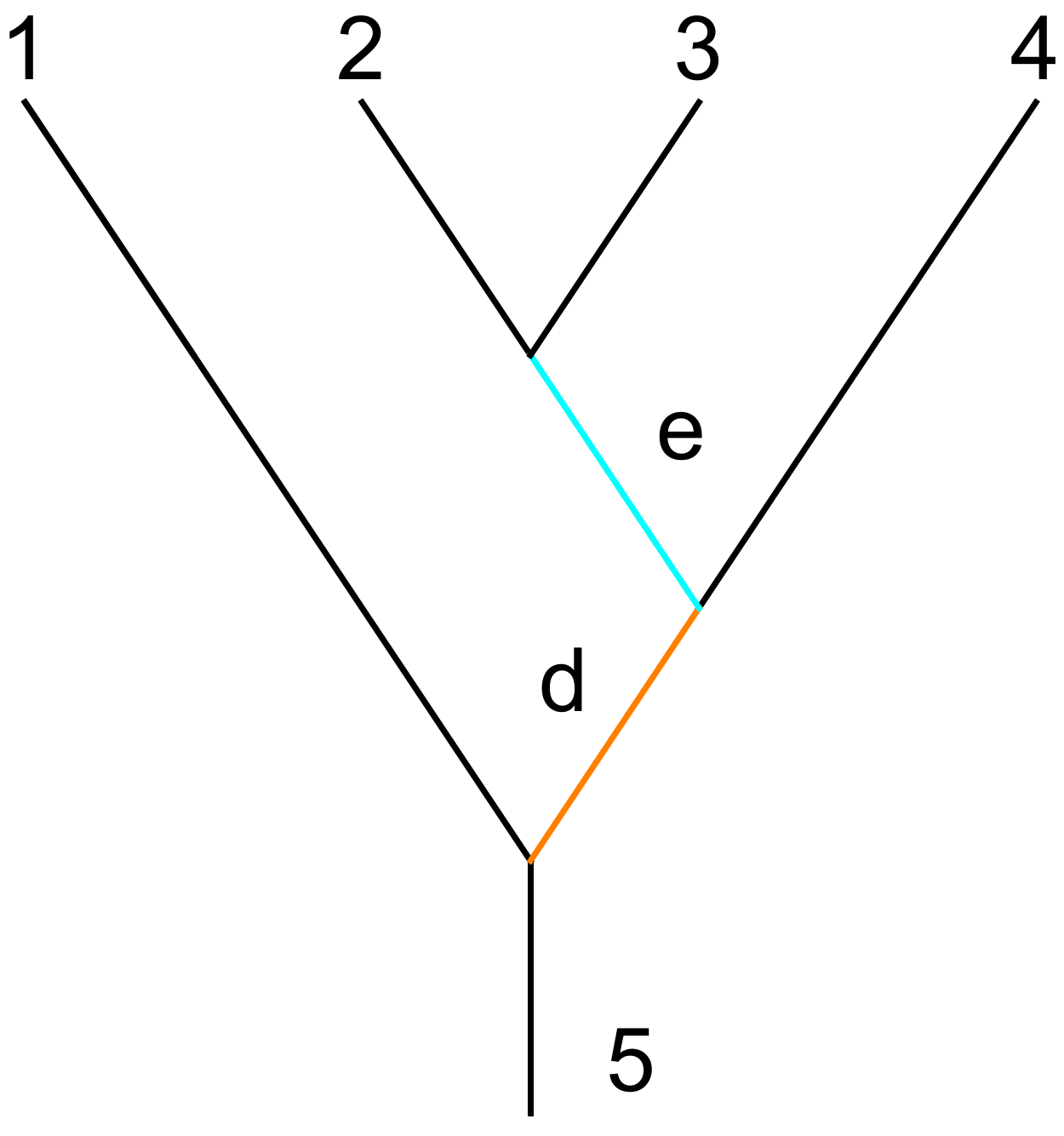}};
\node[inner sep=0pt] (E) at (5.25,3)
{\includegraphics[width=.152\textwidth]{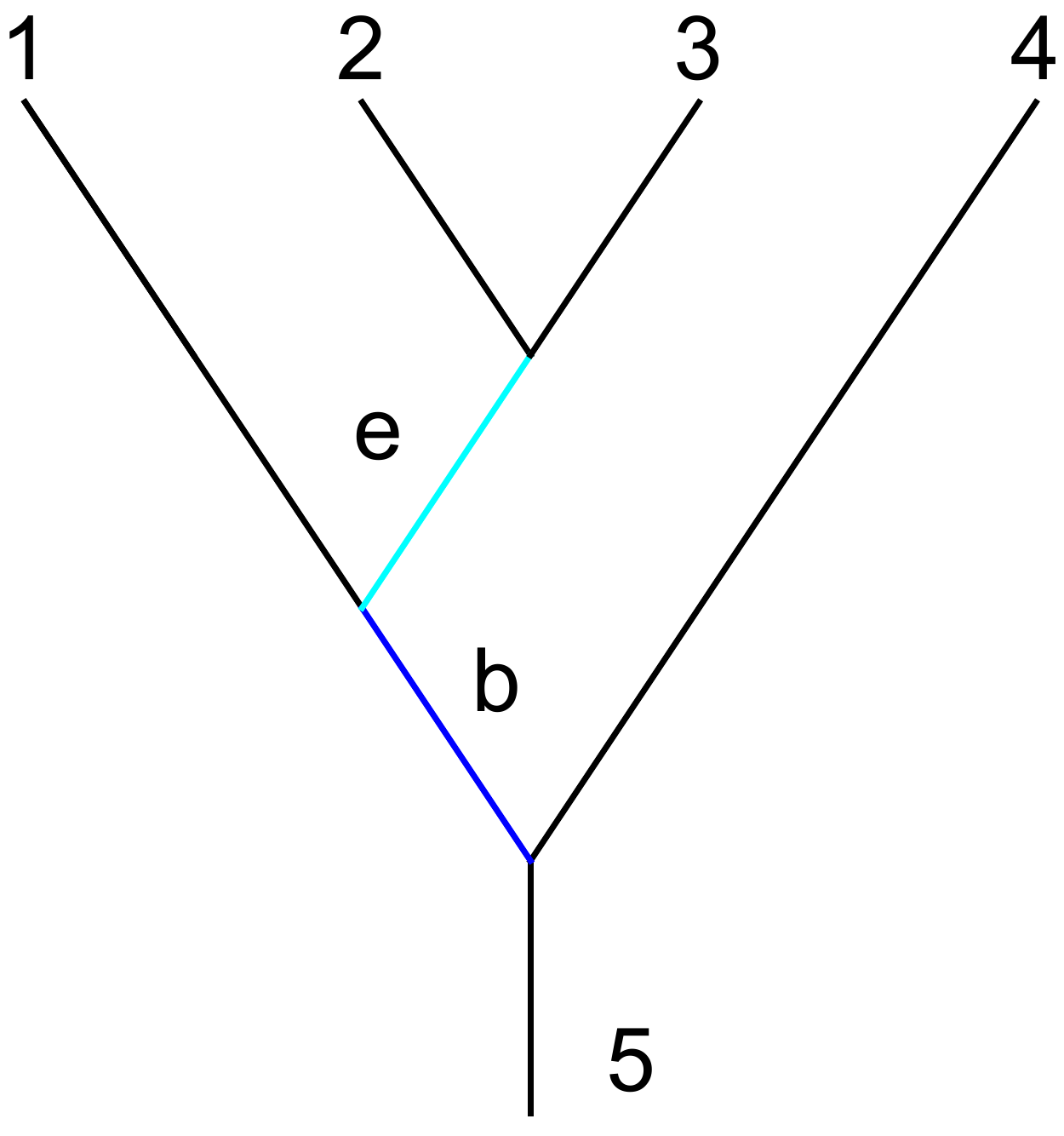}};
\draw[->] (A) to [anchor=west] node [above right=-2] {$(F^{a34}_5)_{bc}$} (B);
\draw[->] (B) to [anchor=west] node [above=-2] {$(F^{12c}_5)_{ad}$} (C);
\draw[->] (C) to [anchor=west] node [above left=-2] {$(F^{234}_d)_{ce}$} (D);
\draw[->] (A) to [anchor=west] node [above=2] {$(F^{123}_b)_{ae}$} (E);
\draw[->] (E) to [anchor=west] node [above=2] {$(F^{1e4}_5)_{bd}$} (D);
\end{tikzpicture}
\caption{Pentagon identity for the F-symbols (the label $c$ is summed over the simple TDLs). }
\label{fig:penta}
 \end{figure}

 They satisfy a consistency condition known as the pentagon equation (see Figure~\ref{fig:penta}), which is obtained from enforcing equality between two different ways of applying the crossing relations to a 5-fold junction. For more details in a closely related context, see \cite{theOGs}.
  		
\begin{figure}
  		    \centering
  		    \includegraphics[width=8cm]{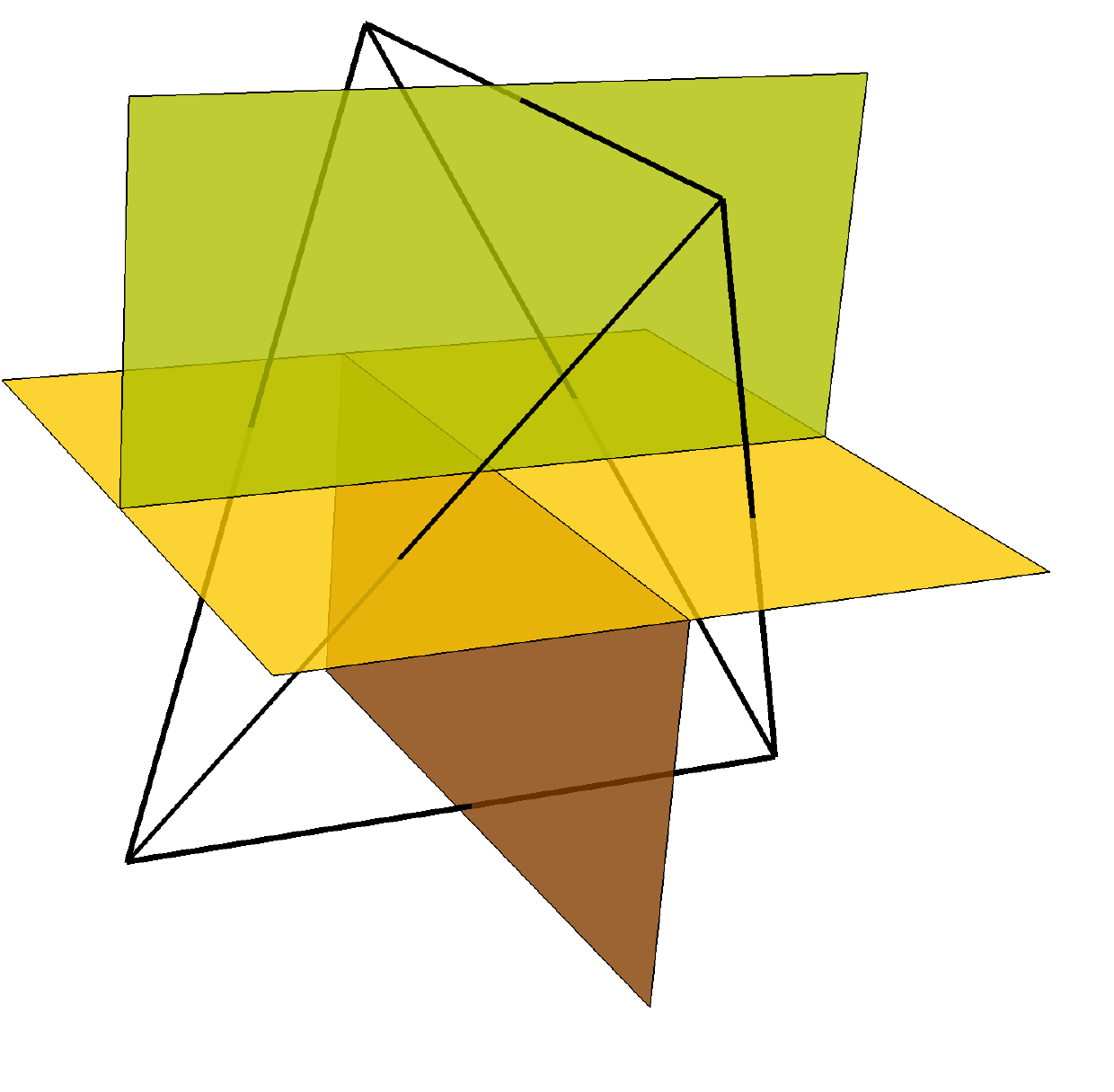}
  		    \caption{A generic 6-fold junction of surfaces. The six $\cA$ labels live along the green and brown half planes and along the four yellow quadrants. The Turaev-Viro state-sum weight is a product over all such junctions, each contributing their $F$-symbol. Fusion labels are included along the four three-fold junctions (legs of the cross on the yellow plane) if there are multiplicities. In the triangulation version of the state-sum, this point-like singularity occurs inside a Poincar\'e dual tetrahedron, drawn in black. The four three-fold junctions are dual to the triangles and the six planar regions are dual to the edges (and intersect their corresponding edge in the figure).}
  		    \label{fig6junction}
  		\end{figure}

  		The most important property of having a gapped boundary condition is that we can use it to reconstruct the entire bulk TQFT. This is done by the Turaev-Viro construction \cite{TuraevViro1992,turaev1992,alex2010turaevviro}, which produces a state-sum formula for the partition function of the TQFT. In this state-sum, we have surfaces with a co-orientation, i.e. a chosen normal direction, labelled by simple objects $a \in \cA$ and line-like 3-fold junctions of surfaces $a_1,a_2,a_3$ labelled by vectors in the fusion space. The partition function is a product of the $F$-symbol over all point-like 6-fold junctions of surfaces. See Fig. \ref{fig6junction}. This state-sum TQFT also arises from the Levin-Wen string-net Hamiltonian \cite{Levin_2005}, which is again defined from the data of the boundary fusion category.
  		
  		As an aside, the category of bulk lines with their braiding may also be reconstructed from the gapped boundary condition. Algebraically it is expressed as the Drinfeld center $\cC = Z(\cA)$. It is known that not all modular tensor categories can be expressed as Drinfeld centers and thus not all 2+1D TQFTs admit gapped boundary conditions, including some with vanishing chiral central charge. See \cite{Davydov_2013,davydov2011structure,Kapustin_2011,Kitaev_2012,Fuchs_2013,Levin_2013} for various perspectives on this problem. Intuitively, for a 2+1D TQFT to admit a gapped boundary condition, ``half" of the anyons must condense on the boundary, confining the ``other half". The simplest situation is where our fusion category is actually braided, and $Z(\cA)$ looks like $\cA \times \bar \cA$, where $\bar \cA$ is $\cA$ with the opposite braiding. For this reason, the theories are also called Drinfeld or quantum doubles. We will only be interested in such TQFTs in this work.
  		
  		In the Turaev-Viro state-sum based on $\cA$, the boundary condition $\bA$ re-appears as the ``fixed boundary condition", defined by restricting the state-sum so that only the ``invisible" surfaces, carrying the label of the vacuum object $\mathbbm{1}$, meet the boundary. This modification produces a topological invariant partition function for manifolds with boundary.
  		
  		In the fixed boundary condition, the boundary defect lines $a \in \cA$ are realized by modifying the condition on boundary edge labels along a curve $\gamma$ so that a bulk surface with label $a$ always ends on $\gamma$ in the state-sum. Then one can compute correlation functions of these boundary operators using the modified state-sum.
  		
\begin{figure}
  		    \centering
  		    \includegraphics[width=5cm]{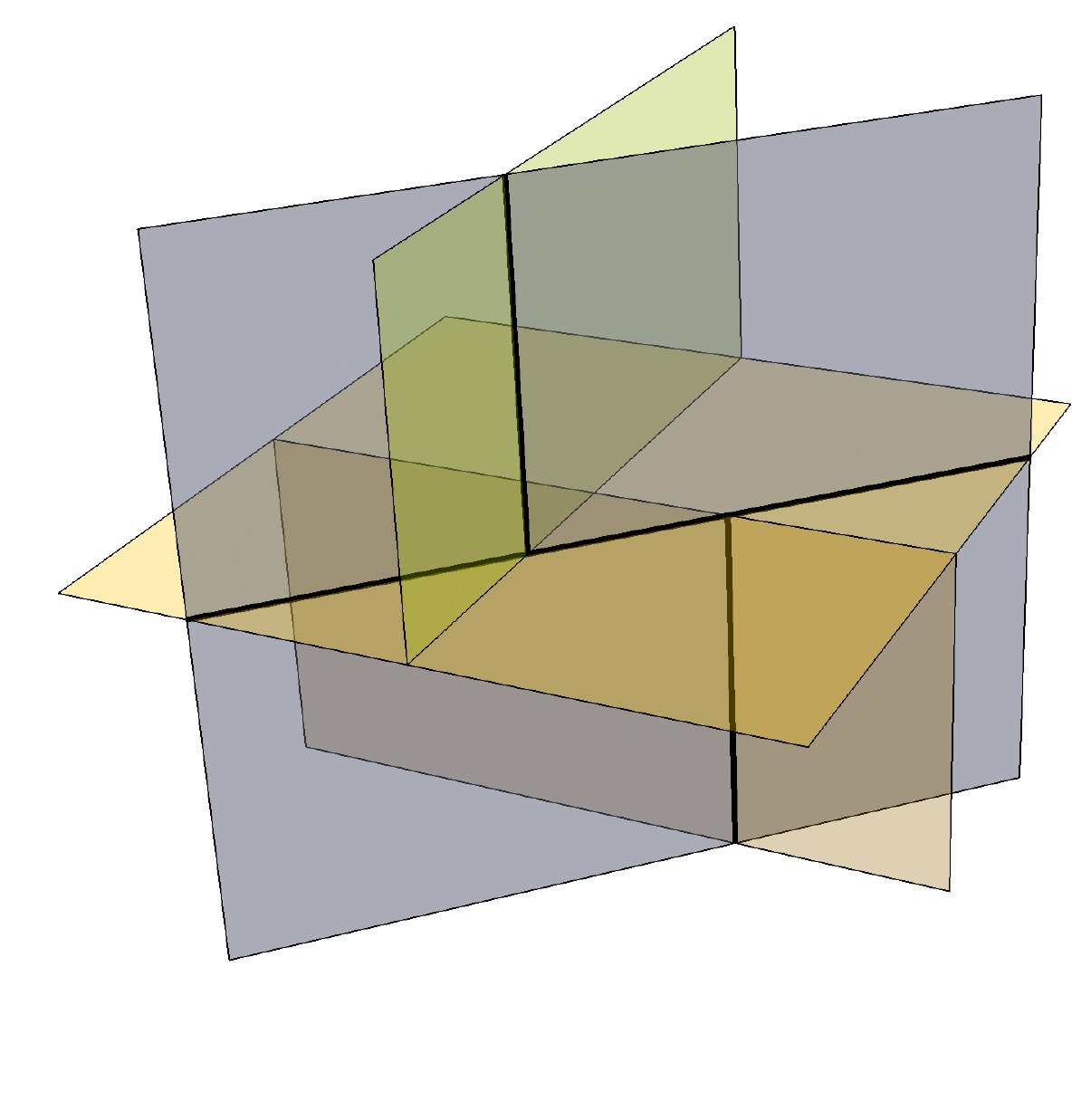}
  		    \includegraphics[width=5cm]{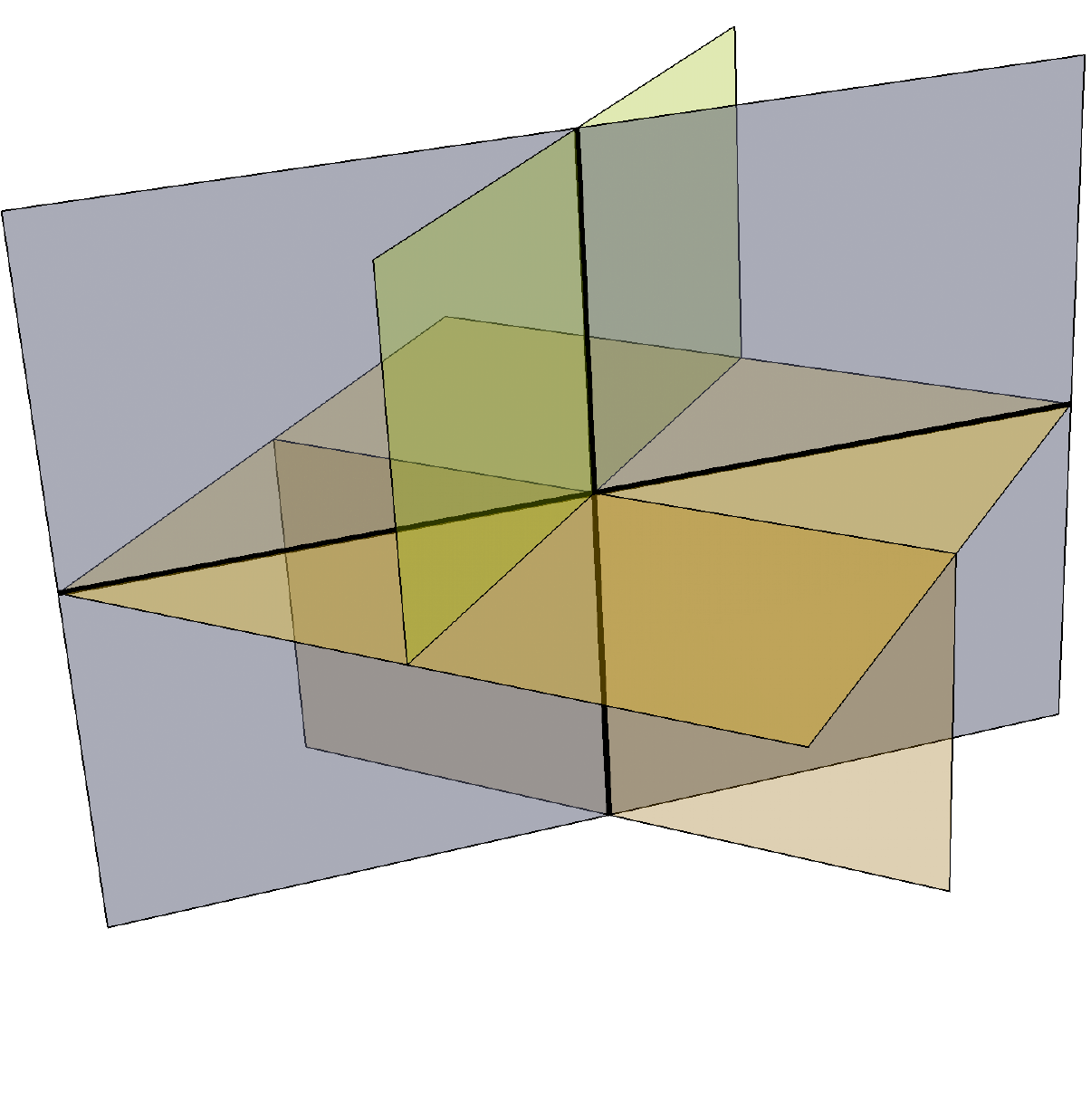}
  		    \includegraphics[width=5cm]{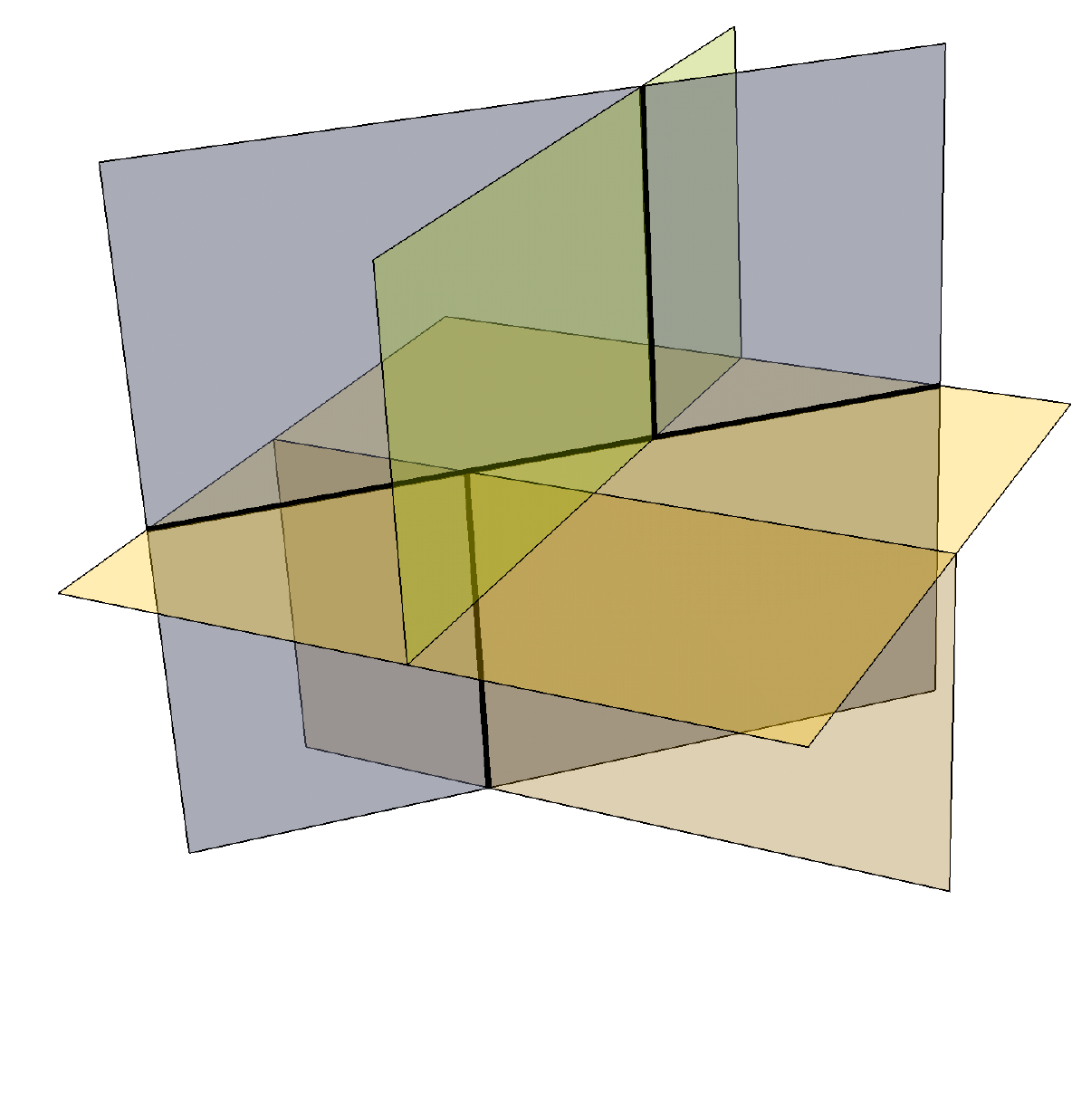}
  		    \caption{When the boundary (grey plane) absorbs a six-fold junction (see Fig. \ref{fig6junction}), the boundary lines (black) perform an $F$-move (compare Eq. \eqref{eqnFmove}). This is also a picture of the pentagon equation for module categories (which classify general gapped boundary conditions, see Section \ref{subsecanomvanishing}) where the left picture contributes two $\mu$'s, the center picture contributes an $F$, and the right picture contributes two more $\mu$'s.}
  		    \label{figboundaryFmove}
  		\end{figure}

  		In contrast one could attempt to define a ``free boundary condition" with an unconstrained state-sum at the boundary. However, the state-sum weight will no longer be a homotopy invariant of the surface configurations, because now the boundary could absorb or emit a 6-fold junction, changing it by a factor of the corresponding $F$-symbol. See Fig. \ref{figboundaryFmove}. We will study below when this ``anomaly" can be cured by a local counter-term on the boundary.
  		
  		\subsection{The Free and Fixed BCs of Dijkgraaf-Witten Theory and Symmetry Breaking}
  		
  		In this section, we argue that the \textit{fixed} boundary condition of the Turaev-Viro state-sum arises from a gapped 1+1D theory where the $\cA$ symmetry is spontaneously completely broken in the ground state. To see this, we first draw an analogy with discrete gauge theory in 2+1D, aka Dijkgraaf-Witten theory.
  		
  		If $G$ is a finite group and $\omega:G^{\times 4} \to \bR/2\pi\bZ$ is a (homogeneous) 3-cocycle \cite{brown2012cohomology}, we can construct a fusion category ${\rm Vec}_G^\omega$, known as the $\omega$-twisted category of $G$-graded vector spaces, which has a simple object for every element $g \in G$ and fusion rules defined by the group multiplication, with one-dimensional fusion spaces. The $F$-symbol with external legs given by $g_1,g_2,g_3,g_4$ has one component since fusion trees are determined by the group law, and this component equals $\exp i\omega(g_1,g_2,g_3,g_4)$.
  		
  		If we feed ${\rm Vec}_G^\omega$ into the Turaev-Viro construction we obtain a theory of co-oriented surfaces labelled by elements of $G$ with a 3-fold junction constraint. We can represent such a configuration in Poincar\'e duality as a $G$-valued 1-cocycle $A \in Z^1(X,G)$, which is equivalently a flat $G$ gauge field (see \cite{kapustin2014anomalies}).
  		
  		Furthermore, considering that the Turaev-Viro state-sum weight is a product over all the $F$-symbols, using the short-hand $\omega(A)$ for the sum of Dirac deltas supported at the 6-fold junctions, each with coefficient $\omega(g_1,g_2,g_3,g_4)$ determined by the surfaces meeting there, we can write the partition function as
  		\begin{equation}Z(X) = \# \sum_{A \in Z^1(X,G)} \exp \left (i \int_X \omega(A) \right),\end{equation}
  		where there is an arbitrary overall normalization. We recognize this as 3D Dijkgraaf-Witten theory defined by the class $[\omega] \in H^3(BG,U(1))$ \cite{dijkgraaf1990}.
  		
  		We see in this description that the fixed boundary condition is the Dirichlet boundary condition
  		\begin{equation}A|_{\partial X} = 0.\end{equation}
  		A gapped spontaneous symmetry-breaking boundary condition flows precisely to this boundary condition in the IR. Indeed, such ground states are characterized by tensionful domain walls where the $g$-surfaces meet the boundary. In the IR where this tension becomes infinite, it forces us into the fixed boundary condition.
  		
  		We can also  study the \textit{free} boundary condition, where $A$ is unconstrained at the boundary. However, as is well-known, the Dijkgraaf-Witten action is not gauge-invariant on spacetimes with boundary. Intuitively, to have gauge-invariance means that we can freely move the surfaces, so long as we do it locally and we preserve the constraint on the 3-fold junctions. In this process, with the free boundary condition, a 6-fold junction may be pushed ``through" the boundary and out of the spacetime, causing a shift of the Turaev-Viro weight by the $e^{i\omega}$ factor associated with that junction.
  		
  		One can cancel this factor by a local counterterm if and only if $\omega$ is an exact cocycle, i.e. $\omega = -\delta \chi$ for some $\chi:G^{\times 3} 
  		\to \bR/2\pi\bZ$ a function of the 3-fold junctions, where $\delta$ is the group coboundary, 
  		\begin{equation}\label{countertermeqn}
  		    \delta\chi(g_1,g_2,g_3,g_4) := \chi(g_2,g_3,g_4) - \chi(g_1,g_3,g_4) + \chi(g_1,g_2,g_4) - \chi(g_1,g_2,g_3) = \omega(g_1,g_2,g_3,g_4).
  		\end{equation}
  		This equation means that if we add up all the $\chi$'s of 3-fold junctions around a 6-fold singularity, keeping care of the co-orientations, the result is $-\omega$ of the 6-fold junction. Then, writing $\chi(A)$ to mean the sum of Dirac deltas at the 3-fold junctions on the boundary $\partial X$ weighted by $\chi$, analogous to $\omega(A)$, we have the gauge invariant, topological partition function
  		\begin{equation}\label{eqnDWtrivboundary}
  		Z(X) = \# \sum_{A \in Z^1(X,G)} \exp \left(
  		i \int_X \omega(A) + i \int_{\partial X} \chi(A)
  		\right).
    		\end{equation}
  		Indeed, one can see geometrically that when the 6-fold singularity with labels $g_1,g_2,g_3,g_4$ is pushed through the boundary, as in Fig. \ref{figboundaryFmove}, the $G$-lines on the boundary perform an $F$-move, and we get a compensating shift by $\exp i \delta \chi(g_1,g_2,g_3,g_4)$. Thus, we obtain a topological and gauge-invariant partition function.
  		
  		It is easy to derive a converse statement for counterterms of this form, i.e. that they must satisfy $\delta \chi = - \omega$ to be gauge invariant. However, one must still show that they are the most general form for such a counterterm. We will discuss the proof in the next subsection, and the converse will follow again from the general anomaly-vanishing condition we develop in Section \ref{subsecanomvanishing}.

  		On the other hand, if we had a gapped, non-degenerate, symmetry-preserving (``trivial") boundary condition of Dijkgraaf-Witten theory, it would flow to some kind of free boundary condition since the domain walls are tensionless, hence one where $A$ is unconstrained. We see that no such boundary condition can exist when $\omega$ is not exact, i.e. when it is nonzero in group cohomology $[\omega] \neq 0 \in H^3(BG,U(1))$.
  		
  		It follows that if some 1+1D theory does define a gauge-invariant boundary condition for this 2+1D gauge theory, then $G$-symmetric deformations of this theory produce other such boundary conditions, and since there is no trivial boundary condition, the theory has no trivial $G$-symmetric deformations. This is the essence of the 't Hooft anomaly for $G$ symmetry.
  		
  		We comment on a certain subtlety of such anomaly-matching arguments. We assumed that if one couples to the bulk theory and then computes the RG flow, one obtains the same result as first computing the RG flow and then coupling to the bulk theory. There are two aspects of this: (1) the coupling to the bulk does not affect the renormalization of the boundary couplings, and (2) the renormalization in the presence of the boundary does not move the bulk theory off of its fixed point. The first point follows from the fact that our perturbation enjoys the global symmetry, meaning that the insertion of topological lines do not affect the computation of the RG flow. Performing a finite sum over such insertions also clearly can be done before or after the flow without any discrepancy. The second point follows from cluster decomposition. Note that these arguments do not rely on being able to ``ungauge" the symmetry.
  		
  		
  		\subsection{The Anomaly-Vanishing Condition}\label{subsecanomvanishing}
  		
  		To proceed further, we will need to systematically understand the topological boundary conditions of Turaev-Viro theory (hence also the IR-fixed points of its gapped boundary conditions), and see if something like the free boundary condition is among them.
  		
  		We use the techniques of \cite{Fuchs_2013}. Suppose we have two topological boundary conditions $\bA$ and $\bB$. We can study the set of topological line junctions between them, denoted
  		\begin{equation}\hom(\bA,\bB) = \{{\rm topological\ line\ junctions\ from\ }\bA{\rm \ to\ }\bB\}.\end{equation}
  		Topological line junctions can be composed by fusion. Further, they have topological point junctions between them which also fuse and therefore $\hom(\bA,\bB)$ forms a category \cite{Kapustin_ICM}.
  		
  		As we have discussed, for any given topological boundary condition $\bA$,
  		\begin{equation}\hom(\bA,\bA) = \cA\end{equation}
  		forms a fusion category. From now on we take $\bA$ to be the \textit{fixed} boundary condition so that $\cA$ defined above is our fusion category of interest.
  		
  		The fusion of line junctions defines an action of $\cA$ on $\hom(\bA,\bB)$ by
  		\begin{equation}\hom(\bA,\bA) \times \hom(\bA,\bB) \to \hom(\bA,\bB).\end{equation}
  		One says that $\hom(\bA,\bB)$ forms a \emph{module category} over $\cA$. We will give a precise definition in a moment. For now let it suffice to say that a module category is generated by a finite list of simple objects $m \in \cM$ such that for any $a \in \cA$ we can define
  		\ie\label{eqnmodcat} a \cdot m = \sum_{m'} V_{am}^{m'} m' \in \cM\fe
  		for some vector spaces $V_{am}^{m'}$ compatible with fusion.\footnote{The dimensions of these vector spaces after summing over $m,m'$ give the $W$-vectors of \cite{Lan_2015}. More general gapped domain walls are given by bimodule categories \cite{Fuchs_2013} which define a $W$ matrix.}
  		
  		It was shown in \cite{Fuchs_2013} that the category $\hom(\bA,\bB)$ along with its action of $\cA$ actually \emph{determines} the boundary condition $\bB$. Thus, we can classify all topological boundary conditions of the Turaev-Viro theory constructed from $\cA$ by classifying module categories over $\cA$. For example, the fusion category $\cA$ is a module category over itself and this module category corresponds to the fixed boundary condition $\bA$.
  		
  		Now let us complete the definition of a module category $\cM$ by seeing how one can define a topological boundary condition of the Turaev-Viro state-sum. First, we extend our labellings to spacetimes with boundary by giving the boundary triangles labels by simple objects in $\cM$. Line junctions at the boundary consist of two boundary triangles, labelled by simples $m,m' \in \cM$, with a bulk triangle, labelled by $a \in \cA$, in between. We label the line junction with a basis vector of $V_{am}^{m'}$, which was defined in \eqref{eqnmodcat}. In the case $\cM = \cA$, these are ordinary fusion junctions, with $V_{ab}^c = \hom(a \otimes b,c)$, and we have the picture that $\cA$ surfaces can pile up at the boundary, but cannot end there without creating a non-trivial defect.
  		
  		To define the boundary contribution to the state-sum weight, we need to assign a value to a point where a three-fold fusion junction meets the boundary. There we have the data of a fusion vertex $v \in \hom(a \otimes b,c)$, objects $m,m',m''$, and $w \in V_{am}^{m''}$, $w' \in V_{am''}^{m'}$, so we want an $F$-symbol-like object
  		 	 \begin{equation}\label{eqnmodcatF}
 	\adjincludegraphics[width=4cm,valign=c]{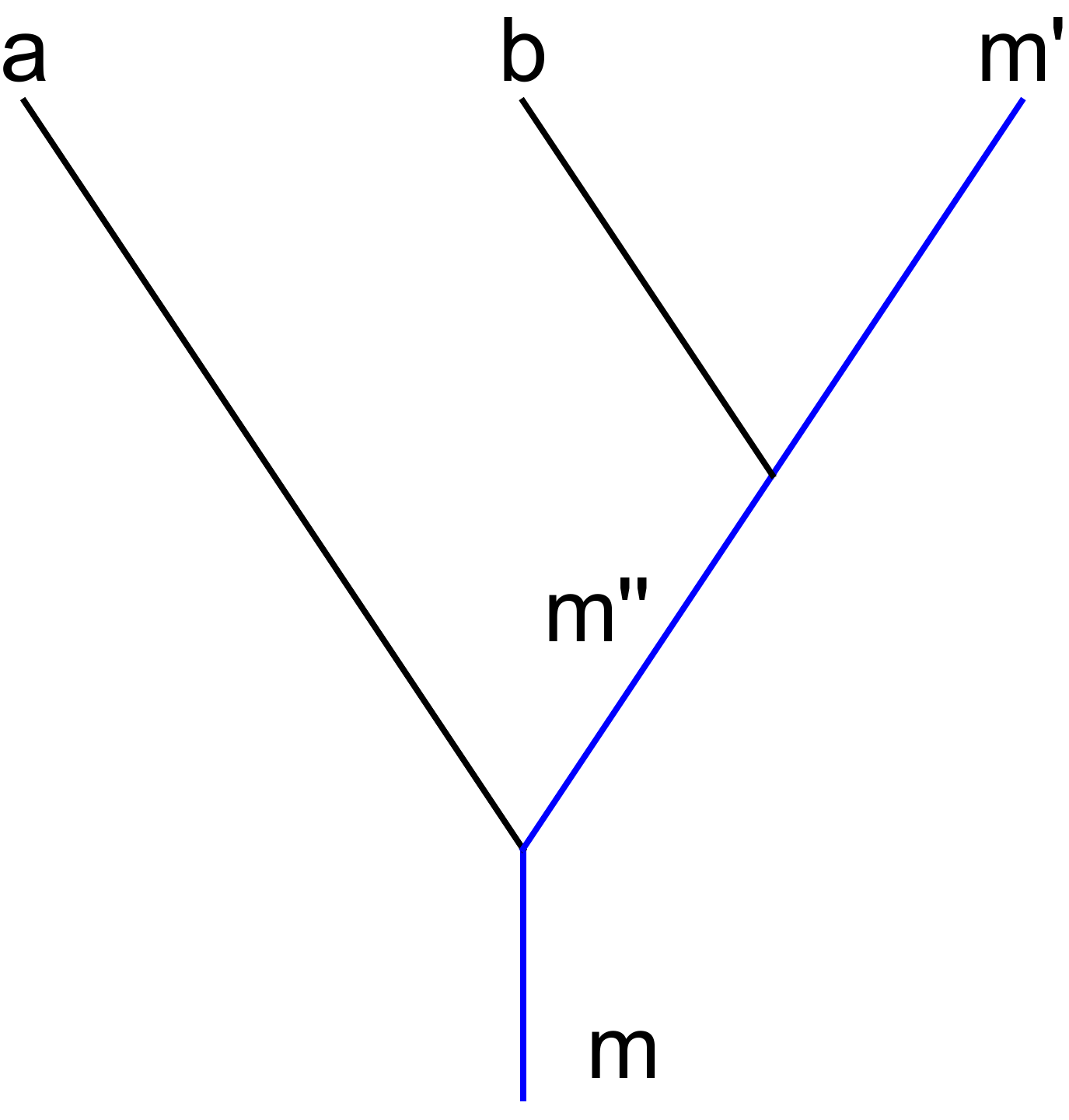} = \sum_c (\mu_{ab}^c)_{mm'm''}\adjincludegraphics[width=4cm,valign=c]{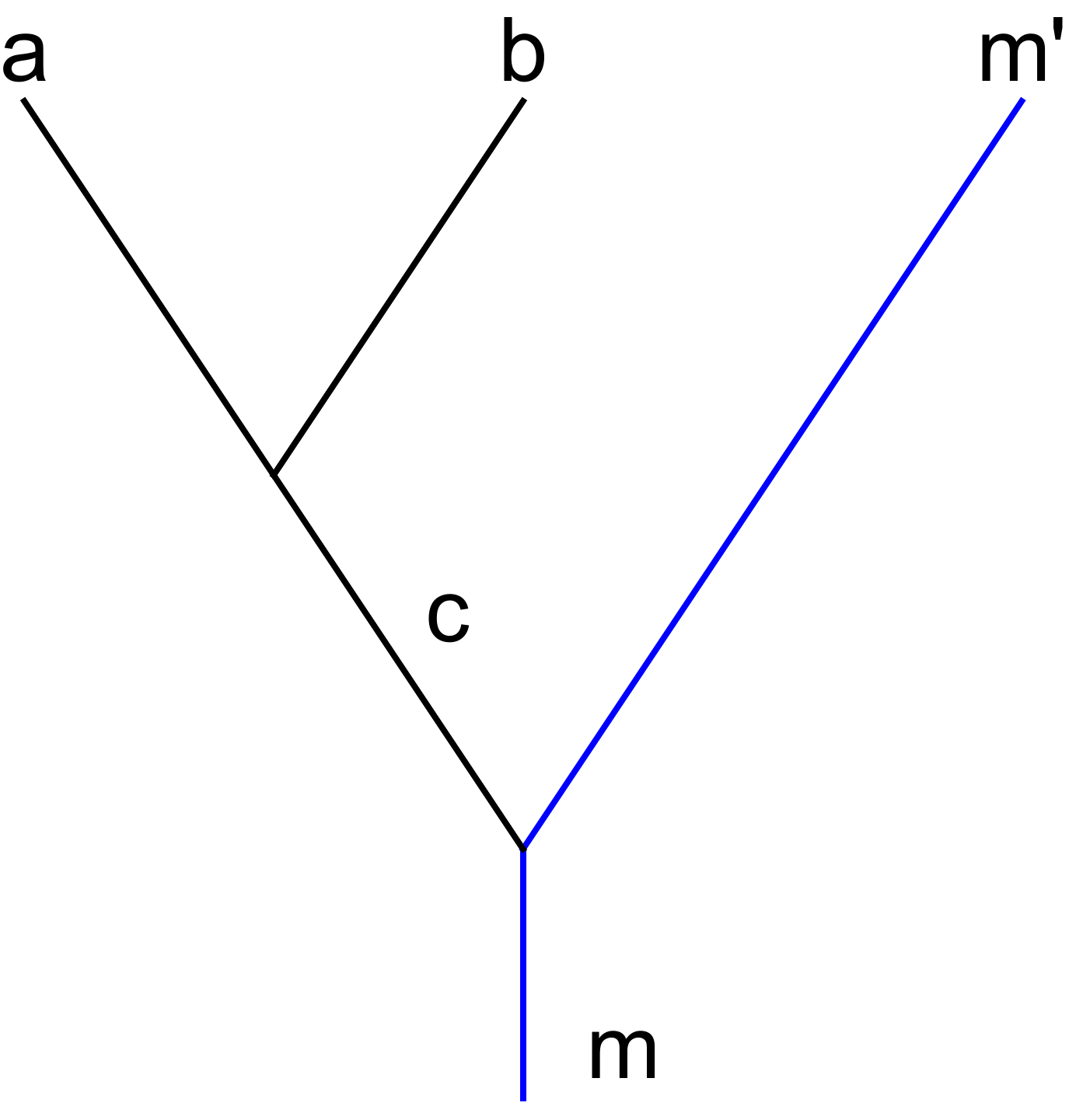},
 	 	\end{equation}
  		where we draw our $\cM$ objects in blue and our $\cA$ objects in black. The boundary contributes a product over all these $\mu$'s. Topological invariance is equivalent to the corresponding pentagon equation for these objects, analogous to Fig. \ref{fig:penta} and Fig. \ref{figboundaryFmove}. The data of the $V_{am}^{m'}$ and the $\mu_{ab}^c$ satisfying the corresponding pentagon equation defines a module category $\cM$. See \cite{Kitaev_2012} for more details. For $\cM = \cA$, $\mu$ is just the $F$-symbol and we recover the fixed boundary condition.
  		
  		What would a topological boundary condition $\bB$ which preserves the $\cA$ symmetry look like in this picture? Such a boundary condition would allow all bulk surfaces to end on it, and such boundary configurations of lines fluctuate freely in the ground state wavefunction, i.e. they are condensed. Meanwhile, the action of $\cA$ on $\hom(\bA,\bB)$ starts in the $\bA$ boundary condition by enforcing an $a$-surface to end on a line $\gamma$ and then fusing that line with a given junction $J \in \hom(\bA,\bB)$.
  		
  		\begin{figure}
  		    \centering
  		    \includegraphics[width=7cm]{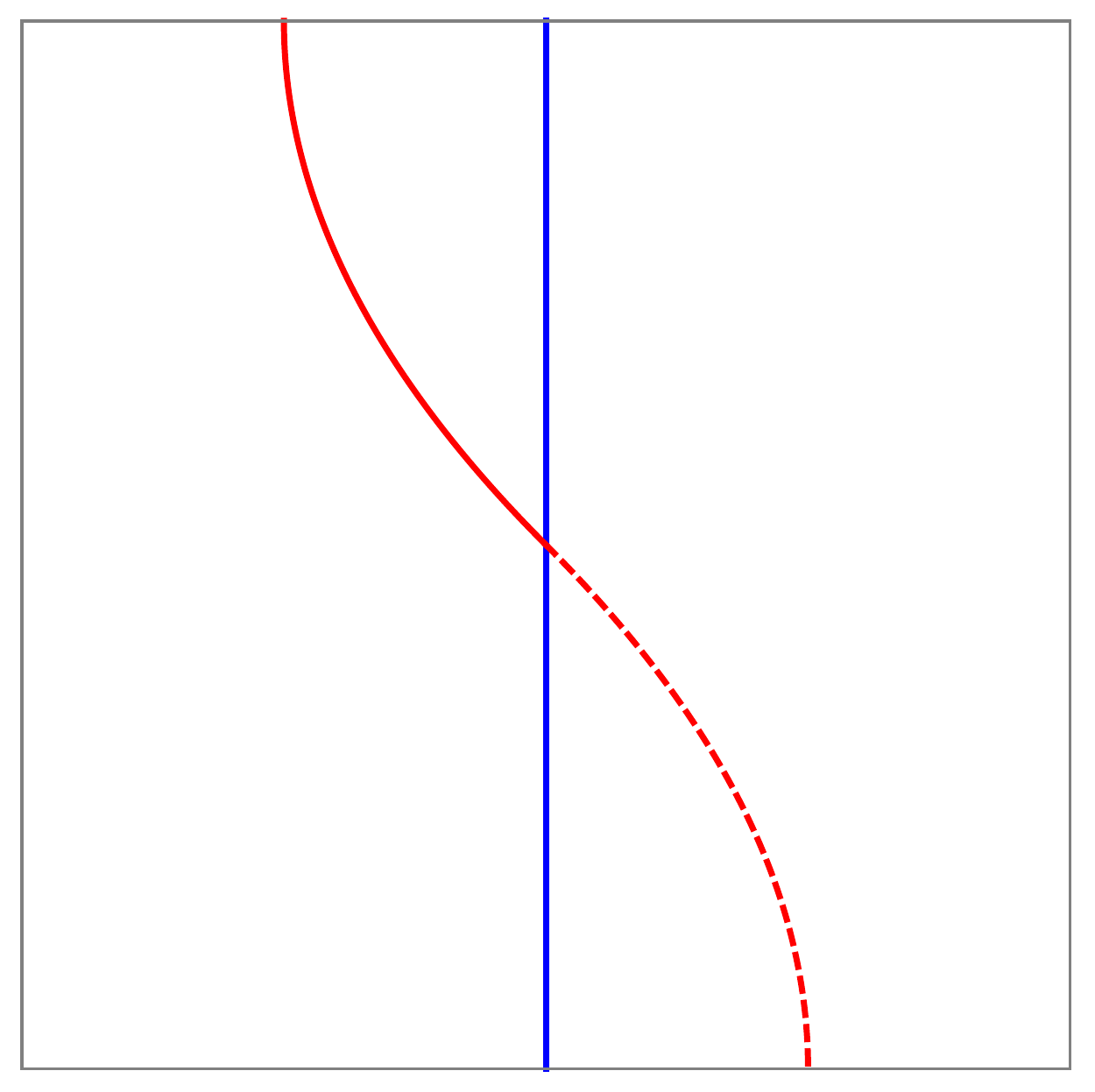}
  		    \caption{An $\cA$ defect $a$ (red curve) passes from the fixed boundary condition on the left to a symmetry-preserving boundary condition on the right, where it becomes invisible. The crossing point gives a map from $a \otimes J \to J$, where $J$ is any boundary-changing junction from the fixed boundary condition to the symmetry-preserving one (blue curve). This proves that the associated module category ${\rm hom}(\bA,\bB)$ of these boundary-changing junctions has one simple object.}
  		    \label{figfiberfuncproof}
  		\end{figure}
  		
  		By drawing $\gamma$ to cross the junction transversely, the line operator becomes absorbed in the condensate of such boundary strings in the $\bB$ side (see Fig. \ref{figfiberfuncproof}). The crossing point thus becomes a morphism $a \otimes J \to J$, so $J$ on its own generates a sub-module category with one simple object, namely $J$. To summarize, we have
  		
  		\begin{thm}\label{thmanomfiberfunc}
  			The Turaev-Viro theory defined by the fusion category $\cA$ admits a \emph{gapped, non-degenerate $\cA$-symmetric boundary condition} iff $\cA$ admits a module category with one simple object, or equivalently a \emph{fiber functor}.
  		\end{thm}
  		
  		If a fusion category does not have a module category with one simple object, we say it is \emph{anomalous}. We will give another proof in the next section by arguing that $\cA$-symmetric gapped phases in general correspond to $\cA$-module categories $\cM$ such that the ground state degeneracy on a circle is the number of simple objects of $\cM$. Specializing to nondegenerate phases yields the theorem.
  		
  		The equivalence in the last clause of the theorem follows from \cite{etingof2015tensor}, example 7.4.6, where it is shown that in general, $\cA$-module categories with one simple object correspond to tensor functors
  		\begin{equation}F:\cA \to {\rm Vec}_\bC,\end{equation}
  		which are known as \emph{fiber functors}, where ${\rm Vec}_\bC$ is the usual tensor category of complex vector spaces, which has the single simple object $\bC$. This can basically be derived by dropping all the $m$ labels in our definition of a module category above, since they are all equal to the unique simple object. Concretely, a fiber functor is an assignment of a vector space  
  		\begin{equation}V_a \quad {\rm \ for\ each\ } a \in \cA,\end{equation}
  		as well as a collection of maps
  		\begin{equation}\mu_{ab}^c(v):V_a \otimes V_b \to V_c\end{equation}
  		for each fusion vertex
  		\begin{equation}v \in \hom(a \otimes b, c),\end{equation}
  		such that $\mu_{ab}^c(v)$ is linear in $v$,
  		\begin{equation}\bigoplus_{c \in a \otimes b} \mu_{ab}^c:V_a \otimes V_b \to \bigoplus_{c \in a\otimes b} V_c\end{equation}
  		is an isomorphism, and satisfies the associativity equation (after proper dualization)
  		\ie\label{eqnfiberfuncpent}
  		     \mu_{ab}^{1} \mu_{cd}^{1} = \sum_2 (F^{bad}_{c})_{12}\mu_{ad}^2\mu_{bc}^2,
  		\fe
  		compare \eqref{eqnfiberfunccrossing}.

  	    Physically, the fiber functor describes an invertible 1+1d TQFT with simple TDLs $\cL_a$ and a defect Hilbert space  
  	\ie
  	\cH_a \equiv  V_a
  	\fe
  	In particular the dimension of $V_a$ is the quantum dimension $\la \cL_a\ra$.
  	The maps $\mu^c_{ab}(v)$ are nothing but defect 3-point functions
  	  \begin{equation}
     	   \m_{ab}^c(v)= \adjincludegraphics[scale=.4,valign=c]{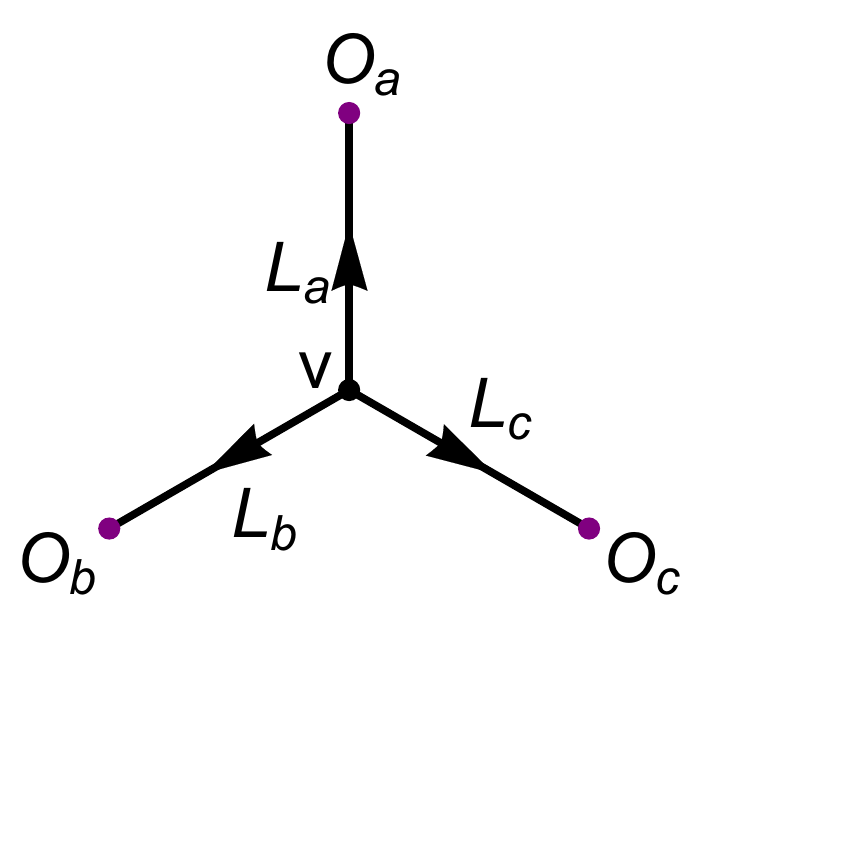}
     	        	   \end{equation}
    with the defect operators $\cO_a$ normalized by their two-point functions as
    \begin{equation}
     	  \D_{ab}=\adjincludegraphics[scale=.25,valign=c]{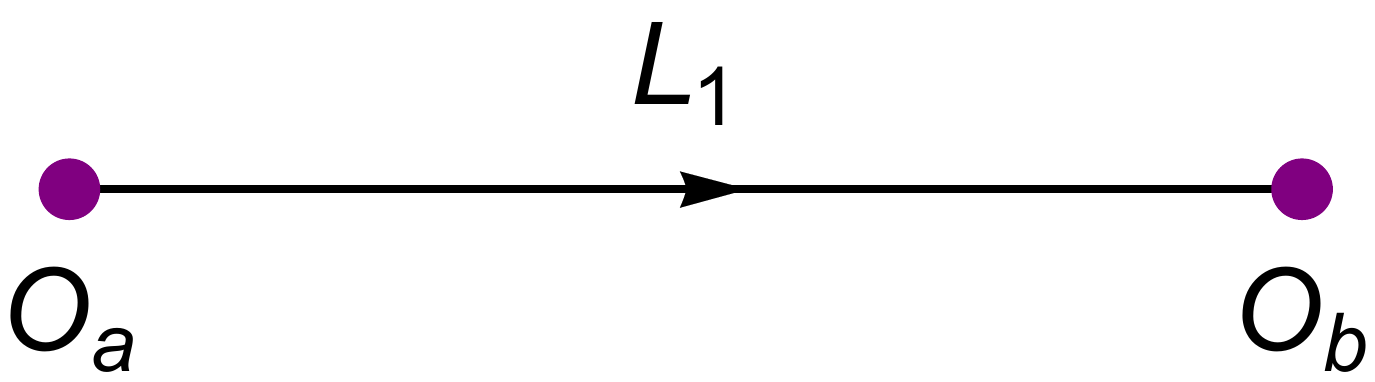}. 
     	        	   \end{equation}
     	        	   Finally the algebraic constraints \eqref{eqnfiberfuncpent} amounts to consistency of 
      cutting and gluing in the extended TQFT
    		\begin{equation}\label{eqnfiberfunccrossing}
    	 \adjincludegraphics[height=3cm,valign=c]{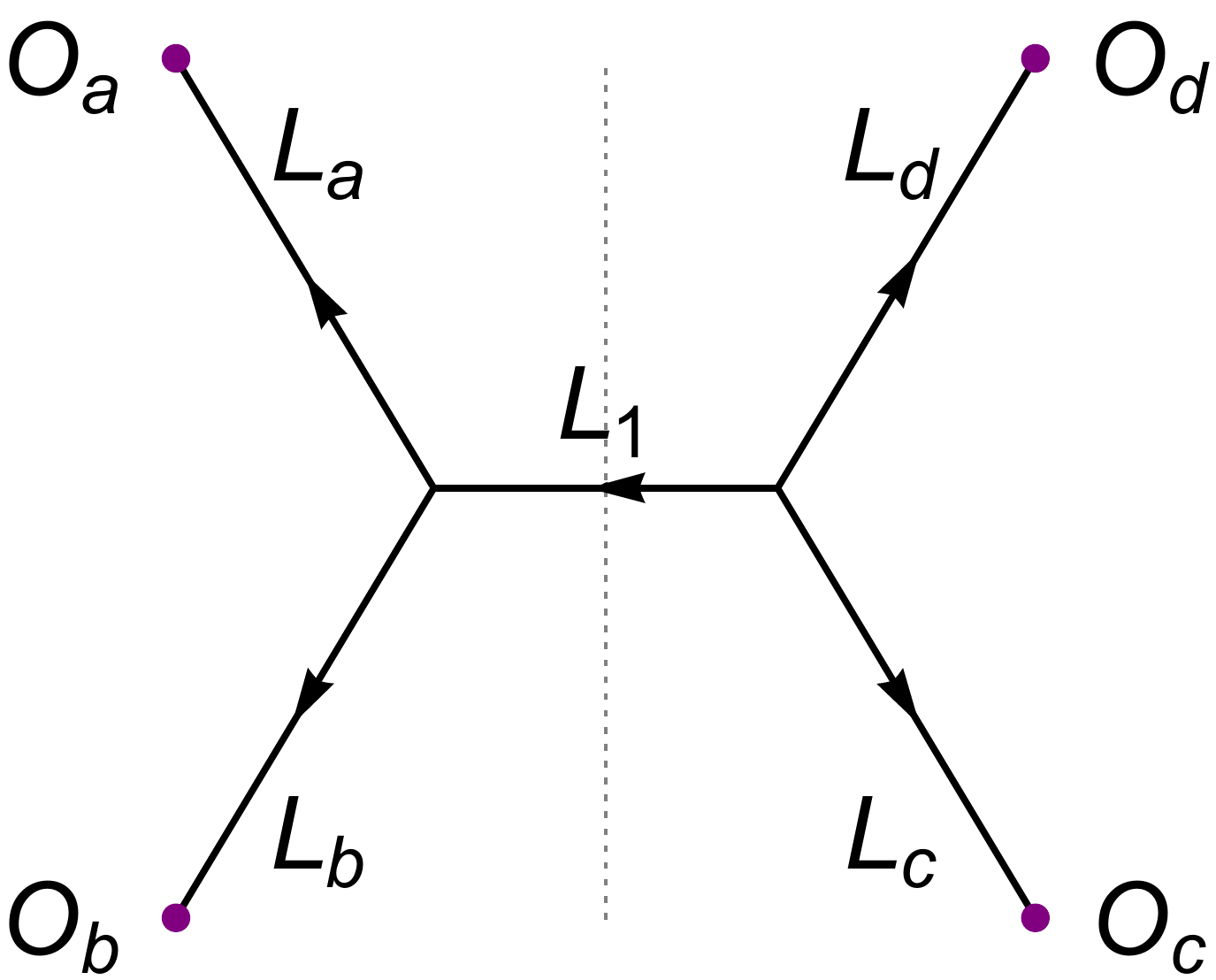}
    	 =
    	 \sum_{{\rm simple}~\cL_2} (F^{bad}_c)_{12}  \adjincludegraphics[height=3cm,valign=c]{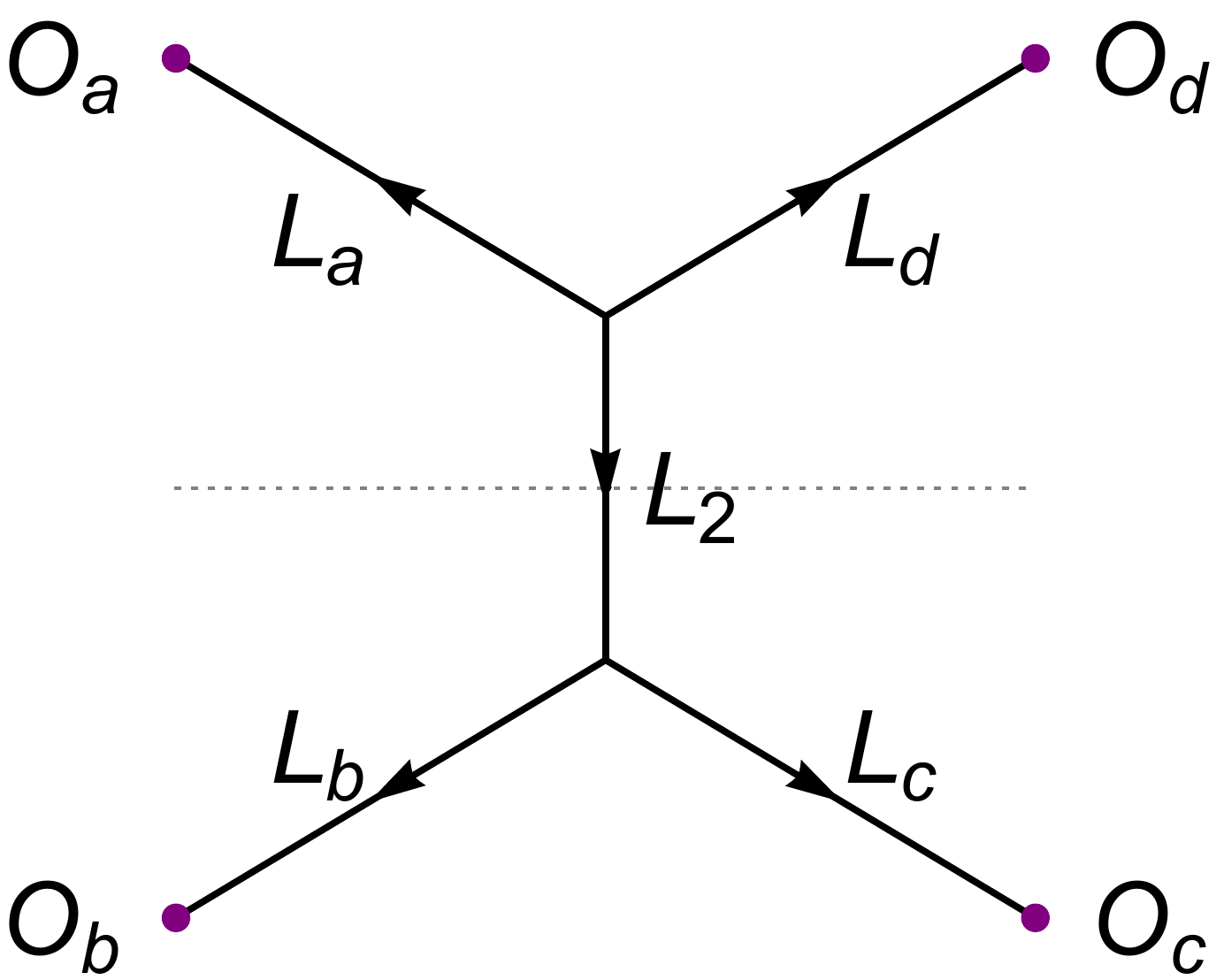}.
    		     		\end{equation}

        Applying our construction of Turaev-Viro boundary conditions based on module categories to fiber functors, we see that there is only one simple to assign to each boundary triangle, while edges where a bulk surface labelled by $a \in \cA$ end are associated with basis vectors in $V_a$. The state-sum weight is the product over $\mu$'s along each boundary triangle as well as $F$'s over each bulk six-fold junction. \eqref{eqnfiberfuncpent} ensures that as a six-fold function passes through the boundary, the $F$-move which is performed there is exactly compensated by the $\mu$'s which are added or dropped to the weight (see Fig. \ref{figboundaryFmove}), yielding a topologically-invariant state-sum weight.
  		
 Since the dimension of $V_a$ is the quantum dimension $\la \cL_a\ra$, a simple corollary of our general theorem is 	
  		\begin{thm}\label{thmnonintquantdim}
  			If $\cA$ has an object with non-integer quantum dimension, then the Turaev-Viro theory defined by $\cA$ does not admit a gapped, nondegenerate, $\cA$-symmetric boundary condition.
  		\end{thm}
  		
  		This captures the anomalies discussed in \cite{theOGs}. However, we will analyze some other examples of anomalous fusion categories which have all integer quantum dimensions.
  		
  		We end this section by describing fiber functors for ${\rm Vec}_G^\omega$. They consist of a 1-dimensional vector space $V_g$ for each element $g \in G$ along with a multiplication map
  		\begin{equation}\mu_{g_1,g_2}^{g_1 g_2}: V_{g_1} \otimes V_{g_2} \to V_{g_1 g_2}.\end{equation}
  		Since these are isomorphisms of 1-dimensional vector spaces, we can identify each $\mu_{g_1,g_2}^{g_1 g_2}$ with a (nonzero) complex number $\chi(g_1,g_2) \in \bC^{\times}$. In terms of $\chi$ the associativity equation \eqref{eqnfiberfuncpent} becomes
  		\begin{equation}\delta \chi(g_1,g_2,g_3) = -\omega(1,g_1,g_2,g_3),\end{equation}
  		which is equivalent to our counter-term equation \eqref{countertermeqn} after extending $\chi$ to a homogeneous cochain. Thus if $\omega$ is nontrivial in group cohomology, ${\rm Vec}_G^\omega$ admits no fiber functor, and we derive the usual 't Hooft anomaly constraint familiar from the theory of SPT phases \cite{CGLW}.

        \subsection{A Converse and Classifying Gapped Phases}\label{secSPTs}
        
        We can prove a converse to Theorem 1, which showed the existence of a fiber functor given an $\cA$-symmetric non-degenerate gapped phase, by constructing such a phase given a fiber functor. Moreover, inequivalent fiber functors correspond to inequivalent non-degenerate gapped phases, in that there are $\cA$-symmetry-protected edge modes between them. These are analogous to 1+1D $G$-SPTs but protected by the fusion category symmetry.
        
        More generally, we recalled that an $\cA$-symmetric gapped phase defines an $\cA$ module category. It is clear that equivalent such phases define equivalent module categories. In this section, we will prove a converse. That is, while we argued that $\cA$-symmetric systems define boundary conditions of Turaev-Viro, given such a boundary condition, we can construct a symmetric system where $\cA$ acts as a global symmetry. This proves that gapped $\cA$-symmetric 1+1D phases are in bijective correspondence with $\cA$-module categories. Moreover, we will show that the number of ground states of the 1+1D phase equals the number of simple objects in the $\cA$-module categories. In particular, non-degenerate gapped $\cA$-symmetric phases (``$\cA$-SPTs") are in bijective correspondence with fiber functors of $\cA$. We summarize this as:
        
        \begin{thm}\label{thmgappedclass}
        Module categories $\cM$ of $\cA$ are in bijection with $\cA$-symmetric gapped phases, such that the ground-states are in bijection with the simple objects of $\cM$.
        \end{thm}
        
        The idea is that we can construct a 1d $\cA$-symmetric phase by studying Turaev-Viro theory on a thin strip with the fixed boundary condition on the top edge and a boundary condition corresponding to the module category $\cM$ on the bottom edge. $\cA$ acts by fusing lines into the top edge.\footnote{For more discussion of degeneracies of topological order in this geometry from a perspective of boundary condensation, see \cite{Hung_2015}.}
        
        \begin{figure}
            \centering
            \includegraphics[width=8cm]{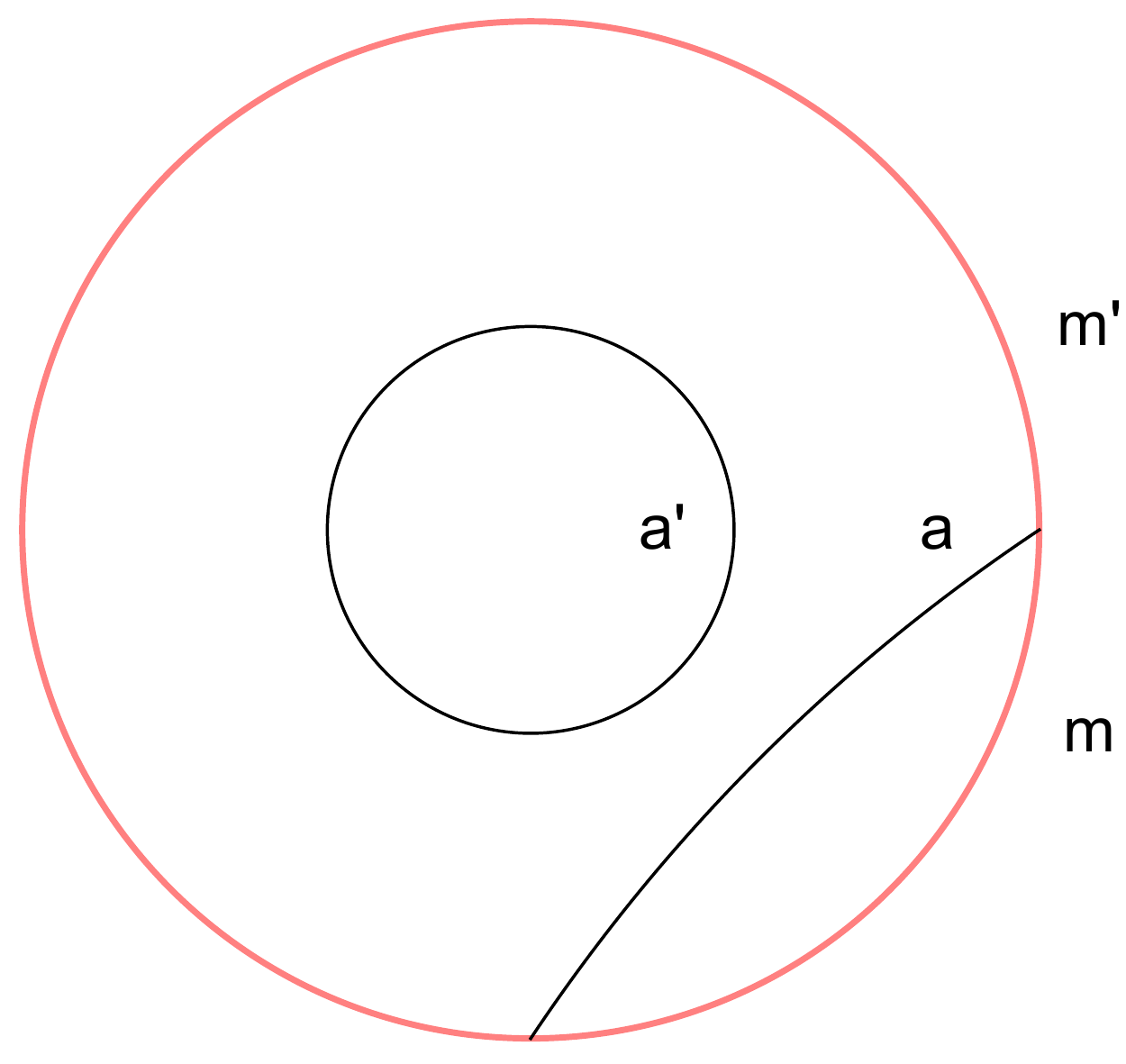}
            \caption{A simple string-net state for the annulus with inner boundary in the fixed boundary condition and outer boundary (red) labelled by simples $m,m' \in \cM$. At the right junction we have a morphism $m \to a \otimes m'$. We clean up this state by fusing the $a$ line into the outer boundary. Then, we lift the $a'$ line off the inner boundary and fuse it to the other boundary, leaving the inner boundary with label $1$ and the outer boundary in a superposition of $\cM$ labels. Such states thus form a basis of the ground states in this geometry.}
            \label{figannulus}
        \end{figure}
        
        We have discussed how to define a boundary condition for the Turaev-Viro state-sum from a module category $\cM$. For computing ground state degeneracy, it is also convenient to have a Hamiltonian model of the boundary condition. The Levin-Wen Hamiltonian \cite{Levin_2005} gives a construction of Turaev-Viro theory on closed surfaces in terms of $\cA$. Its ground states are ``string-nets": networks of strings labelled by $\cA$ simple objects with junctions labelled by fusion vertices, modulo isotopies, introducing small $\cA$-string loops, and $F$-moves. Kitaev-Kong \cite{Kitaev_2012} extended this Hamiltonian to surfaces with boundary by having the boundary strings take labels in $\cM$. The point-junctions between bulk and boundary strings are labelled by the $\cA$ action on $\cM$. In the fixed boundary condition, $\cA$ lines can pile up on the boundary but they cannot end there.
        
        We consider the string-net ground-states on an annulus with the inner boundary given by the fixed boundary condition, and the outer boundary given by $\cM$. In any string-net state, the $\cA$ lines can only end on the outer boundary. We can use $F$-moves and isotopies to push all of the lines into the outer boundary, the result is a state with identity labels everywhere, including on the inner boundary, except for one simple label $m \in \cM$ which runs all the way around the outer boundary. States which result in different $m \in \cM$ on the outer boundary this way clearly cannot be connected by $F$-moves, isotopies, or introducing small $\cA$-string loops. Thus, the string-net ground-states on this annulus in bijection with the simple objects of $\cM$. See Fig. \ref{figannulus}.
        
        Specializing to module categories with one simple object, i.e. fiber functors, on general abstract grounds, $\cA$-SPTs associated with inequivalent fiber functors of $\cA$ have degenerate edge modes between them protected by the fusion category symmetry. Indeed, suppose otherwise. Then there would be an invertible boundary-changing operator between the two fiber functors as boundaries of $\cA$-Turaev Viro theory. This would yield an equivalence between their associated module categories, so they would be equivalent fiber functors, a contradiction. We will see a simple example of a ``projective representation" of $\cA$ acting on these edge modes in Section \ref{secZ2Z2fiberfuncs}.
    
        In the next section, we will give a physical method for computing this classification in the case that $\cA$ is a Tambara-Yamagami category. These consist of topological lines associated with elements of an abelian group $G$ as well as a duality line $\tau$ associated with a Kramers-Wannier-like duality for $G$-symmetric systems. We will find inequivalent $\cA$-SPTs which as $G$-symmetric systems are in the same $G$-SPT phase, but which nonetheless have different symmetry fractionalization patterns for the duality line. This detailed study will give us some insight into these new 1+1D phases.

  		\section{Anomalies and Gapped Phases from Gauge Theoretic Techniques}\label{secgappedphases}
  		
  		In this section, we describe a simple method for classifying general $\cA$-symmetric (possibly degenerate) gapped phases for $\cA$ a Tambara-Yamagami category or even more generally an iterated group extension category, aka a group-theoretical category. Currently, all known integral fusion categories, i.e. those with only integer quantum dimensions, are Morita equivalent to a category in the latter class. Thus, conjecturally what we describe here is a general method for determining whether any fusion category is anomaly-free (recall all non-integral fusion categories are anomalous, by Theorem \ref{thmnonintquantdim} and by the arguments of \cite{theOGs}). However, there are also iterated group extension categories such as the Ising category which are not integral, and our method for determining all the symmetric gapped phases also works for those.
  		
  		\subsection{Iterated Group Extensions and a General Strategy for Fiber Functors}\label{secstrategy}
  		
  		A $G$ symmetry of Turaev-Viro theory defined by a fusion category $\cA$ may be described by the action of $G$ on the anyons as well as fractionalization data which describes the fusion junctions of $G$ domain walls \cite{Barkeshli_2019}. There are certain anomaly-vanishing conditions for such a symmetry which were worked out in \cite{Etingof_2010}. An anomaly-free $G$ symmetry for Turaev-Viro theory is one for which we can also assign consistent $F$-symbols to the $G$ domain walls. That is, we construct a $G$-graded fusion category
  		\begin{equation}\cA^G = \bigoplus_{g \in G} \cA_g\end{equation}
  		such that $\cA_g$ describe the $g$-domain walls in the complete symmetry-breaking boundary condition where both $\cA$ and $G$ symmetries are broken. So in particular $\cA_1 = \cA$. The fusion product in this category factorizes according to the $G$-grading:
  		\begin{equation}\cA_{g_1} \otimes \cA_{g_2} \to \cA_{g_1 g_2}.\end{equation}
  		
  		Conjecturally \cite{Etingof_2011}, all integral fusion categories $\cA$ are Morita equivalent (i.e. have the same set of symmetric gapped phases) to one which may be described by a sequence of groups $G_j$ and categories $\cA_j$, $1 \le j \le l$, with
  		\begin{equation}\cA_1 = {\rm Vec}_{G_1}^\omega\end{equation}
  		\begin{equation}\cA_{j+1} = \cA_j^{G_j}\end{equation}
  		\begin{equation}\cA_l = \cA\end{equation}
  		for some anomaly-free $G_j$ action on $\cA_j$. We see that since $\cA_j$ is a subcategory of $\cA_{j+1}$, namely it is the component with grading $1 \in G_j$, a fiber functor of $\cA_{j+1}$ defines a fiber functor for $\cA_j$ by restriction. Thus, if $\cA_j$ is anomalous, so are all $\cA_k$ with $k > j$, in particular $\cA = \cA_l$.
  		
  		We have already discussed how ${\rm Vec}_{G}^\omega$ is anomaly-free if and only if $[\omega] = 0 \in H^3(BG,U(1))$. To determine whether a general fusion category described as above has a fiber functor, we just need to study the problem where $\cA$ is a fusion category with a fiber functor and an anomaly-free $G$ action and we ask whether $\cA^G$ admits a fiber functor. Then, we can iteratively determine above whether each $\cA_j$ has a fiber functor.
  		
  		A fiber functor of $\cA$ defines a boundary condition for the $\cA$ Turaev-Viro theory, and the $G$-action on this theory acts on its boundary conditions. For $\cA^G$ to have a fiber functor, $\cA$ needs to have a fiber functor which is fixed by this $G$-action. Such a fixed fiber functor may or may not define a fiber functor for $\cA^G$ or could even define multiple fiber functors, as we will see in the next sections.
  		
  		The general problem of when a $G$-fixed point fiber functor of $\cA$ determines a fiber functor of $\cA^G$ was studied in \cite{Meir_2012}. We will give a physical interpretation of some of their results, which can be applied to classifying all gapped phases of $\cA^G$. This will come down to studying the twisted sectors for the $G$-lines in the $\cA$ fiber functor. We will focus on Tambara-Yamagami categories for which $G = \bZ_2$ and reproduce some results of Tambara \cite{Tambara2000}, but the extension to other abelian groups is immediate.
  		
  		
  		
  		\subsection{Gapped Phases for Tambara-Yamagami Categories}\label{secTYphases}
  		
  		Let us derive the gapped phases for $G$-Tambara-Yamagami (TY) categories. These categories are $\bZ_2$ extensions of ${\rm Vec}_G$ where $G$ is an abelian group. They arise from self-duality under gauging $G$. In particular, given a non-degenerate bicharacter $\chi:G \times G \to U(1)$, then we obtain an assignment of $G$ charges to $G$ twisted sectors of the $G$ gauge theory such that the $g$ charge of the $h$ twisted sector is $\chi(g,h)$ (charges of operators in the untwisted sectors are trivial). Thus, with respect to a bicharacter, $G$-gauging becomes a Fourier-like transformation from $G$-symmetric theories to $G$-symmetric theories. As shown in \cite{tambarayam}, the remaining piece of data to define a fusion category associated with this transformation is a sign $\epsilon = \pm 1$ which will be interpreted as the Frobenius-Schur (FS) indicator of the duality line\footnote{This sign appears as the $H^3(B\bZ_2,U(1)) = \bZ_2$ (trivialized) torsor from the general extension study of \cite{Etingof_2010}.}. We will see how this data appears in the self-dual phases in Sections \ref{secselfdualspt} and \ref{secisingdualities}, but for now we will take all the data as given and try to classify the symmetric gapped phases.
  		
  		Let us describe the category. Its simple objects consist of invertible lines $g \in G$ with group fusion rules corresponding to an ordinary global symmetry plus a ``duality line" $\cN$ which absorbs $G$ lines
  		\ie  g \otimes \cN = \cN\fe 
  		and squares to a projection
  		\ie \cN^2 = \sum_{g \in G} g,\fe 
  		from which we see the quantum dimension of $\cN$ is $\sqrt{|G|}$. Thus, all such categories are anomalous if $|G|$ is not a perfect square.
  		
  		The bicharacter appears in the $F$-symbol of the crossing relation
  		        \begin{equation}
        \adjincludegraphics[width=3cm,valign=c]{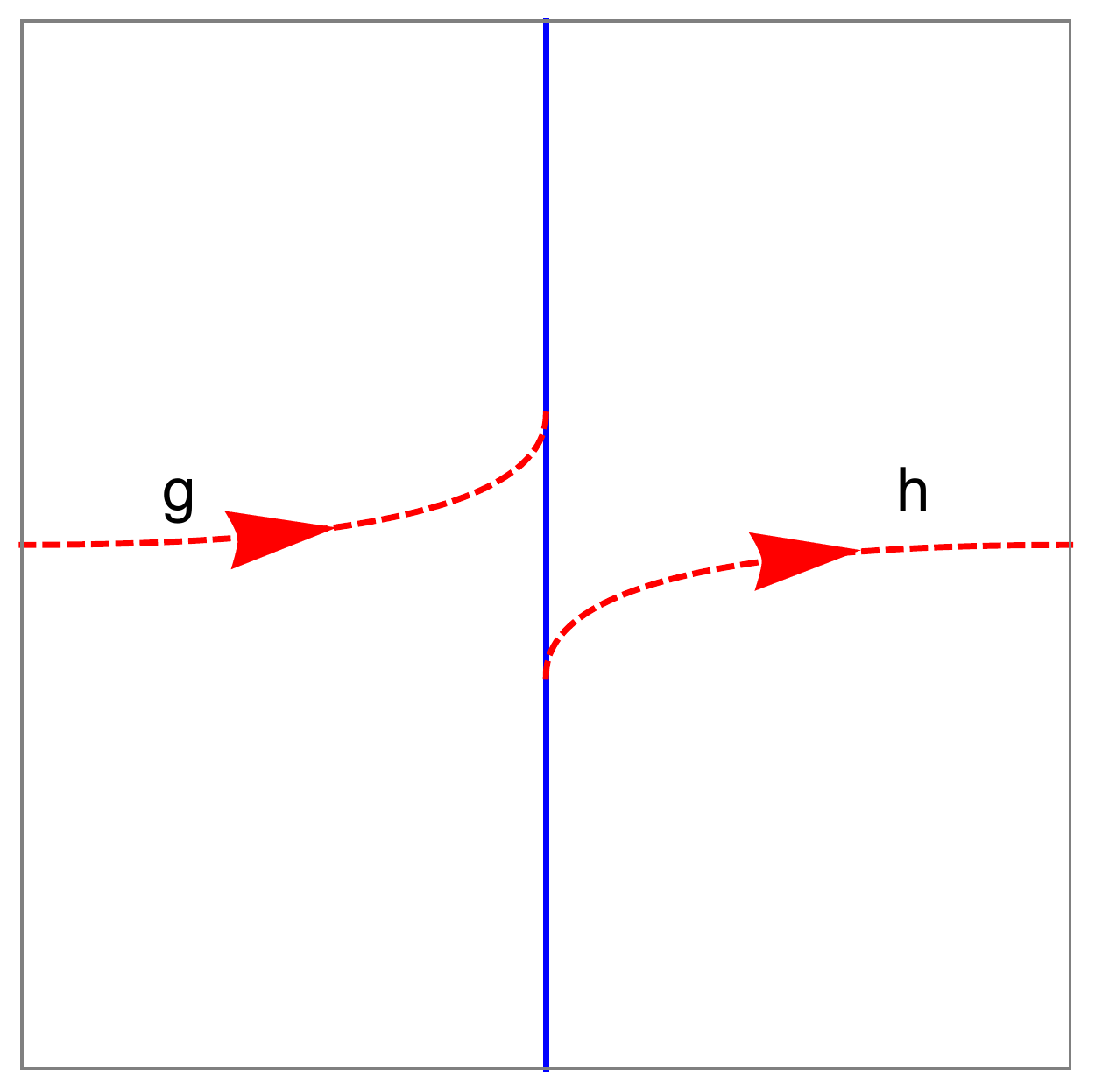}=\chi(g,h)\adjincludegraphics[width=3cm,valign=c]{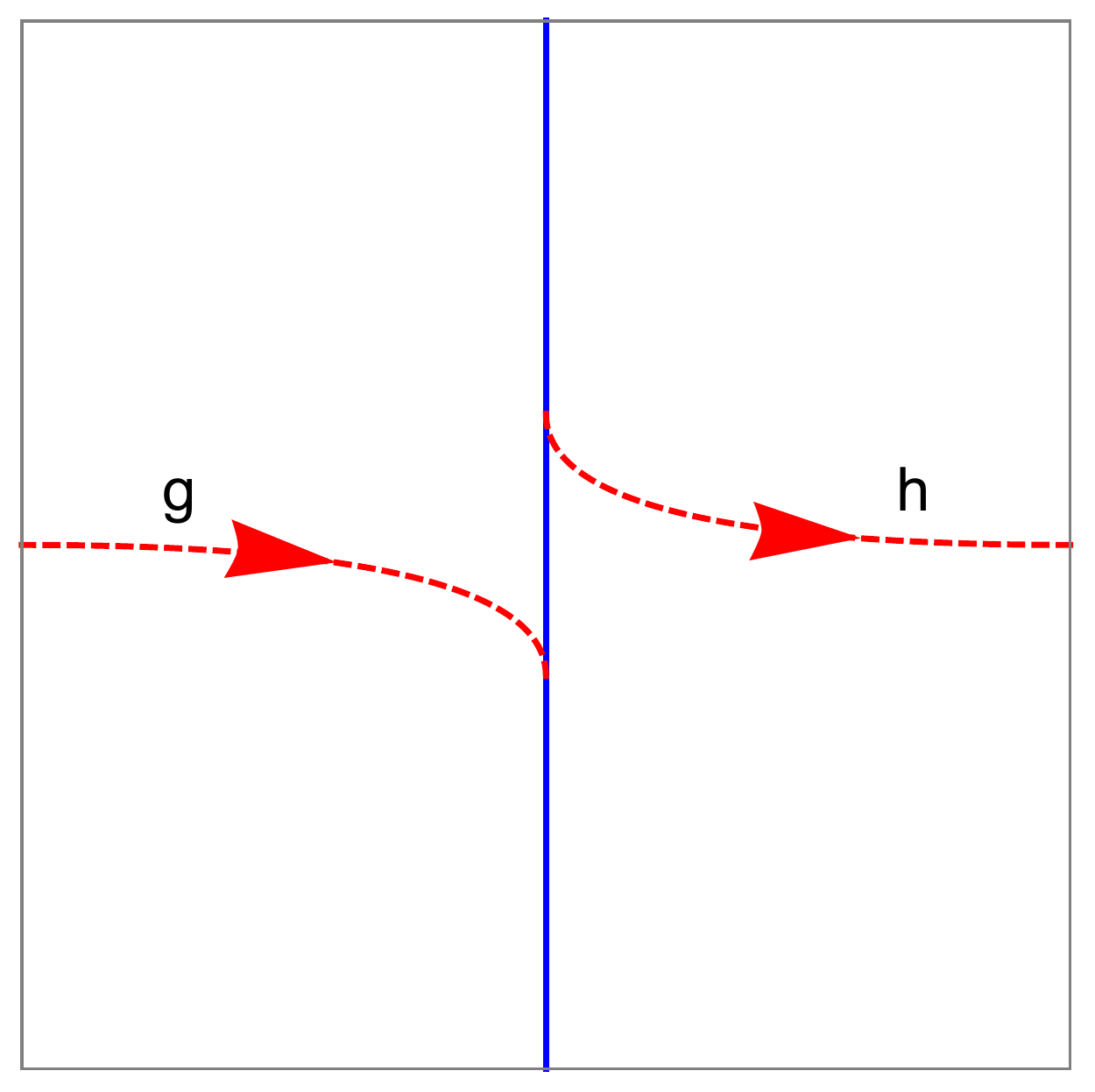},
        \label{TYbichar}
        \end{equation}
  		where the blue line is the duality line $\cN$. The other important crossing relation is
  		\begin{equation}
        \adjincludegraphics[width=3cm,valign=c]{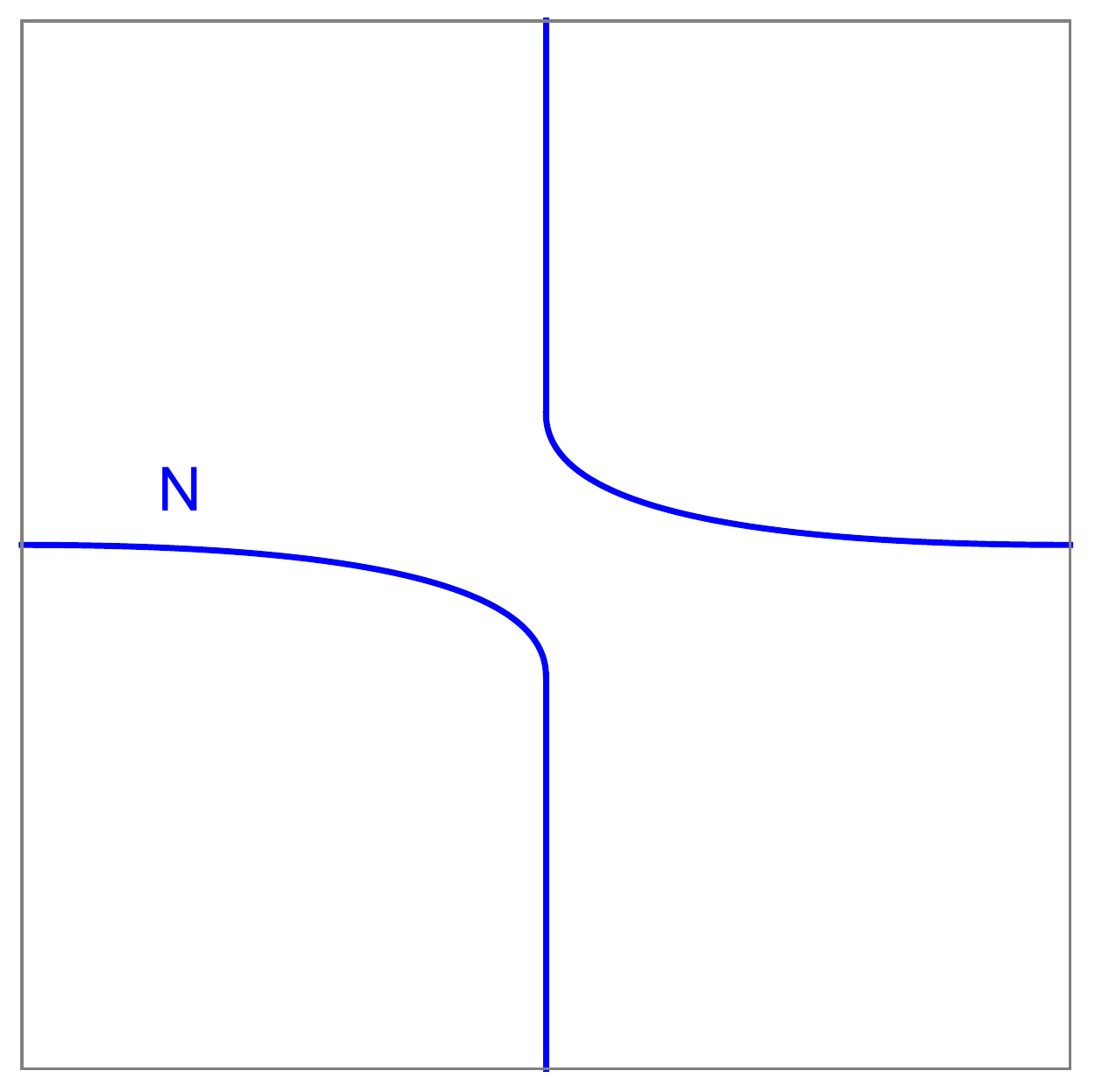}= \frac{\epsilon}{\sqrt{|G|}}\sum_g \adjincludegraphics[width=3cm,valign=c]{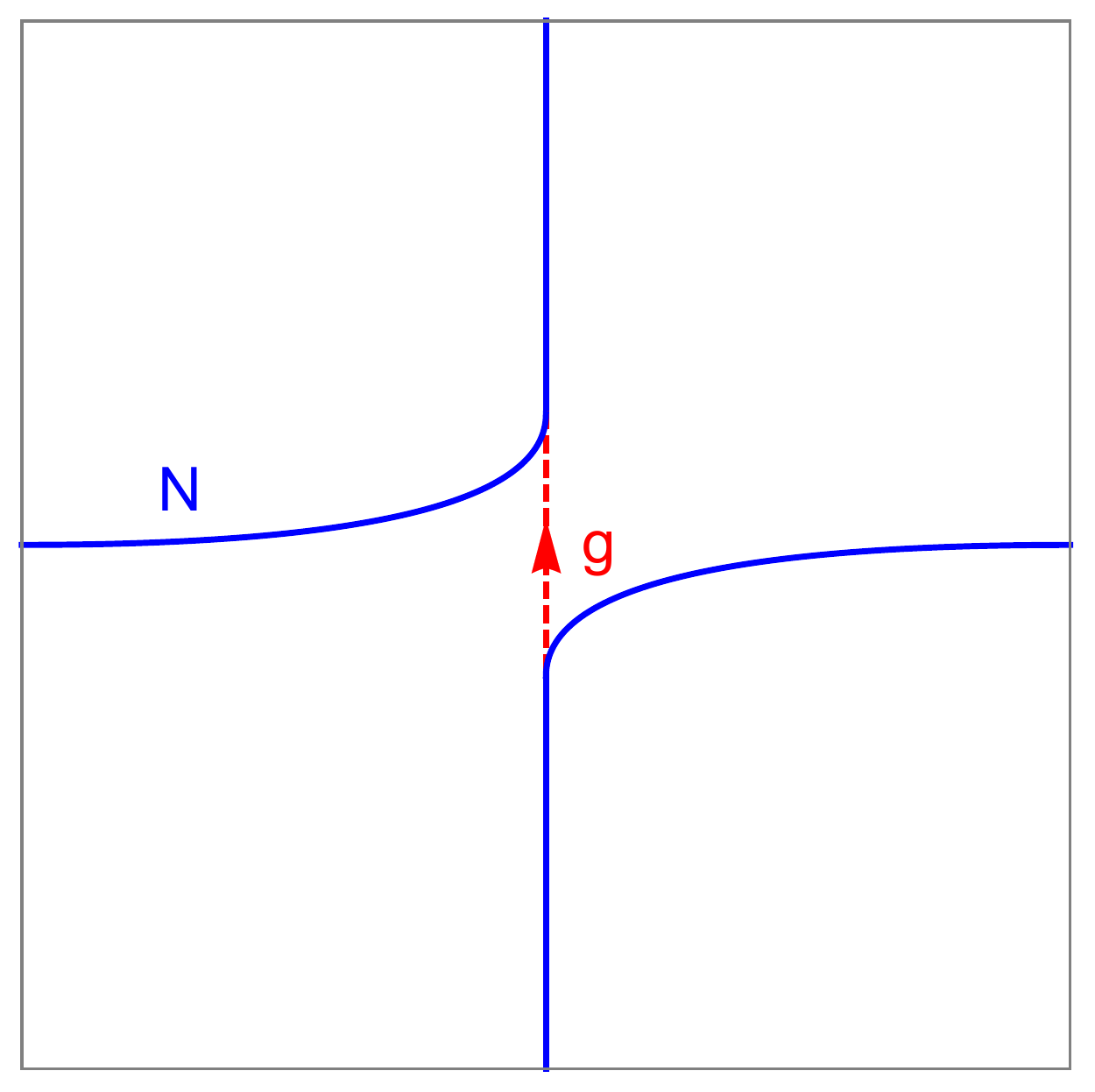},
        \end{equation}
        where we see the FS indicator of $\cN$ appearing.
  		
  		\subsubsection{Gapped $G$-Symmetric Phases}
  		
  		For classifying gapped phases of $G$ TY categories ($G$ is abelian), it is useful to first classify $G$-symmetric gapped phases. The most general phase of such type combines spontaneous symmetry breaking (SSB) and SPT order. It is determined by the subgroup
  		\begin{equation}G_{\rm unbroken} \subset G\end{equation}
  		of unbroken symmetries and an SPT cocycle
  		\begin{equation}[\alpha] \in H^2(BG_{\rm unbroken},U(1))\end{equation}
  		capturing how $G_{\rm unbroken}$ acts on a given ground state.\footnote{If $G$ is nonabelian, then each ground state may have a different subgroup of unbroken symmetries, known as the stabilizers, which may be fractionalized, so we should choose an SPT class for each of these groups. However, because $G$ acts transitively, they are all conjugate in $G$, and the SPT classes should be conjugate as well, so it suffices to specify the SPT class in just one of these ground states.} See \cite{Ostrik_2003} which reproduces this result by classifying module categories for ${\rm Vec}_G$.
  		
  		The partition function for this phase may be written
  		\begin{equation}Z(X,A) = \delta(\pi(A)) e^{2\pi i \int_X \alpha(A)},\end{equation}
  		where $A$ is the background $G$-gauge field (see \cite{kapustin2014anomalies} for a primer on how to compute using these), $\pi(A)$ is its image in the quotient $G/G_{\rm unbroken}$, $\delta$ is a Dirac delta function which projects $\pi(A) = 0$ so that $Z(X,A) = 0$ unless $A$ is valued in $G_{\rm unbroken}$, and in this case the value is the usual SPT partition function $\exp 2\pi i \int_X \alpha(A)$. The delta function comes from the fact that the twisted sectors for a spontaneously broken symmetry are lifted in energy above the ground-states of the untwisted sector, so these states are projected out in the low energy limit.
  		
  		For example, with $G = \bZ_2$, there are only two stable gapped phases: the trivial and the SSB phase. They have the partition functions
  		\begin{equation}\label{Z2phases}Z_1(X,A) = 1\end{equation}
  		\begin{equation}Z_0(X,A) = \delta(A).\end{equation}
  		
  		A more interesting class is $G = \bZ_2 \times \bZ_2$, which has six stable gapped phases: one trivial, one SPT, three that break a $\bZ_2$ subgroup of $G$, and one that breaks all of $G$. Writing the $G$ gauge field as two $\bZ_2$ gauge fields $A_1, A_2$, their partition functions are
  		\begin{equation}Z_1(X,A_1,A_2) = 1\end{equation}
  		\begin{equation}Z_{SPT}(X,A_1,A_2) = (-1)^{\int_X A_1 \cup A_2}\end{equation}
  		\begin{equation}Z_{Ix}(X,A_1,A_2) = \delta(A_1)\end{equation}
  		\begin{equation}Z_{Iz}(X,A_1,A_2) = \delta(A_2)\end{equation}
  		\begin{equation}Z_{Iy}(X,A_1,A_2) = \delta(A_1 + A_2)\end{equation}
  		\begin{equation}Z_0(X,A_1,A_2) = \delta(A_1)\delta(A_2).\end{equation}
  		
  		Tensor product (i.e. stacking) of $G$-symmetric gapped phases corresponds to multiplication of these partition functions. We see that this algebra is generated by invertible elements, which are SPT phases, and projectors, which are spontaneous symmetry breaking phases. The trivial phase acts as the multiplicative unit $1$ and the completely symmetry broken phase acts as the absorptive element $0$.
  		
  		\subsubsection{Permutation Action of the Duality Line}\label{secpermaction}
  		
  		Having the partition functions available allows us to determine how our $G$-gauging procedure given by the bicharacter $\chi$ acts on the $G$-symmetric phases. We have
  		\begin{equation}\label{eqnTYaction}
  		 Z^\star(X,A) = \frac{1}{|G|^g}\sum_{B \in H^1(X,G)} Z(X,B) \exp \left( 2\pi i \int_X \chi(A,B) \right),
  		\end{equation}
  		where   	$\chi$ is extended to a pairing on gauge fields using the cup product, and $g$ is the genus of $X$. Note that this action doesn't depend on the FS indicator $\epsilon$.
  		
  		For $G = \bZ_2$, there is only one choice of $\chi$:
  		\begin{equation}Z^\star(X,A) = \frac{1}{2^g}\sum_{B \in H^1(X,\bZ_2)} Z(X,B) (-1)^{\int A \cup B}.\end{equation}
  		This transformation exchanges $Z_0$ and $Z_1$ in \eqref{Z2phases}. Thus, there is no TY-symmetric stable $G$-phase, as we expected from Theorem \ref{thmnonintquantdim} since the quantum dimension of the duality line is not an integer. This implies that with either FS indicator $\epsilon$ (the usual Ising category corresponds to $\epsilon = +1$), we have only one stable gapped phase, with partition function
  		\begin{equation}Z_0 + Z_1,\end{equation}
  		which we recognize as the 1st order transition between the trivial symmetric and SSB phases, with 3 ground states.
  		
  		Slightly generalizing the Ising category for systems with global $G = \bZ_n$ symmetry, for every $k$ co-prime to $n$ we have a Kramers-Wannier-like transformation
  		\ie  Z^\star (X,A) = \frac{1}{n^g} \sum_{B \in H^1(X,\bZ_n)} Z(X,B) e^{\frac{2\pi i k}{n} \int A \cup B}.\fe
  		The stable gapped $\bZ_n$ phases are all symmetry-breaking, and labelled by an integer $m$ dividing $n$, which describes the subgroup $\bZ_m$ of unbroken symmetries. We see $\bZ_n$ gauging exchanges $m$ with $n/m$, so none of these categories have stable non-degenerate gapped phases, i.e. they are all anomalous, although the duality line has integer quantum dimension $\sqrt{n}$ whenever $n$ is a perfect square. We will see an example of such an anomaly for $n = 4$ in Section \ref{secisingdualities}.
  		
  		For $G = \bZ_2 \times \bZ_2$ with $\chi$ yielding the diagonal pairing $A_1 \cup A_2 + B_1 \cup B_2$ in \eqref{eqnTYaction}, we compute the orbits:
  		\begin{equation}Z_1 \leftrightarrow Z_0\end{equation}
  		\begin{equation}Z_{SPT}\ {\rm fixed}\end{equation}
  		\begin{equation}Z_{Ix} \leftrightarrow Z_{Iz}\end{equation}
  		\begin{equation}Z_{Iy} \ {\rm fixed}.\end{equation}
  		For $G = \bZ_2 \times \bZ_2$ with $\chi$ yielding the off-diagonal pairing $A_1 \cup B_2 + B_1 \cup A_2$, we get
  		\begin{equation}Z_1 \leftrightarrow Z_0\end{equation}
  		\begin{equation}Z_{SPT}\ {\rm fixed}\end{equation}
  		\begin{equation}Z_{Ix} \ {\rm fixed}\end{equation}
  		\begin{equation}Z_{Iy} \ {\rm fixed}\end{equation}
  		\begin{equation}Z_{Iz} \ {\rm fixed}.\end{equation}
  		
  		Whenever there is a non-trivial orbit of the $G$-symmetric phases under gauging, we can consider the phase which looks like a first-order transition among the elements of the orbit and this phase will be stabilized by the fusion category symmetry. On the other hand, for fixed-point phases under gauging, it is possible for the self-duality to be spontaneously broken by some order parameter that has trivial $G$ quantum numbers but is duality-odd. These two families of phases correspond to the induced module categories in the mathematical language \cite{Meir_2012}. We will discuss the phases with unbroken self-duality below.
  		
  		\subsubsection{Generalizations to $n$-ality Categories}\label{secnality}
  		
  		This method of constructing transformations of $G$-symmetric phases by studying the partition function has a nice generalization to gauging with discrete torsion. That is we modify our transformation to
  		\ie Z^\star (X,A) = \frac{1}{|G|^g} \sum_{B \in H^1(X,G)} Z(X,B) \exp  \left( 2\pi i \int \chi(A,B) + \omega(B) \right),\fe
  		where $\omega \in H^2(BG,U(1))$ (one could also add such a term for $A$). Such a transformation is typically no longer a duality, as the following example illustrates, with $G = \bZ_2 \times \bZ_2$, $\chi(A,B) = A_1 B_2 + A_2 B_1$. It acts on the phases by
  		\ie Z_0 \to Z_1 \to Z_{SPT} \to Z_0\fe
  		\ie Z_{Ix}, Z_{Iy}, Z_{Iz} \ {\rm fixed}.\fe
  		This fusion category is therefore anomalous, since there are no non-degenerate fixed point phases.
        
        Since this triality maps the trivial phase to the symmetry broken phase, doing it three times projects out any local operators with nontrivial $\bZ_2 \times \bZ_2$ charge. This implies that any associated ``triality TDL'' $\cN$ must satisfy
        \ie \cN^3 = \sum_{g \in \bZ_2 \times \bZ_2} g.\fe
  		For the same reason we must have
  		\ie g \otimes \cN = \cN.\fe
  		We expect the $F$ symbols are mostly fixed by the transformation rule above, except for the Frobenius-Schur indicator of $\cN$, which may be a third root of unity, corresponding to the $H^3(B\bZ_3,U(1)) = \bZ_3$ torsor of \cite{Etingof_2010}, which we also believe is trivialized. It would be very interesting to describe fusion categories with these fusion rules. Note the dimension of $\cN$ above is $4^{1/3}$.
  		
  		This triality is related by $\bZ_2$ gauging to a $\bZ_3$ outer automorphism of the symmetry group $\bZ_2 \times \bZ_2$, which acts by
  		\ie Z^\star(X,A_1,A_2) = Z(X,A_2,A_1 + A_2),\fe
        and on the phases as
  		\ie Z_{Ix} \to Z_{Iy} \to Z_{Iz} \to Z_{Ix}\fe
  		\ie Z_0, Z_{1}, Z_{SPT} \ {\rm fixed}.\fe
        Such outer automorphisms are associated with ordinary global symmetries.
        
        We will see examples of CFTs with self-triality in a follow-up work \cite{toappear}. See also Section \ref{secisingdualities} below.
  		
  		\subsubsection{Self-Dual SPT Phases}\label{secselfdualspt}
  		
  		We have explained how to classify the phases where the self-duality is broken. More interesting are the phases where it is preserved by the ground states. Let us first restrict our attention to the self-dual $G$-SPT phases and try to promote them to fiber functors, i.e. non-degenerate symmetric phases for the full $G$-TY category. We follow \cite{Tambara2000}.
  		
  		 For our $G$-SPT to be self-dual under $G$ gauging, a necessary condition is that the SPT 2-cocycle $\alpha \in H^2(BG,U(1))$ has to be \emph{nondegenerate}, in the sense that the torus partition function gives a non-degenerate pairing $G \times G \to U(1)$ between the holonomies around the spatial and temporal cycles. Otherwise, there is a trivial subsector of the theory which will get dualized to a symmetry-breaking phase. In a sense these SPTs are as topological as possible.
  		 
  		 A necessary and sufficient condition for the SPT to be self-dual is that there exist an involution (i.e. a homomorphism squaring to the identity)
  		 \begin{equation}\sigma:G \to G\end{equation}
  		 such that
  		 \begin{equation}\label{eqninvolcond}\frac{\alpha(g,\sigma(h))}{\alpha(\sigma(h),g)} = \chi(g,h).\end{equation}
  		 Note that the left hand side is the torus partition function for the SPT in the gauge background with $g$ around one cycle and $\sigma(h)$ around the other. Because of the non-degeneracy of this pairing, $\sigma$ is determined uniquely if it exists. Using this identity, we may prove that for all closed surfaces
  		\begin{equation}\int \alpha(A + \sigma(B)) = \int \alpha(A) + \chi(A,B) + \alpha(B),\end{equation}
  		from which the self-duality of the SPT immediately follows. We will show the necessity of $\sigma$ below, and interpret it.
  		 
  		 
        To proceed, we must study the duality-twisted sector of the SPT by placing the duality line $\cN$ along a time-like cycle. It is instructive to cut the torus along a time-like cycle opposite the duality line, creating a cylinder. This introduces two boundaries, and we choose some $G$-symmetric boundary condition for the SPT on each of these boundaries. A canonical boundary condition has a basis state $v_g$ for every element $g\in G$ with the twisted left action
  		\begin{equation}g \cdot v_h = \alpha(g,h) v_{gh},\end{equation}
  		for the boundary condition on the right, and the twisted right action
  		\begin{equation}v_h \cdot g = \alpha(h,g) v_{hg}\end{equation}
  		for the boundary condition on the left. The states of the $\cN$-twisted sector are obtained by setting the states on the two boundaries to be equal.
  		
    \begin{figure}
        \centering
        \includegraphics[width=7cm]{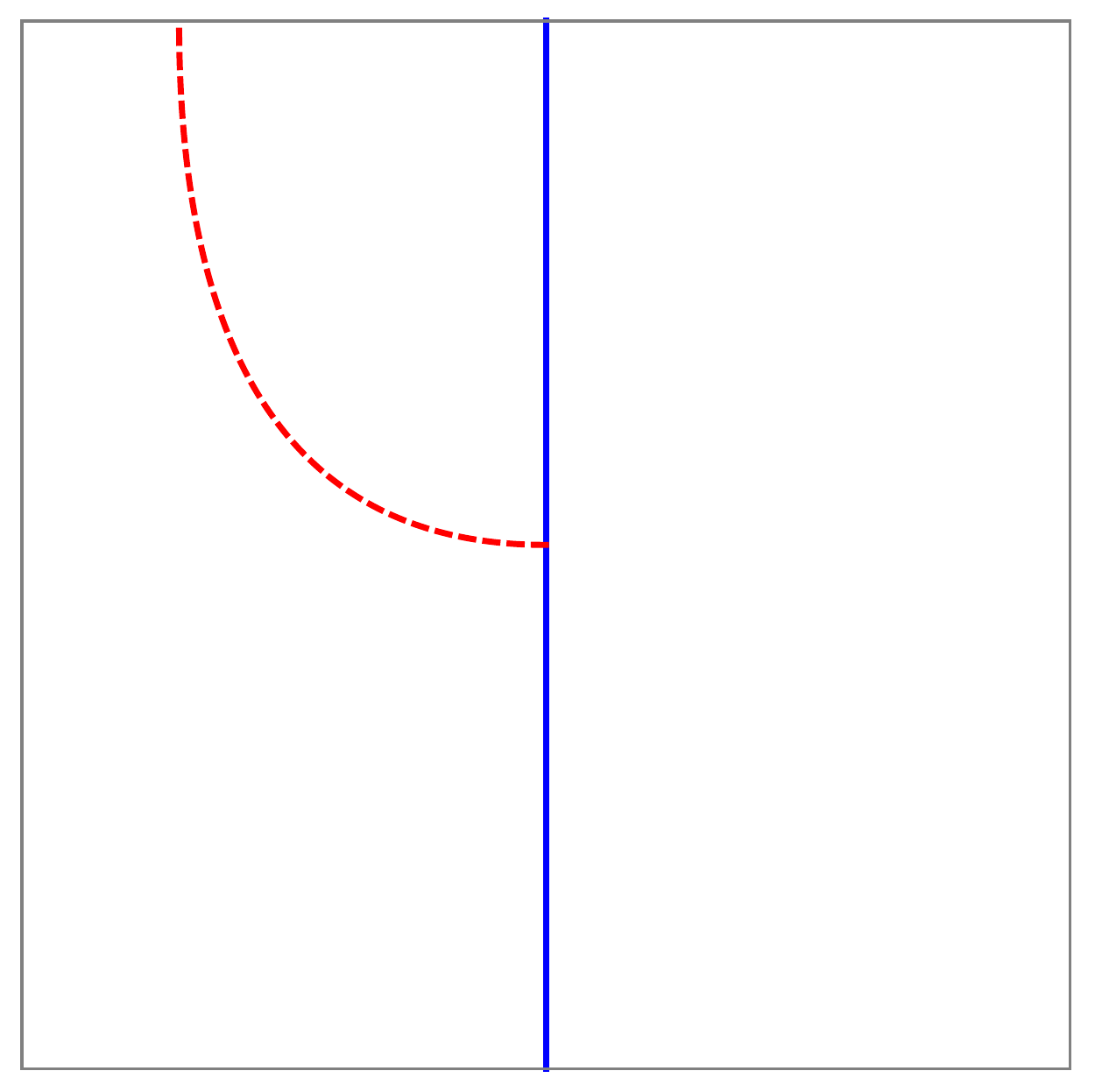}
        \caption{In the duality-twisted sector, with a duality line (blue) inserted along a time-like cycle, there is both a left and a right (projective) $G$ action. Here we see a left action where a $G$ line (red) fuses onto the left of the duality line. Because of the crossing relations, the left and right actions commute only up to the bicharacter $\chi$. Below we find the left and right actions must be anti-isomorphic in a non-degenerate TY-symmetric phase.}
        \label{figleftaction}
    \end{figure}
  		
  		If we begin with such a ``diagonal" state and apply a $g$-line to the left or right half of the system, then we obtain a state in the $g \otimes \cN$ or $\cN \otimes g$ sector. However, we know $g \otimes \cN = \cN \otimes g = \cN$, so this simply gives another state in the $\cN$-twisted sector. This gives the twisted sector the structure of both a left and right projective $G$-action. See Fig. \ref{figleftaction}.
  		
         Because of the non-degeneracy of $\alpha$, the twisted group algebra $\bC^\alpha[G]$ is simple, and has only one irreducible representation. It follows that the right action of $G$ on the duality-twisted sector is anti-isomorphic to the left action of $G$, so up to isomorphism, one of them is the canonical one (the regular representation of the twisted group algebra on itself) and the other one is described by an isomorphism
  		\begin{equation}f:\bC^\alpha[G] \to \bC^\alpha[G]^{op}\end{equation}
  		  which is order 2 in the sense that
  		\begin{equation}f^{-1} = f^{op}.\end{equation}
  		
  		We can decompose this isomorphism as an involution $\sigma:G \to G$ (the notation will soon make sense) along with a numerical factor $\nu:G \to \bC^\times$ such that
  		\begin{equation}\label{eqnquadratic}\nu(g)\nu(h)\nu(gh)^{-1} = \alpha(g,h) \alpha(\sigma(h),\sigma(g))^{-1},\end{equation}
  		\begin{equation}\nu(g)\nu(\sigma(g)) = 1.\end{equation}
  		  		Pictorially this means
  		 \begin{equation}
\adjincludegraphics[width=4cm,valign=c]{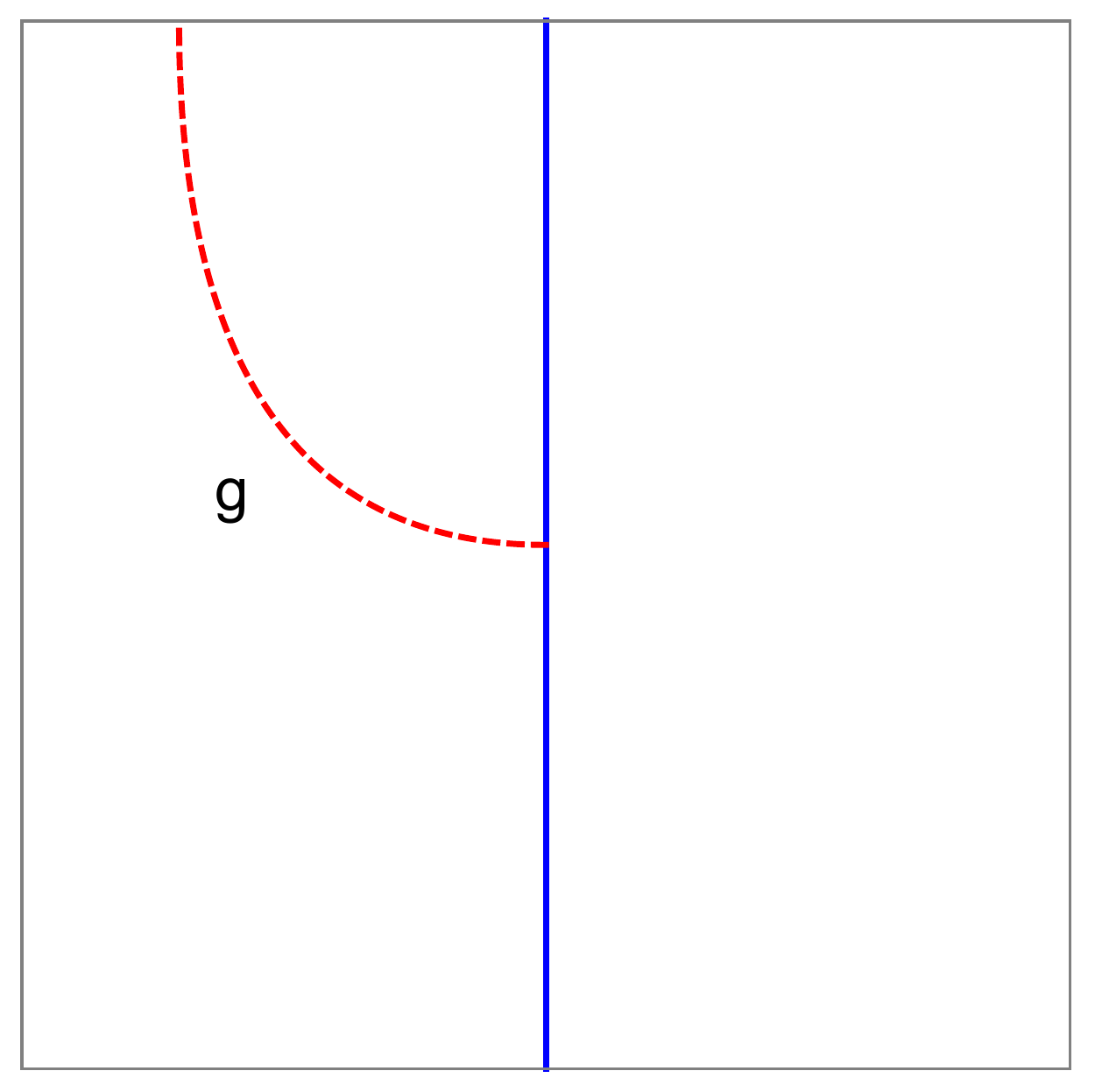}\quad = \quad \nu(g)\ \adjincludegraphics[width=4cm,valign=c]{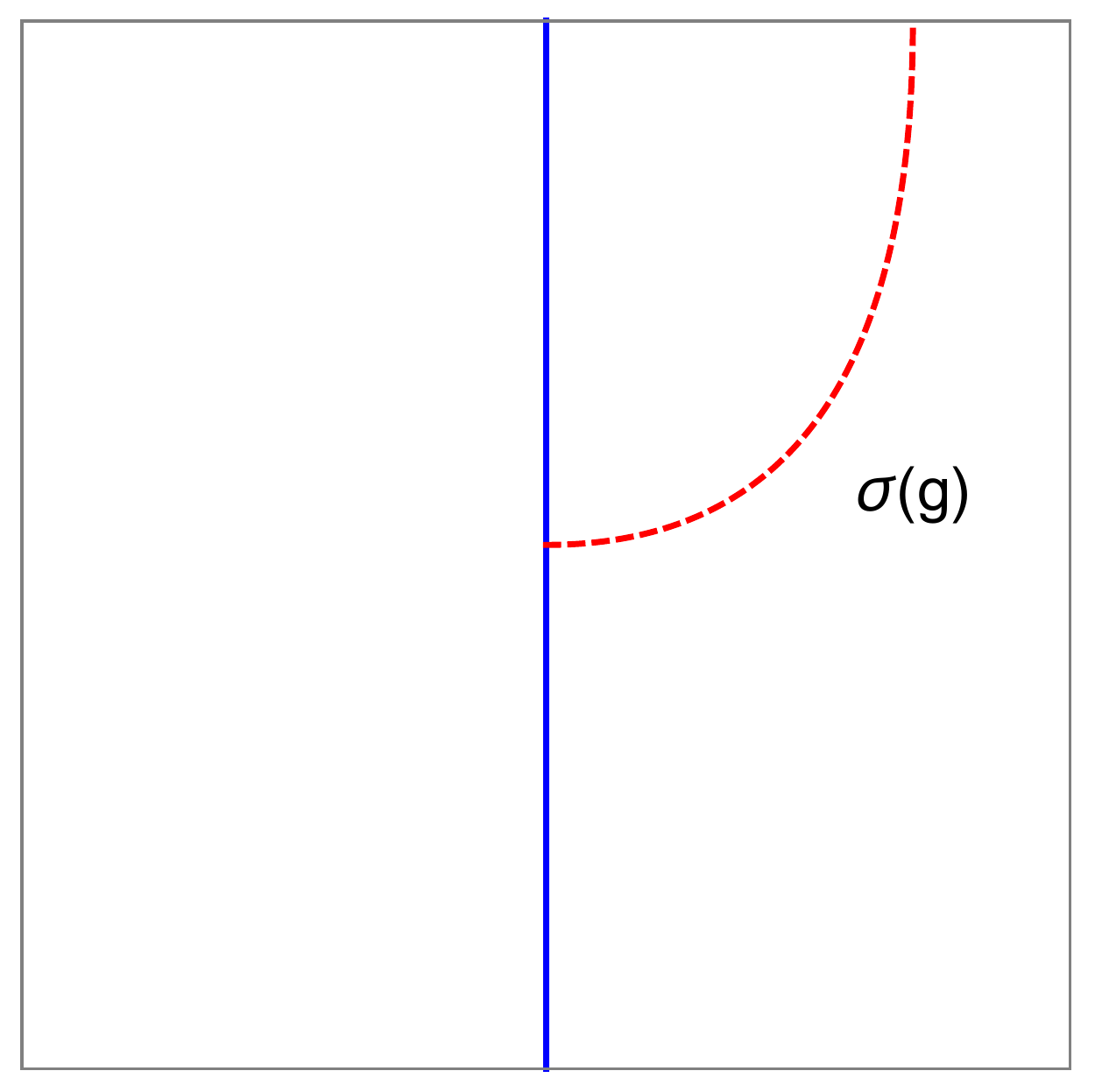}.
\end{equation}
  		The fact that the left and right action commute up to the bicharacter $\chi$ yields the further constraint
  		\begin{equation}\frac{\alpha(g,\sigma(h))}{\alpha(\sigma(h),h)} = \chi(g,h),\end{equation}
  		which we recognize as the defining condition for $\sigma$ we had above, so indeed they are the same.
  		
  		There is one more condition $\nu$ must satisfy, which comes from studying the fusion rule $\cN^2 = \sum_g g$. This gives us a pairing of $\cN$-twisted sector states to $g$-twisted sector states for each $g$. Let us denote this pairing $[-,-]_g$. By studying the crossing relations near the $\cN\otimes\cN \to h$ fusion vertex, we find some simple relations
        \begin{equation}g[x,y]_h = [gx,y]_{gh}\end{equation}
        \begin{equation}[x,yg]_h = [x,yg]_{hg}\end{equation}
        \begin{equation}[xg,y]_h = \chi(g,h)[x,gy]_h.\end{equation}
        All of this data may be packaged into the form defined by $[x,y]_1 = \gamma(x,y) \cdot 1$, which must satisfy
        \begin{equation}\gamma(gx,y) = \gamma(x,yg)\end{equation}
        \begin{equation}\gamma(xg,y) = \gamma(x,gy).\end{equation}
        
        Bringing in another duality line we find the important associativity relation
        \begin{equation}x[y,z]_g = \epsilon \sum_h \chi(g,h)^{-1}[x,y]_h z,\end{equation}
        where recall $\epsilon$ is the FS indicator of $\cN$. By some simple manipulations this is equivalent to the (anti-)symmetry condition
        \begin{equation}\gamma(y,x) = \epsilon \gamma(x,y).\end{equation}
        Our conditions for $\nu,\sigma$ above yield
        \begin{equation}\gamma(gx,y) = \nu(g) \gamma(x,\sigma(g)y).\end{equation}
        This compatibility condition is actually very strong. It allows us to determine the (anti-)symmetry of $\gamma$ in terms of $f$ and $\nu$, and we find the anomaly-matching condition
        \begin{equation}{\rm sign}\left(\sum_{g \in G\ |\ \sigma(g) = g} \nu(g)\right) = \epsilon.\end{equation}
        Note that on this subgroup of $\sigma$-fixed elements of $G$, $\nu$ gives a quadratic refinement of the torus partition function pairing, and the left-hand-side above is the Arf invariant of this quadratic form.
        
        To summarize, self-dual $G$-SPT phases are given by a 2-cocycle $\alpha$ which is non-degenerate, equivalently for which there is an involution $\sigma:G \to G$ satisfying \eqref{eqninvolcond}. To promote this to a symmetric phase for the associated TY category, we need a choice of function $\nu:G \to U(1)$ satisfying \eqref{eqnquadratic} and such that its associated quadratic form on $\sigma$-fixed $G$ elements has Arf invariant $\epsilon$. The ambiguity in $\nu$ is that $\nu$ gets shifted by a 1-cochain $\lambda$ whenever $\alpha \mapsto \alpha + \delta \lambda$. In particular without shifting $\alpha$ we can shift $\nu$ by a homomorphism, i.e. a 1-cochain $\lambda$ with $\delta \lambda = 0$. There are no other ambiguities.
        
        \subsubsection{$\bZ_2 \times \bZ_2$ Examples, Edge Modes between Fiber Functors}\label{secZ2Z2fiberfuncs}
        
        As we have discussed above, there are two choices of non-degenerate bicharacter $\chi$ for $G = \bZ_2 \times \bZ_2$, either the diagonal or off-diagonal, and two choices of FS indicator $\epsilon = \pm 1$ for the duality line, giving rise to four $\bZ_2 \times \bZ_2$ TY categories. Meanwhile, besides the trivial state, which is not duality-invariant, there is one $\bZ_2 \times \bZ_2$-SPT phase which is duality-invariant for either action, and with the analysis of the previous section we can now determine when it gives rise to a TY-symmetric state, hence a fiber functor.
        
        For the diagonal bicharacter, yielding $\chi(A_1,B_1,A_2,B_2) = A_1 A_2 + B_1 B_2$, one can check with the SPT cocycle $\alpha(A,B) = AB$, the associated involution is $\sigma(A,B) = (B,A).$  The phase factor $\nu$ may be either $+1$ or $-1$ on the fixed element $(1,1)$. We see only the $+1$ choice can be associated with a fiber functor, for the choice of sign $\epsilon = +1$, corresponding to the TY category known as ${\rm Rep}(H_8)$\footnote{$H_8$ is a non-grouplike Hopf algebra of dimension 8 known as the Kac-Paljutkin algebra \cite{kats1966finite}.}, while the TY category with this $\chi$ but $\epsilon = -1$ admits no fiber functor, and is therefore anomalous.
        
        For the off-diagonal $\chi$ for $\bZ_2 \times \bZ_2$, the self-dual SPT phase is associated with the identity involution $\sigma$. We thus get to choose a quadratic form on $\bZ^x_2 \times \bZ^y_2$ which refines the pairing $xy$. There are three different ones with positive Arf invariant, corresponding to three different fiber functors of the TY category with this $\chi$ and $\epsilon = +1$, aka ${\rm Rep}(D_8)$. Meanwhile there is one quadratic form with negative Arf invariant, yielding a single fiber functor for the TY category with this $\chi$ and $\epsilon = -1$, aka ${\rm Rep}(Q_8)$. This seems to violate the intuition that the nontrivial FS indicator of the duality line implies an anomaly analogous to the familiar $\bZ_2$ anomaly, but since the TY category gives a nontrivial extension of the $\bZ_2$ twisted group fusion category, it is logically possible for this anomaly to be cancelled, as we see it is in this example. We discuss both of these examples from the perspective of gauge theory in Section \ref{secfinitegauge} below.

        The three inequivalent ${\rm Rep}(D_8)$ fiber functors are especially interesting because they are all associated with the same $\bZ_2 \times \bZ_2$-SPT, yet we expect to find degenerate edge modes between them. Indeed, let $\nu,\nu'$ be the quadratic forms associated with two inequivalent fiber functors. Then $\nu/\nu':\bZ_2 \times \bZ_2 \to U(1)$ is a non-trivial homomorphism. A quick calculation shows that $g \in \bZ_2 \times \bZ_2$ and the duality line anti-commute at the junction when $\nu(g)/\nu'(g) = -1$:
        \begin{equation}
        \adjincludegraphics[width=3cm,valign=c]{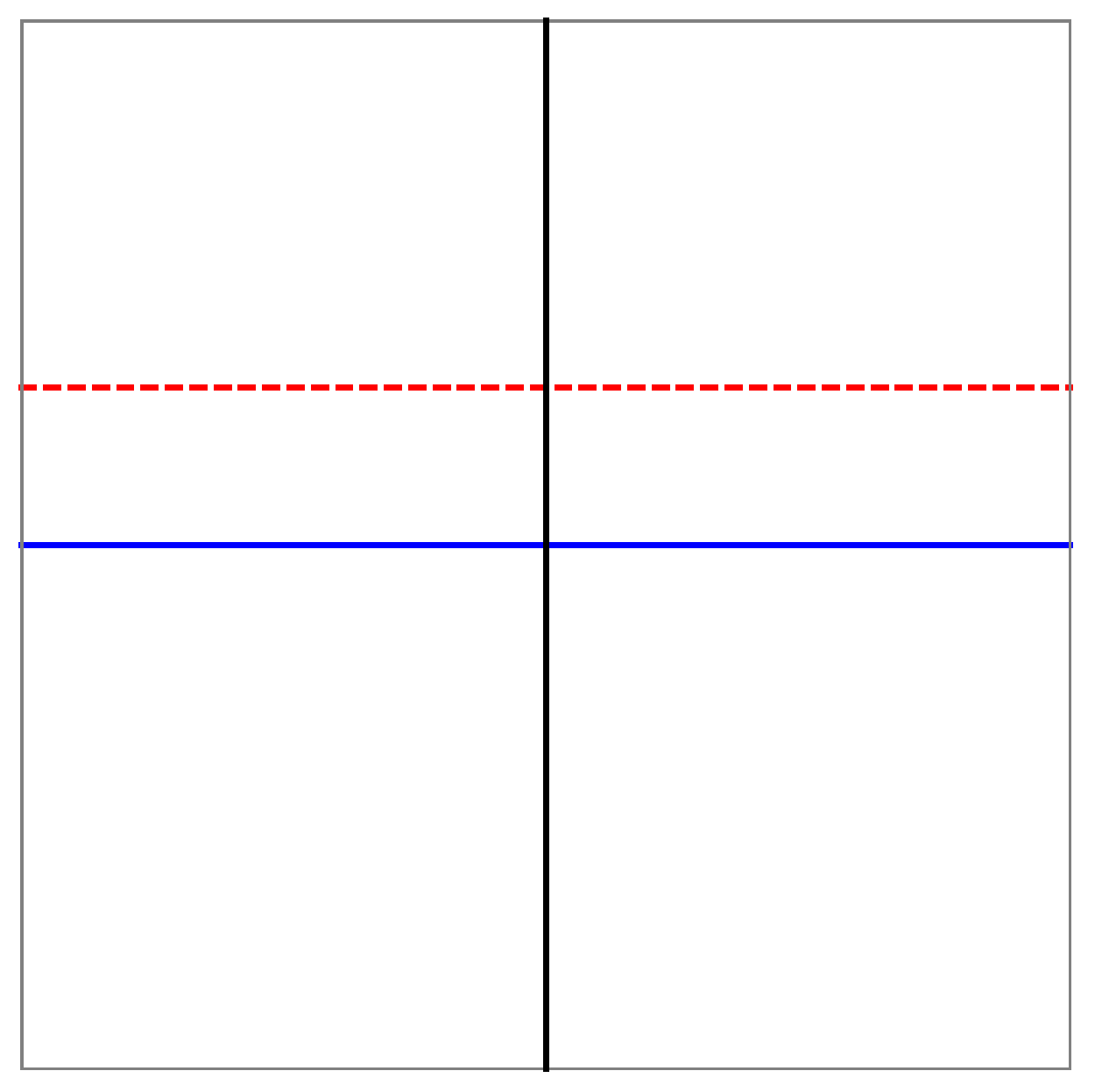}=\adjincludegraphics[width=3cm,valign=c]{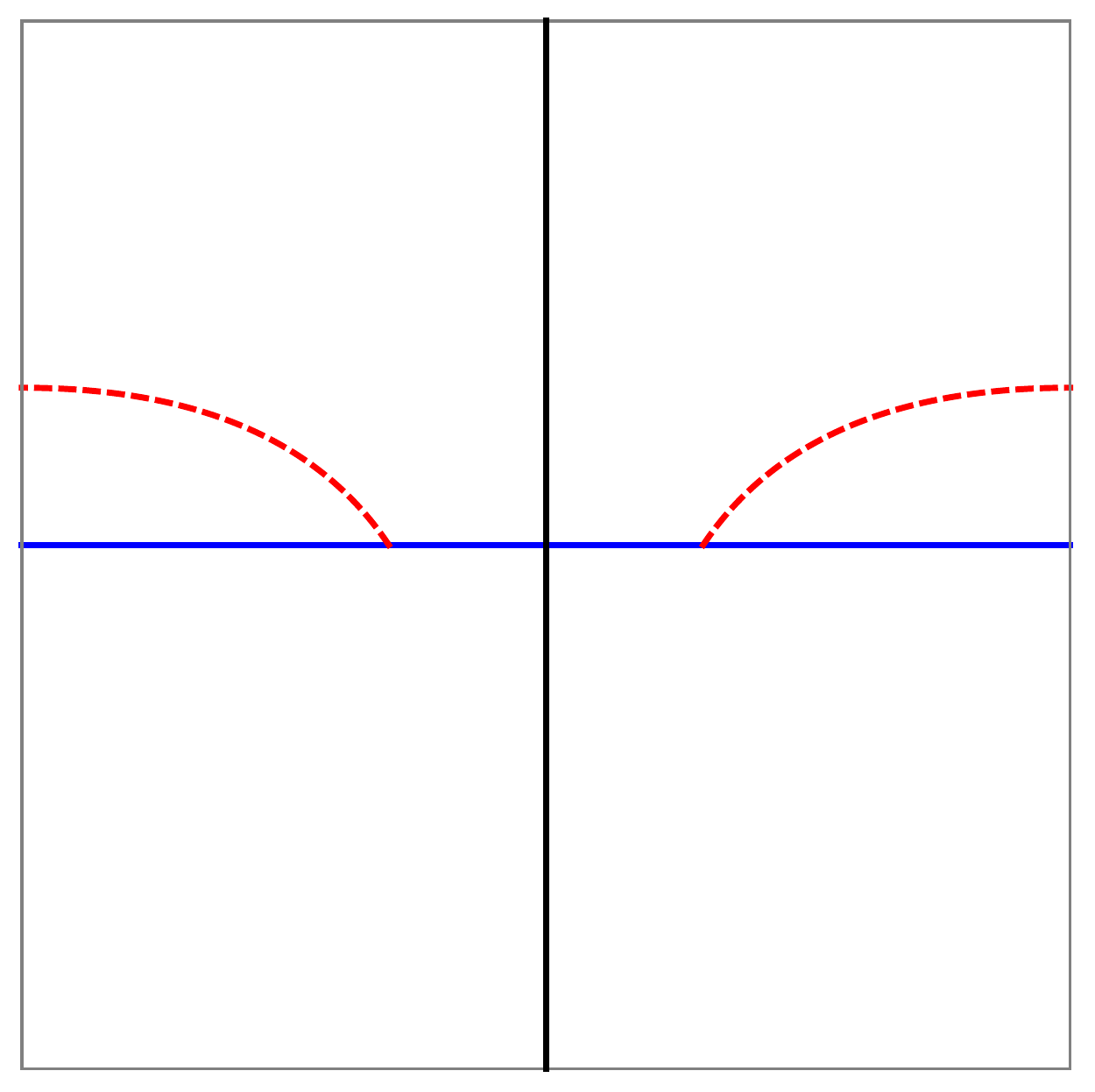}=\frac{\nu(g)}{\nu'(g)}\adjincludegraphics[width=3cm,valign=c]{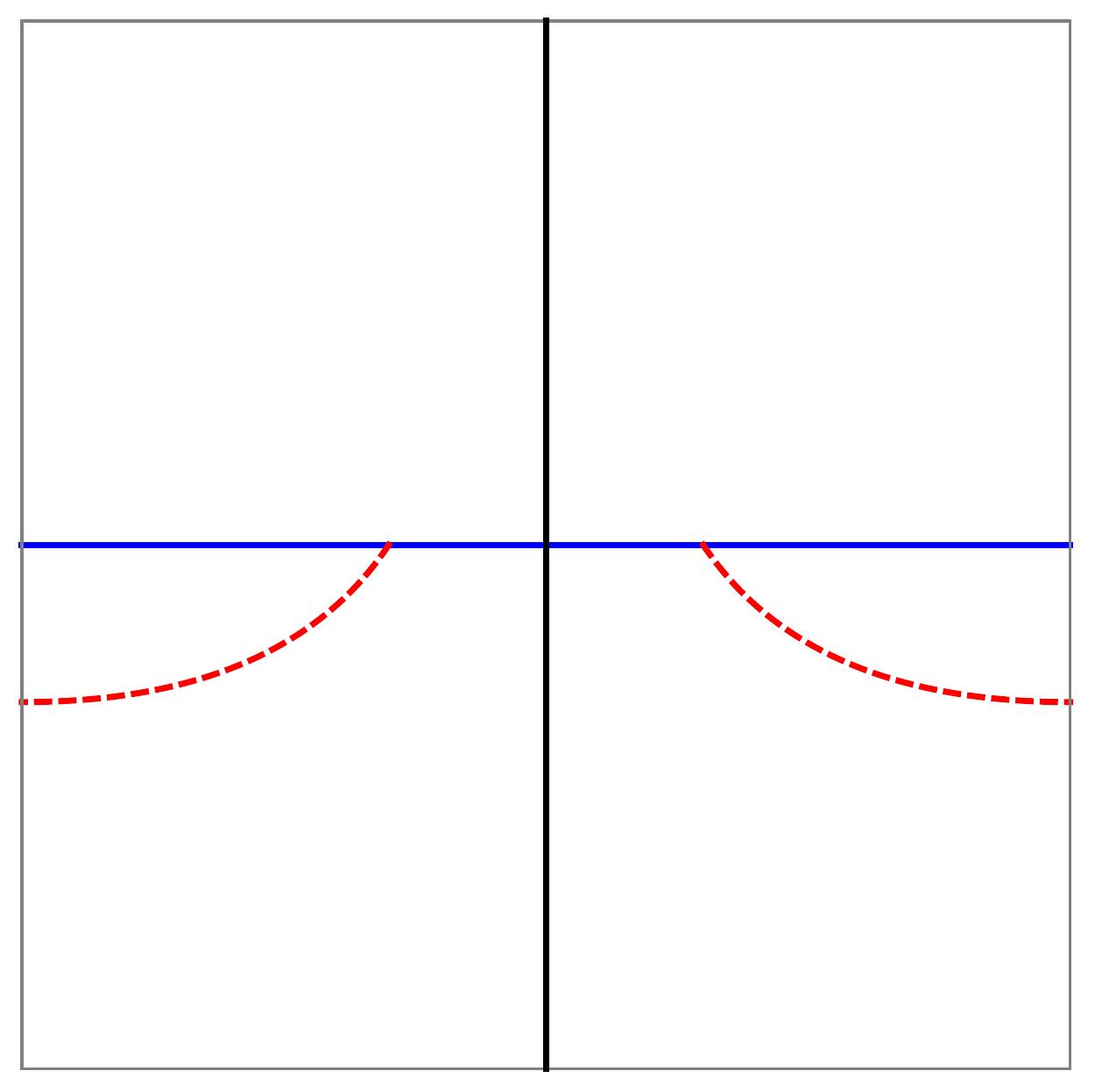}=\frac{\nu(g)}{\nu'(g)}\adjincludegraphics[width=3cm,valign=c]{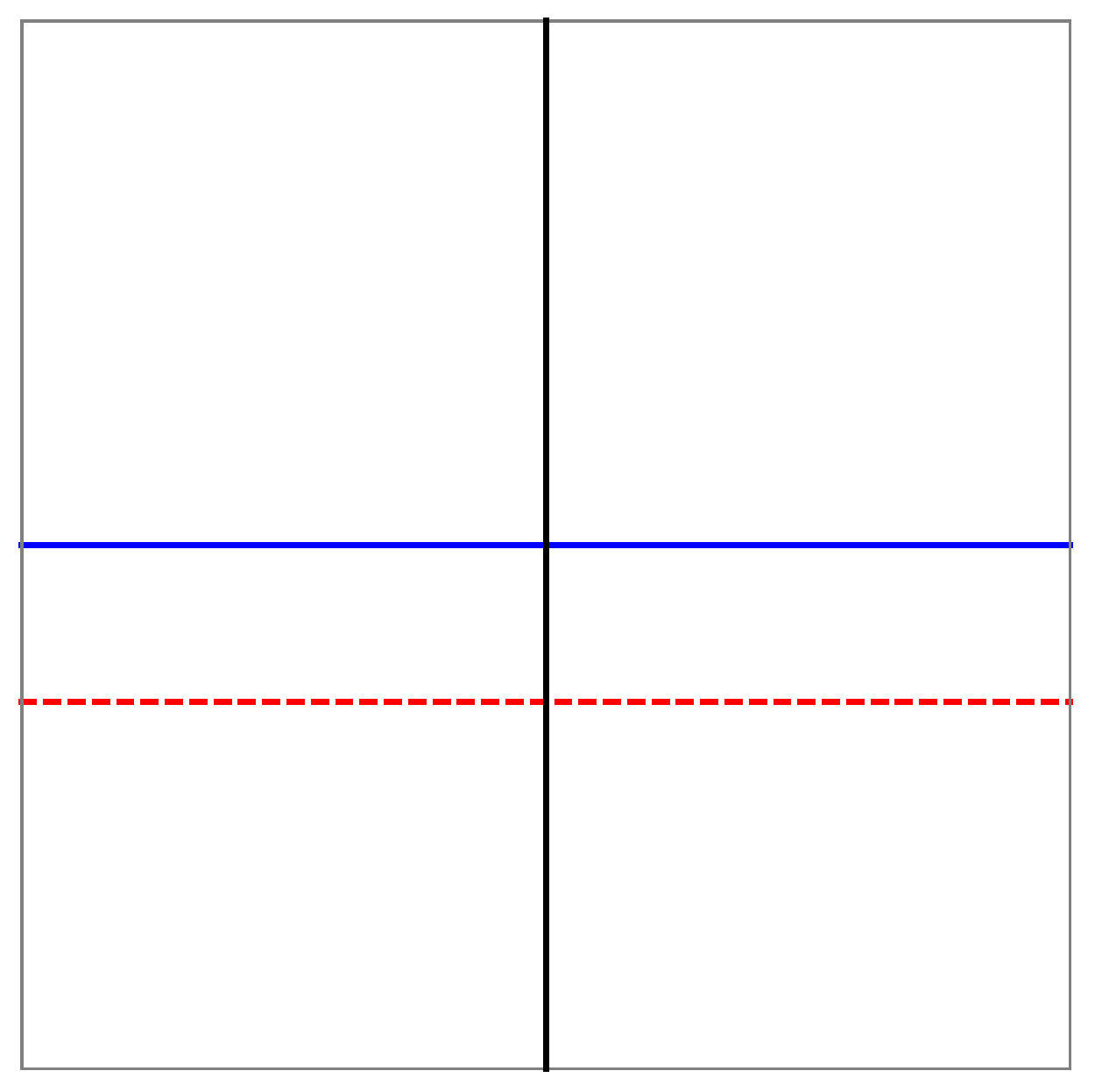},
        \end{equation}
        where on the right we have the fiber functor with $\nu$ and on the left the fiber functor with $\nu'$, the black line is the junction between them, the blue line is the duality defect, and the red line is the symmetry defect $g$. This implies a degenerate junction, even though $\bZ_2 \times \bZ_2$ acts linearly there.

        \subsubsection{General Self-Dual Phases}\label{secgenselfdual}
        
        More generally we may have also have degenerate phases with unbroken self-duality, but where $G$ is partially broken with only a subgroup $H<G$ preserved. Symmetry fractionalization in partial symmetry broken phases can be analyzed in a single ground state---the other states are then determined by the action of $G$. Since $G$ is abelian, the action of $G$ on $H$ is trivial, so the $H$-symmetry fractionalization in all ground states will be determined by a single SPT class $\alpha \in H^2(BH,U(1))$. We will observe a similar logic applies to the study of the duality line. Indeed, we will construct a different Tambara-Yamagami action on a single ground state and thereby reduce to the conclusions of the previous section. For a rigorous approach, which proves sufficiency of our data, see \cite{Meir_2012}.

        The $H$-SPT class $\alpha \in H^2(BH,U(1))$ which characterizes the fractionalization of the remaining symmetry satisfies a weaker non-degeneracy condition than we studied above to ensure self-duality. Let $H^\perp$ denote the subgroup
        \begin{equation}H^\perp = \{g \in G \ |\ \chi(g,h) = 0 \quad \forall h \in H\} < G.\end{equation}
        We need the co-isotropy condition $H^\perp < H$, and moreover defining the radical ${\rm Rad}(\alpha)$ to be the subgroup of $H$ elements which are invisible in the torus partition function, self-duality is equivalent to
        \begin{equation}H^\perp = {\rm Rad}(\alpha).\end{equation}
        
        Observe that the sum over $G$ gauge fields in the duality transformation
        \begin{equation}Z(X,B) = \# \sum_{A \in H^1(X,G)} Z(X,A) e^{i \int \chi(A,B)}\end{equation}
  		restricts to a sum over $A \in H^1(X,H)$, since $Z(X,A)$ vanishes for other gauge backgrounds, being a symmetry-breaking phase. Further, $Z(X,A)$ is insensitive to a shift of $A$ by a gauge field valued in ${\rm Rad}(\alpha)$, by definition, which is also $H^\perp$, by assumption, so we can define the partition function $Z(X,\bar A)$ where $\bar A \in H/H^\perp$. We have the induced duality transformation
  		\begin{equation}Z(X,\bar B) = \# \sum_{\bar A \in H^1(X,H/H^\perp)} Z(X,\bar A) e^{i\int \bar\chi(A,B)},\end{equation}
  		where now $\chi$ defines a \emph{nondegenerate} form $\bar \chi$ on $H/H^\perp$.
  		
  		This transformation is associated with the same duality line as the original transformation, but now defines a $(H/H^\perp,\bar\chi)$ Tambara-Yamagami action on a single ground state of our system, which represents a fiber functor for this associated category. We can now repeat the arguments of the previous section and find that our self-dual symmetry-breaking phase is classified by an involution
  		\begin{equation}\sigma:H/H^\perp \to H/H^\perp\end{equation}
  		satisfying
  		\begin{equation}\frac{\alpha(h,\sigma(h'))}{\alpha(\sigma(h'),h)} = \bar \chi(h,h'),\end{equation}
  		as well as a $\nu:H/H^\perp \to U(1)$ with
  		\begin{equation}\nu(h)\nu(\sigma(h)) = 1\end{equation}
  		\begin{equation}\nu(h)\nu(h')\nu(hh')^{-1} = \frac{\alpha(h,h')}{\alpha(\sigma(h'),\sigma(h))},\end{equation}
  		and the anomaly-vanishing condition
  		\begin{equation}{\rm sign}\left(\sum_{h \in H \ | \ \sigma(h) = h} \nu(h)\right) = \epsilon.\end{equation}

  		\subsection{Gapped Phases of Finite Gauge Theories}\label{secfinitegauge}
  		
  		An interesting special case of what we have described applies to finite gauge theories. In a gapped phase of $G$ gauge theory where $G$ is a finite (possibly nonabelian) group, the Wilson lines are topological operators generating an integral fusion category ${\rm Rep}(G)$ (the quantum dimension of a Wilson line is the dimension of its representation). When $G$ is abelian, then ${\rm Rep}(G) = {\rm Vec}_{G^*}$ where $G^* = \hom(G,U(1))$, and the symmetries generated by the Wilson line are grouplike symmetries known as the magnetic symmetries of the gauge theory.
  		
  		The nonabelian case is more interesting, since now ${\rm Rep}(G)$ contains simple objects which are not invertible, corresponding to irreps of dimension greater than one, so one needs the fusion category perspective to analyze the symmetry properly. A discussion of these nonabelian magnetic symmetries may be found in \cite{bhardwaj2017finite}. Note that all fusion category symmetries obtained this way are anomaly-free because every ${\rm Rep}(G)$ admits a fiber functor.
  		
  		In a similar spirit as \cite{gukov2013topological,Kapustin_2017}, we can identify the different gapped phases of the gauge theory by the action of ${\rm Rep}(G)$ on its ground states. For example, the completely ${\rm Rep}(G)$-symmetry broken phase, corresponding to the fixed boundary condition of ${\rm Rep}(G)$-Turaev-Viro theory corresponds to the deconfined phase of the 1d gauge theory. Indeed, the ground states of this phase are labelled by the holonomy of the gauge field around the spatial cycle, which can be any conjugacy class in $G$. The action of the Wilson line operators is diagonal in this basis (from a symmetry-breaking perspective these are the cat states) and in the state corresponding to some conjugacy class the eigenvalue of the Wilson line of representation $R$ is the trace of that conjugacy class in $R$.
  		
  		Meanwhile the fiber functor itself describes the Higgs phase of the gauge theory, where it can only have the trivial holonomy around the spatial cycle, and the Wilson lines are again diagonal, and their eigenvalues are the dimensions of the representations. More complicated module categories of ${\rm Rep}(G)$ correspond to partial Higgs phases of the gauge theory.
  		
  		One can apply a similar reasoning to general $d$ space dimensions, but where now since the Wilson lines are codimension $d-1$, they generate $d-1$-form symmetries \cite{Kapustin_2017,Gaiotto_2015}. For $G$ abelian, these symmetries are described by a $d$-group with $G^*$ in degree $d$ and a trivial group in all other degrees. For general $G$, these symmetries will be described by a suitable kind of fusion $d$-category, with only identity objects and $k$-morphisms for $1\le k\le d-2$ and a $d-1$ morphism for each object of ${\rm Rep}(G)$ with composition given by fusion of representations.
  	 
  	    \subsubsection{Dihedral Group $D_8$}
  	    
  	    We can express the dihedral group $D_8$ of 8 elements by two generators $r$, an order 4 rotation, and $s$ a reflection:
  	     \ie
            D_8=\la s,r | s^2=r^4=(rs)^2=1\ra .
         \fe
  	   This has a faithful 2-dimensional real representation generated by the Pauli matrices $s = \sigma^x$ and $sr = \sigma^z$. In this representation, the central element $r^2 = \sigma^x \sigma^z \sigma^x \sigma^z$ acts as the scalar matrix $-1$. The conjugacy classes of subgroups of $D_8$ are:
  	    \ie D_8^*\fe
  	    \ie \langle s,r^2\rangle^*, \langle sr, r^2\rangle^*, \langle r \rangle\fe
  	    \ie \langle s \rangle, \langle sr \rangle, \langle r^2 \rangle\fe
  	    \ie 1.\fe
  	    The stable gapped phases of $D_8$ gauge theory are given by Higgsing the gauge field down to one of the subgroups above, and then for the starred subgroups $H$ we also get to choose a topological term in $H^2(H,U(1)) = \bZ_2$. This yields a total of 11 phases.
  	    
  	    As a fusion category, ${\rm Rep}(D_8)$ is equivalent to the $\bZ_2 \times \bZ_2$ TY category obtained using the off-diagonal bicharacter $\chi$ and taking positive Frobenius-Schur indicator $\epsilon = 1$, whose phases we have identified above. The bijection between that classification and the gauge theory classification is rather nontrivial, complicated by the fact that $\bZ_2\times \bZ_2$ has an automorphism group $S_3$ generated by
  	    \begin{equation}S:a,b \mapsto b,a\end{equation}
  	    \begin{equation}T:a,b \mapsto a+b,b\end{equation}
  	    which gives an $S_3$ action on the fusion category, but where only $S$ is realized as an automorrphism of $D_8$ (${\rm Out}(D_8) = \bZ_2$).\footnote{It is also known that ${\rm Rep}(D_8)$ is also Morita equivalent to the $\bZ_2^3$ group category with cubic twist \cite{Mu_oz_2018,maya2018classification}, for which the $S_3$ action is the one which permutes the $\bZ_2$ factors.} The $\bZ_3$ generated by $ST$ permutes the three fiber functors, the three 2 ground state self-dual symmetry breaking phases $Ix, Iy, Iz$, their three 4 ground state associated duality breaking phases, while the 2 ground state duality breaking phase associated with the $SPT$ and the 5 ground state phase of the orbit of the trivial phase are fixed by all automorphisms.
  	    
  	    However, we can make progress by evaluating the Wilson lines in each of the phases above. For example, the phase associated with the $\bZ_2$ subgroup $\langle r^2 \rangle$ has 2 ground states, corresponding to a $D_8$ gauge holonomy which is either trivial or the central element. We can identify the $\bZ_2 \times \bZ_2$ subcategory of the TY presentation with the abelian representations of $D_8$, in all of which $C$ acts trivially. Thus, $\langle r^2 \rangle$ corresponds to a $\bZ_2 \times \bZ_2$-symmetric phase, which we identify as the spontaneous duality-breaking phase associated with the $\bZ_2 \times \bZ_2$-SPT. Indeed, duality is broken in this phase because the Wilson line of the two dimensional $D_8$ irrep has a nontrivial action of $r^2$.
  	    
  	    Meanwhile, we see that the phases with holonomy restricted to $\langle s\rangle, \langle sr \rangle$ correspond to the self-dual phases $Ix$ and $Iz$, exchanged by the automorphism $S$. The third element in this $S_3$ orbit actually comes from the ``deconfined" phase associated to $D_8$ but with the nontrivial topological term. Indeed, the topological term makes the gauge field self-confine, by giving nontrivial gauge charge to certain holonomies. We find that the only nontrivial allowed holonomy is $r$, which we see gives the correct action of the Wilson lines in this phase. This illustrates how nontrivial the triality is from the perspective of the gauge theory.
  	    
  	    The three associated duality-breaking four-ground-state phases are given by $\langle s,r^2\rangle$, $\langle sr,r^2\rangle$, $\langle r\rangle$ (with trivial topological terms) just by counting. Note that the subgroup structure of these is $\bZ_2 \times \bZ_2$ for the first two and $\bZ_4$ for the last one. Nonetheless, they form a single $S_3$ orbit. Observe that duality is broken in all three phases as all three allow a non-trivial holonomy in the center of $D_8$.
  	    
        Finally we have the three fiber functors, which also form an $S_3$ orbit. One of them is associated with the complete Higgs phase where the $D_8$ gauge holonomy is restricted to be trivial. The other two are associated with the $\bZ_2 \times \bZ_2$ subgroups $\langle s,r^2\rangle$, $\langle sr, r^2\rangle$ with the non-trivial (and non-degenerate in the sense above) topological term, which self-confines the gauge holonomy to be trivial.
        
        As we have discussed above, these fiber functors host edge modes between them. For the boundary between the complete Higgs phase and the other two fiber functors, we can think of this as a symmetry-breaking boundary condition of a 1d $\bZ_2 \times \bZ_2$ gauge theory with topological term. This is equivalently described as the boundary of a nontrivial $\bZ_2 \times \bZ_2$-SPT for the magnetic symmetry, so there is a doubly degenerate edge mode.
        
        More interesting is the boundary between the two $\bZ_2 \times \bZ_2$ gauge theories. We again fold the theory at the junction. We can think about the junction as the edge of an SPT producted by three commuting magnetic $\bZ_2$ symmetries in the bulk, one which measures the $s$ or $sr$ holonomy of the gauge field on the left or right of the junction, respectively, and one which measures the $r^2$ holonomy along the whole system (which cannot be separated into left and right pieces, since by a gauge transformation in the center, which is everywhere unbroken, we can move contributions to the holonomy from one side to the other). The total SPT class among these three $\bZ_2$ symmetries (from the two SPT classes on each side of the junction) is $\frac{1}{2}(A_1 \cup B + A_2 \cup B) \in H^2(B\bZ_2^3,\bR/\bZ)$, where $A_{1,2}$ couple to the former $\bZ_2$'s and $B$ to the latter. This realizes a two-dimensional projective representation where $A_1,A_2$ both couple to the Pauli matrix $\sigma^x$, while $B$ couples to the Pauli matrix $\sigma^z$, which anticommutes with $\sigma^x$. This corresponds to a group extension
        \ie \bZ_2 \to D_8 \times \bZ_2 \to \bZ_2^3,\fe 
        where the $\bZ_2$ factor in $D_8 \times \bZ_2$ acts trivially in the 2d representation.
        
        In terms of the $\bZ_2 \times \bZ_2$ TY fiber functors and their junctions, according to Section \ref{secZ2Z2fiberfuncs}, this third one corresponds to a junction where the ration of quadratic functions $\nu/\nu'$ is odd on both generators (hence even on the diagonal element, being a homomorphism), reflecting the term $\frac{1}{2}(A_1 \cup B + A_2 \cup B)$, since the duality line acts as the magnetic symmetry for the center. Likewise one matches the other junctions with ones where $\nu/\nu'$ is odd on a single generator. Thus we can match the quadratic forms $\nu$ with the subgroups of $D_8$:
        \ie \langle s,r^2\rangle \sim +-++,\fe 
        \ie \langle sr, r^2 \rangle \sim ++-+,\fe 
        \ie 1 \sim +++-,\fe 
        where we label the quadratic forms by their values on the elements of $\bZ_2^2$ in the order $(0,0)$, $(1,0)$, $(0,1)$, $(1,1)$. This matches the proposed $S$ action on the fiber functors.

  	    \subsubsection{Quaternion Group $Q_8$}
  	    
  	    The quaternion group $Q_8$ of 8 elements can be presented by generators $I,J,K$ with the relations $I^2 = J^2 = K^2 = C$, a central $\bZ_2$ element, $IJ = K$, $JI = CK$. The conjugacy classes of subgroups are
  	    \ie Q_8\fe
  	    \ie \langle I\rangle, \langle J \rangle, \langle K \rangle\fe
  	    \ie \langle C \rangle\fe
  	    \ie 1.\fe
  	    There are no available topological terms for these subgroups, so there are only 6 stable gapped phases.
  	    
  	    The corresponding TY category is given by $\bZ_2 \times \bZ_2$ with the off-diagonal bicharacter $\chi$ and nontrivial twist $\epsilon = -1$. As we have discussed, this category has a single fiber functor, corresponding to the complete Higgs phase above. The three four ground state phases associated with $Ix$, $Iy$, $Iz$ are given by the partial Higgs phases $\langle I\rangle$, $\langle J\rangle$, $\langle K\rangle$. The two gound state phase given by the spontaneous duality breaking $\bZ_2 \times \bZ_2$-SPT corresponds to $\langle C\rangle$, and the five ground state phase corresponds to the deconfined phase of $Q_8$, which has five conjugacy classes.
  	    
  	    Observe that while it was possible to construct a self-dual SPT phase, according to our analysis above, the symmetry breaking phases $Ix$, $Iy$, $Iz$ do not extend to self-dual phases because $\epsilon = -1$ and there are no quadratic forms $\nu:\bZ_2 \to U(1)$ with negative Arf invariant.
  
 \section{Six Self-Dualities of The Ising$^2$ Model}\label{secisingdualities}
 
 In this section, we will discuss the relationship between fusion category symmetry, especially the Tambara-Yamagami categories, and self-duality under gauging finite subgroups, especially the well-known Kramers-Wannier duality of the critical Ising model. We will see that the fusion ring is determined by the global symmetry data, while determining the full category with its $F$-symbols requires studying the duality twisted sectors, which we approach by modular bootstrap techniques (see also \cite{Ji_2019}). We find that the $c= 1$ Ising$^2$ CFT is a stable gapless edge for four Turaev-Viro/Levin-Wen theories, including one gauge-theory-like model based on the Hopf algebra $H_8$, which is anomaly-free.
 
 \subsection{Ising Kramers-Wannier Duality}
 
 First, we discuss Kramers-Wannier (KW) duality of the 1+1D critical Ising model. This $c = 1/2$ theory has two non-vacuum primary fields $\sigma$, the order parameter, and $\epsilon$, the energy operator. The former is odd under the global $\bZ_2$ symmetry while the latter is even.
 
 Because it is the unique $c = 1/2$ CFT, this theory is self-dual under $\bZ_2$ gauging. We would like to promote this to a fusion category symmetry. First, we find that because $\bZ_2$ gauging exchanges the ordered and disordered phases, it must square to a projector, so the fusion rules are determined to be
  \ie \cN^2 = 1 + g\fe
 \ie g\cN = \cN.\fe
 
 There are two fusion categories Ising$_\pm$ with these fusion rules, distinguished by the Frobenius-Schur indicator of the duality line $\cN$. To determine this, we will need to compute some twisted partition functions.
 
 First, by the fusion rules the allowed eigenvalues of $\cN$ are $\pm \sqrt{2},0$. Since $\sigma$ is charged under the global symmetry, it is sent to zero by duality. Meanwhile, the energy operator $\epsilon$ has eigenvalue $-\sqrt{2}$, since if we perturb the theory by this operator, depending on the sign, we find either the ordered or disordered $\bZ_2$ phases, which are exchanged by the duality. Therefore, the twisted partition function with the duality line $\cN$ inserted around the spatial cycle may be written
 \ie Z_{1\cN} = \sqrt{2} |\chi_0|^2 - \sqrt{2} |\chi_{1 \over 2}|^2,\fe
  where $\chi_h$ are the Virasoro characters at $c={1\over 2}$:
 \ie
 \chi_0=&{1\over 2}\left(
 \sqrt{\theta_3\over \eta}+\sqrt{\theta_4\over \eta}
 \right)
 \\
 \chi_{1\over 2}=&{1\over 2}\left(
 \sqrt{\theta_3\over \eta}-\sqrt{\theta_4\over \eta}
 \right)
 \\
 \chi_{1\over 16}=&{1\over \sqrt{2}} 
 \sqrt{\theta_2\over \eta}.
 \label{IsingChars}
 \fe
Using the modular $S$-matrx, 
\ie
S={1\over 2} \begin{pmatrix}
1 & 1 & \sqrt{2}
\\
1 & 1 & -\sqrt{2}
\\
\sqrt{2} & -\sqrt{2} & 0
\end{pmatrix}
\fe
we find by an $S$ transformation the duality-twisted partition function
 \ie Z_{\cN 1} = \chi_0 \bar \chi_{1 \over 16} + \chi_{1 \over 2} \bar \chi_{1 \over 16} + c.c.\fe

We can now determine the Frobenius-Schur indicator using modularity. Indeed, the Frobenius-Schur indicator gives a spin-selection rule for $Z_{\cN 1}$. If we perform two modular $T$ transformations, we find
 \begin{equation}\label{eqnspinsel}
\adjincludegraphics[width=2cm,valign=c]{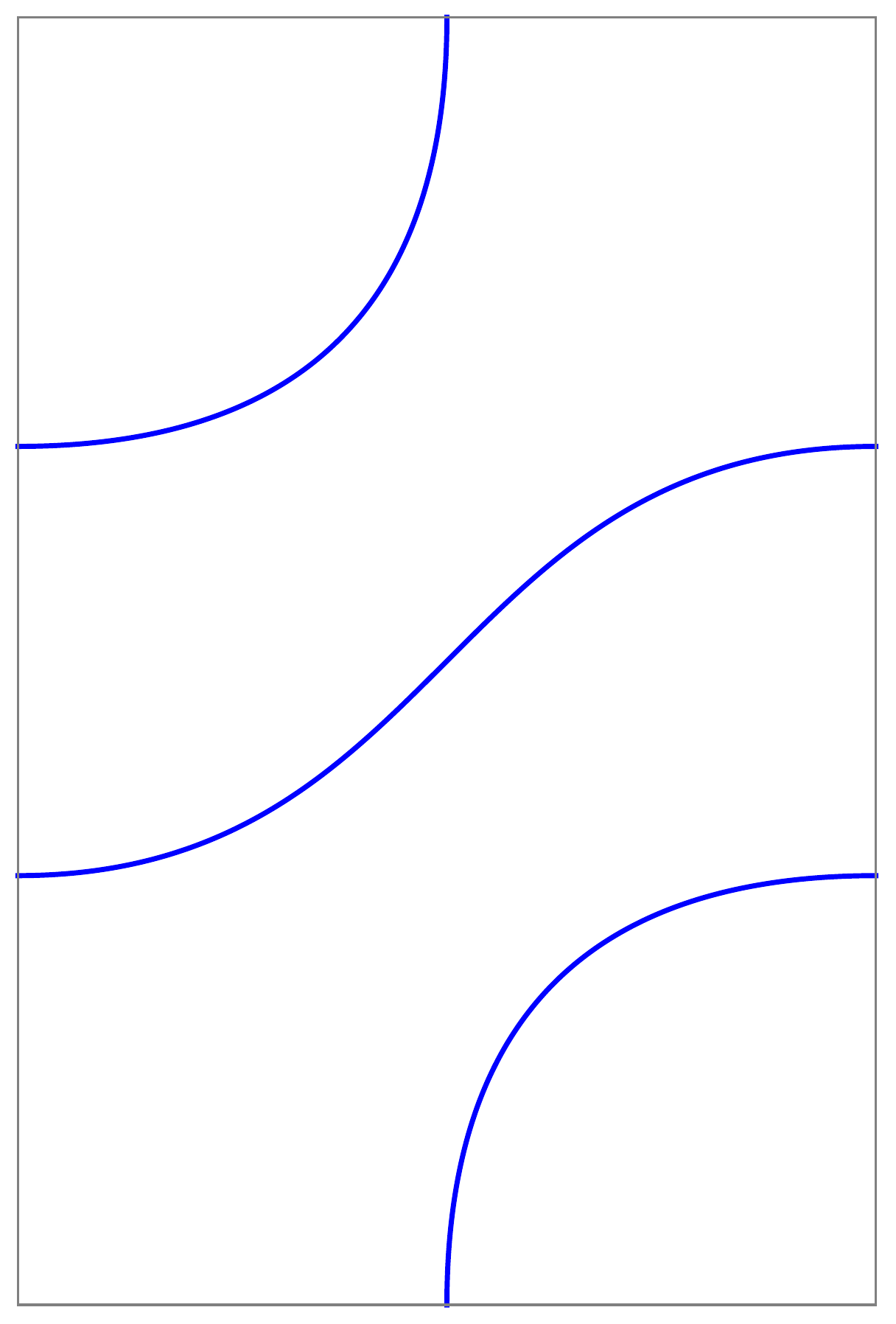} = \frac{\epsilon}{\sqrt{2}}\adjincludegraphics[width=2cm,valign=c]{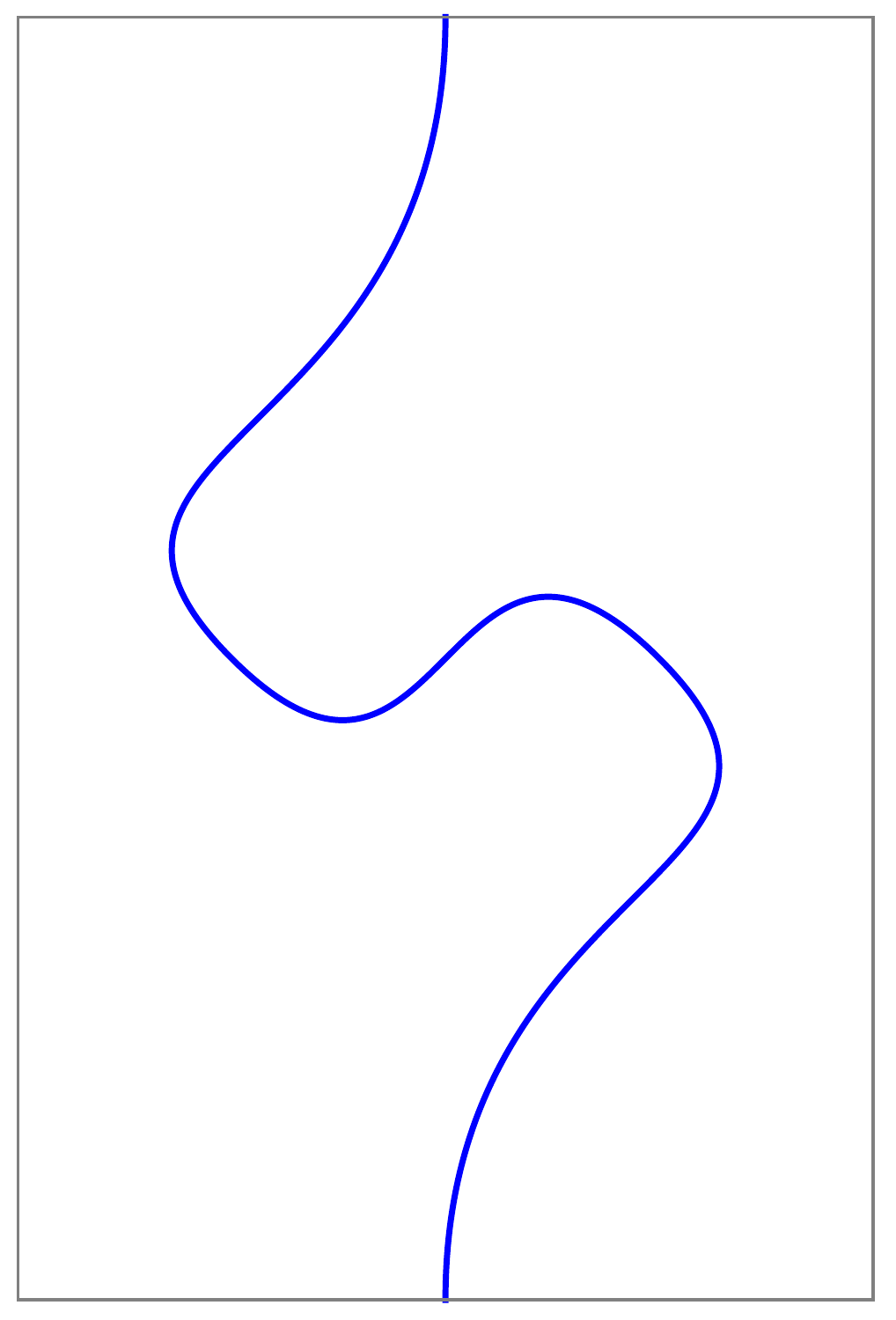} + \frac{\epsilon}{\sqrt{2}} \adjincludegraphics[width=2cm,valign=c]{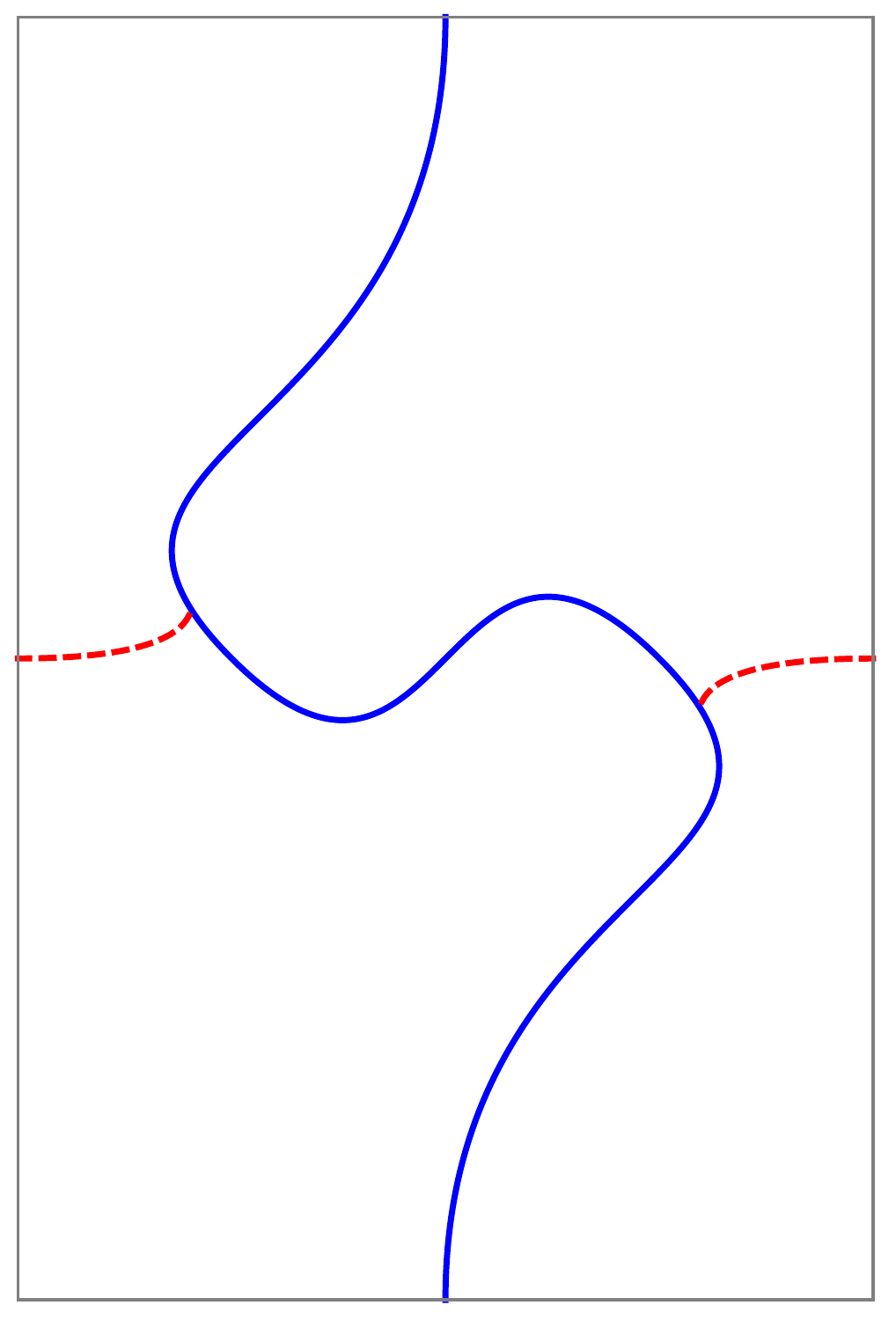},
\end{equation}
or equivalently
 \ie\label{eqnspinrel}
 Z_{\cN 1}(\tau + 2) = \frac{\epsilon}{\sqrt{2}} (Z_{\cN 1}(\tau) + Z_{\cN g+}(\tau)),
 \fe
 where $\epsilon = \pm 1$ is the Frobenius-Schur indicator and $Z_{\cN g+}$ is the twisted partition function with $\cN$ around the time cycle and $g$ around the spatial cycle, with the crossing resolved as above.
 
 We can compute that the $\bZ_2$ symmetry charges in the duality-twisted sector must be $\pm i$, since two duality defects can fuse to an order parameter. We find the only choice choice consistent with \eqref{eqnspinrel} is
 \ie Z_{\cN g+} = i (\chi_0 \bar \chi_{1 \over 16} + \chi_{1 \over 2} \bar \chi_{1 \over 16}) + c.c.,\fe
 \ie \frac{1}{\sqrt{2}} (Z_{\cN 1} + Z_{\cN g+}) = e^{2\pi i/8} Z_{\cN 1},\fe
 in particular the Frobenius-Schur indicator must be $\epsilon = +1$.
 
 A CFT with Ising$_-$ symmetry may be constructed as a product of the Ising CFT with a $c = 1$ compact boson, where we take $\cN$ to act as KW duality on the Ising factor and as an anomalous $\bZ_2$ symmetry on the compact boson factor, which flips the sign of the Frobenius-Schur indicator.
 
 \subsection{Dualities in the Ising$^2$ CFT}
 
 If we take two decoupled critical Ising chains, we obtain an interesting $c = 1$ CFT we refer to as Ising$^2$. The CFT is rational and the corresponding chiral algebra has two generators of spin 2
 \ie
 T=T_1+T_2,\quad T'=T_1-T_2
 \fe
 given by combinations of the individual stress tensors $T_i$ of each Ising CFT. The chiral primaries in the {Ising$^2$ CFT are  given by the tensor product of the 3 primaries $\{1,\epsilon_i,\sigma_i\}$ in each Ising factor.

The CFT has a global $D_8$ symmetry presented by
 \ie
D_8=\la s,r | s^2=r^4=(rs)^2=1\ra .
\fe
This acts as on the fields as
\ie 
&s: (\sigma_1,\sigma_2,\epsilon_1,\epsilon_2) \mapsto (-\sigma_1,\sigma_2,\epsilon_1,\epsilon_2)
\\
&r: (\sigma_1,\sigma_2,\epsilon_1,\epsilon_2) \mapsto ( \sigma_2,-\sigma_1,\epsilon_2,\epsilon_1).
\label{IsingD8}
 \fe
 The theory is self-dual under gauging the $\bZ_2$ subgroups generated by $s$ and $sr^2$ as these are the KW dualities of each Ising factor, and each generates an Ising$_+$ fusion category symmetry.
 
 The Ising$^2$ CFT has an alternative realization which we will find useful: the $c=1$ compact boson CFT orbifolded by the $\bZ_2^C$ charge conjugation symmetry at the bosonization radius $R=2$. In this description, we introduce (normalized) non-chiral compact boson $\theta\equiv {X_L+X_R\over R}$ and its T-dual $\phi\equiv{R(X_L -X_R)\over 2}$ (normalized such that both $\theta$ and $\phi$ have unit radii) where $\bZ_2^C$ acts as
  \ie
  \mZ_2^C: (\theta, \phi)  \to  (-\theta, -\phi)
  \fe
  The operator spectrum in  orbifold description consists of two sectors, the untwisted sector and the $\mZ_2^C$ twisted sector. The latter is charged under the magnetic symmetry $\mZ_2^{\tilde C}$.
  
  The twisted sector consists of two Virasoro primaries
  \ie
  \rho_1{\rm ~and~}\rho_2,\quad (h,\bar h)=\left( {1\over 16}, {1\over 16}\right)
  \fe
  that correspond to the ground states at the two fixed points of $S^1/\mZ_2^C$, as well the first excited states (Virasoro primaries)
   \ie
  \tau_1{\rm ~and~}\tau_2,\quad (h,\bar h)=\left( {9\over 16}, {9\over 16}\right).
  \fe
  The untwisted sector includes reflection-invariant momentum-winding operators  $V^+_{n,w}$  defined by 
  \ie
  V^+_{n,w}={V_{n,w}(\theta, \phi)+V_{n,w}(-\theta, -\phi)\over \sqrt{2}}
  \fe
  where $V_{n,w}$ denotes the usual momentum-winding operators in the unorbifolded theory. 
  \ie
  V_{n,w}\equiv e^{in \theta +i w\phi} ,\quad (h,\bar h)=\left({1\over 2}\left({n\over R}+{w R\over 2}\right)^2,{1\over 2}\left({n\over R}-{w R\over 2}\right)^2\right).
  \fe
The rest of the  Virasoro primaries in the untwisted sector are built from $\mZ_2^C$ invariant normal-ordered Schur polynomials in  the $U(1)$ currents $d\theta, d\phi$ and their derivatives in the unorbifolded theory \cite{Dijkgraaf:1989hb}
  \ie
j_{n^2}j_{m^2}~{\rm with}~{m-n\in 2\mZ},~(h,\bar h )=(n^2,m^2)
  \fe
For example,
\ie
j_{1}=\pa X_L \qquad j_4=:j_1^4:-2:j_1\pa^2 j_1:+{3\over 2} :(\pa j_1)^2:
\fe
The exactly marginal operator that gives rise to the orbifold branch (which includes the Ising$^2$ point) is $j_1\bar j_1$. The spin-1 $U(1)$ currents themselves are projected out in the orbifold, but there is a spin $4n^2$ left-moving (and right-moving) current for every $n \in \mZ_+$ at general $R$ and they generate a W-algebra of type $W(2,4)$ that extends the Virasoro algebra. At special radii, this chiral algebra can be enhanced to an even larger W-algebra such that the CFT becomes rational, with a finite number of primaries and conformal blocks for this chiral algebra \cite{Dijkgraaf:1989hb}. Here the Ising$^2$ CFT at $R=2$ is precisely one such rational point. In this case, the two descriptions of the CFT are related at the operator level by
\ie
\sigma_1=i(\rho_1-\rho_2),\quad \sigma_2=\rho_1+\rho_2,\quad \epsilon_1=V_{2,0}^++V_{0,1}^+,\quad \epsilon_2=V_{2,0}^+-V_{0,1}^+.
\label{Ising2id}
\fe
In the orbifold description, the $D_8$ global symmetry acts as
 \ie\label{eqnD8raction}
r: (\theta, \phi)\mapsto (\theta+\pi, \phi+\pi),~s:(\theta, \phi)\mapsto (\theta+\pi, \phi).
\fe
on the operators in the untwisted sector, and  
\ie
&r: (\rho_{1},\rho_2) \mapsto (i\rho_{1},-i\rho_2),~(\tau_{1},\tau_2) \mapsto (i\tau_{1},-i\tau_2)
\\
&s: (\rho_{1},\rho_2) \mapsto (\rho_{2},\rho_1),~(\tau_{1},\tau_2) \mapsto (\tau_{2},\tau_1)
 \fe
for the operators in the twisted sector. Note that $r^2$ is the magnetic symmetry. One can check that this is consistent with the action of $D_8$ in the Ising variables \eqref{IsingD8} using \eqref{Ising2id}.

 \subsubsection{$\bZ_2 \times \bZ_2$ TY Categories}
 
 Let us consider self-dualities upon gauging the $\bZ_2 \times \bZ_2$ subgroup generated by both $s$ and $sr^2$. Each KW duality preserves the stress tensor of each Ising factor, so we can describe the action in terms of the $c = 1/2$ characters. Let us consider the diagonal KW duality $\cN_{\rm diag}$, which generates a $\bZ_2 \times \bZ_2$ Tambara-Yamagami category. A priori there are four such categories (four solutions to the pentagon equations) \cite{tambarayam}, 
 \ie
 \TY(\bZ_2 \times \bZ_2,\chi_{a,s},\epsilon)
 \fe
determined by two choices of bicharacter:
\ie
\chi_s(m_1,n_1,m_2,n_2) = (-1)^{m_1 m_2 + n_1 n_2} 
,\quad \chi_a(m_1,n_1,m_2,n_2) = (-1)^{m_1 n_2 + n_1 m_2}
\label{d4bichar}
\fe
(see also Section \ref{secpermaction}), and two choices of Frobenius-Schur indicator $\epsilon = \pm 1$. It is clear from its presentation as the diagonal TDL in Ising$_+\times$Ising$_+$ that $\cN_{\rm diag}$ generates the TY category with the diagonal bicharacter $\chi_s$ and $\epsilon = 1$, also known as ${\rm Rep}(H_8)$ which we discussed in Section \ref{secZ2Z2fiberfuncs}. However, let us also verify this from the twisted partition functions.
 
 By the fusion rules, this operator $\cN_{\rm diag}$ has eigenvalues $\pm 2, 0$ acting on the bulk Hilbert space. We find all operators involving the order parameters $\sigma_j$ are charged under $\bZ_2 \times \bZ_2$ and therefore sent to zero. Further, by perturbing to the nearby ordered/disordered phases, we see both energy operators $\epsilon_1 , \epsilon_2$ have eigenvalue $-2$.\footnote{Strictly speaking, we are assuming that $\epsilon_1 + \epsilon_2$ drives the theory into the trivial $\bZ_2 \times \bZ_2$ symmetric phase, rather than the SPT. However, this depends on how $\bZ_2 \times \bZ_2$ acts on the disorder operators. See Section \ref{secisingstar} below. For the product Ising$_+\times$Ising$_+$ symmetry, however, it is clear that both energy operators are odd under the diagonal duality.} Finally, using integrality and positivity of the partition functions, we see the exactly marginal operator $\epsilon_1\epsilon_2$ has eigenvalue $+2$, so
 \ie Z_{1\cN_{\rm diag}} = 2|\chi_0|^4 - 4|\chi_0 \chi_{1 \over 2}|^2 + 2|\chi_{1 \over 2}|^4.\fe 
 By a modular $S$-transformation we have 
 \ie Z_{\cN_{\rm diag} 1} = 4( \chi_{1 \over 16}^2 (\bar \chi_0 + \bar \chi_{1 \over 2})^2 + |(\chi_0 + \chi_{1\over 2})\chi_{1 \over 16}|^2).
 \label{ZN1}
 \fe 
 which determines the defect Hilbert space $\cH_{\cN_{\rm diag}}$. Below we will show that knowledge of the defect Hilbert space for the duality and symmetry TDLs is sufficient to determine the corresponding fusion category (i.e. the $F$-symbols) completely in this case.
 
To determine the bicharacter and FS indicator, we study the spin-selection rules associate to the duality twisted partition function $Z_{\cN 1}$ as in \eqref{eqnspinsel}. 
Performing a $T^2$ transformation, followed by crossing and fusion, we obtain for a general $\mZ_2^2$ duality defect $\cN$
 \ie
\quad Z_{\cN 1}(\tau +2) = \frac{\epsilon}{2}(Z_{\cN 1}(\tau) + Z_{\cN s+}(\tau) + Z_{\cN sr^2+}(\tau) + Z_{\cN r^2+}).
\label{T2transf}
\fe
The twisted partition functions that appear on the RHS are constrained by the modular bootstrap equations such as
\ie
Z_{\cN s_+}(-1/\tau)=Z_{s \cN_-}(\tau)
\fe
where the two sides are decomposed into Virasoro characters for \textit{primaries} in the defect Hilbert space $\cH_\cN$ and $\cH_s$ respectively as
\ie
\sum_{(h,\bar h)\in \cH_\cN^{\rm prim}} (s_+)_{h,\bar h} \chi_h(q')\chi_{\bar h }(\bar q')
=
\sum_{(h,\bar h)\in \cH_s^{\rm prim}} ( \cN_-)_{h,\bar h} \chi_h(q)\chi_{\bar h }(\bar q),
\fe
similarly for the twisted partition  functions involving the other symmetry TDLs, $Z_{\cN {sr^2}_+}$ and $Z_{\cN {r^2}_+}$. Now taking $Z_{\cN 1}=Z_{\cN_{\rm diag} 1}$, we find that the modular bootstrap equations fixes these twisted partition functions to be of the form
\ie
Z_{\cN s_+} 
=&\,
2\A_1
|(\chi_0-\chi_{1\over 2}) \chi_{1\over 16}|^2
+\B_1
\left
(
\chi_{1\over 16}^2 (\bar \chi_0+\bar \chi_{1\over 2})^2
-
{\rm c.c}
\right)
\\
Z_{\cN {sr^2}_+} 
=&\,
2\A_2
|(\chi_0-\chi_{1\over 2}) \chi_{1\over 16}|^2
+\B_2
\left
(
\chi_{1\over 16}^2 (\bar \chi_0+\bar \chi_{1\over 2})^2
-
{\rm c.c}
\right)
\\
Z_{\cN {r^2}_+} 
=&\,
\A_3
\left
(
\chi_{1\over 16} (\bar \chi_0+\bar \chi_{1\over 2})
-
{\rm c.c}
\right)^2
\fe
where $\A_{1,2,3}$ and $\B_{1,2}$ are undetermined constants. Next using \eqref{T2transf} and \eqref{ZN1},  we get a set of linear equations for these constants as well as the FS indicator $\epsilon$, which demands
\ie
\A_1+\A_2=2i,~\B_1+\B_2=0,~\A_3=-1
\label{abconst}
\fe
and  in particular $\epsilon=1$.

By fusion and crossing, it's easy to see that the symmetry TDLs acting on $\cH_\cN$ in general satisfies the algebra
\ie
g_+ h_+ =\chi(g,h) (gh)_+
\fe
and we find the further constraints on the bicharacters using the explicit defect Hilbert space in this case,
\ie
&\A_1 \A_2 =\B_1 \B_2= \chi(s,sr^2) \A_3, 
~\A_1^2=\B_1^2=\chi(s,s),~\A_2^2=\B_2^2=\chi(sr^2,sr^2),~\A_3^2=\chi(r^2,r^2).
\fe
From defining property of the bicharacter $\chi(s,s)=\pm 1$, and to be compatible with \eqref{abconst}, we have
\ie
\chi(s,s)=-1
\fe
similarly the other components of the bicharacter is determined and the result coincides with the symmetric bicharacter $\chi_s$ in \eqref{d4bichar}. Thus we have determined that $\cN_{\rm diag}$ generates $\TY(\mZ_2^2,\chi_s,+)$ or equivalently ${\rm Rep}(H_8)$ as promised. 

There is one more $\bZ_2^2$ duality enjoyed by the Ising$^2$ CFT, which is to gauge each $\bZ_2$ spin-flip symmetry ($s$ and $sr^2$ in $D_8$) and then apply the swapping symmetry which exchanges the two chains ($sr$ in $D_8$). This is clearly also a self-duality, giving rise to a TDL $\cN_{D_8}$ with
\ie
\cN_{D_8}=sr \cN_{\rm diag},\quad\cN_{D_8}^2=1+s+r^2+sr^2
\fe
which generates another $\mZ_2^2$ TY fusion category symmetry.\footnote{
One can stack any of the $D_8$ symmetry TDLs with $\cN_{\rm diag}$ (in either order). Since the $\mZ_2^2$ TDLs are absorbed by $\cN_{\rm diag}$ and $D_8/\bZ_2^2=\bZ_2$ contains a single nontrivial element, $\cN_{D_8}$ is only other duality defect generated this way.
} It is clear from the construction that to obtain the $\bZ_2^2$ bicharacter associated with this TY category, we simply compose the bicharacter of $\cN_{\rm diag}$, namely $\chi_s$, with the swapping symmetry, while the Frobenius-Schur indicator does not change, since $sr$ is anomaly-free. More explicitly, we can derive the new bicharacter $\chi$ of $\cN_{D_8}$ using fusion and multiple F-moves as in Figure~\ref{fig:stackf}, yielding
\ie
\chi(s,g)=\chi_s(g,sr^2)
\fe
which corresponds to the other bicharacter $\chi_a$ in \eqref{d4bichar}. Therefore this $\bZ_2^2$ gauging followed by swapping symmetry describes the fusion category $\TY(\mZ_2^2,\chi_a,+)$ or equivalently ${\rm Rep}(D_8)$ \cite{Tambara2000}. Note that since $sr$ interchanges the two Ising factors, it does not commute with the full chiral algebra, in particular $T'$ is odd under $sr$. We will include the various twisted partition functions in part II \cite{toappear}.

In fact, these self-dualities persist beyond the Ising$^2$ CFT. Recall the tensor product operator $\varepsilon_1\varepsilon_2$  of the thermal operators $\varepsilon_{1,2}$ in the two Ising factors has weight $(1,1)$ and is exactly marginal. It is also uncharged under the entire $D_8$ symmetry. Since $\varepsilon_1\varepsilon_2$ does not commute with the individual Ising $\mZ_2$ dualities, the $\mZ_2$ duality defects are no longer topological when we move away from the Ising$^2$ point. However it does commutes with the  duality defect $\cN_{\rm diag}=\cN_{H_8}$, 
\ie
\cN_{H_8} \cdot  \varepsilon_1\varepsilon_2= \la \cN_{H_8} \ra \varepsilon_1\varepsilon_2
=2  \varepsilon_1\varepsilon_2
\fe
thus the $\mZ_2^2$ duality defects $\cN_{H_8}$ and $\cN_{D_8}$ and the associated fusion category symmetries ${\rm Rep}(H_8)$ and ${\rm Rep}(D_8)$ are present along the entire orbifold branch of $c=1$ CFT.

  	 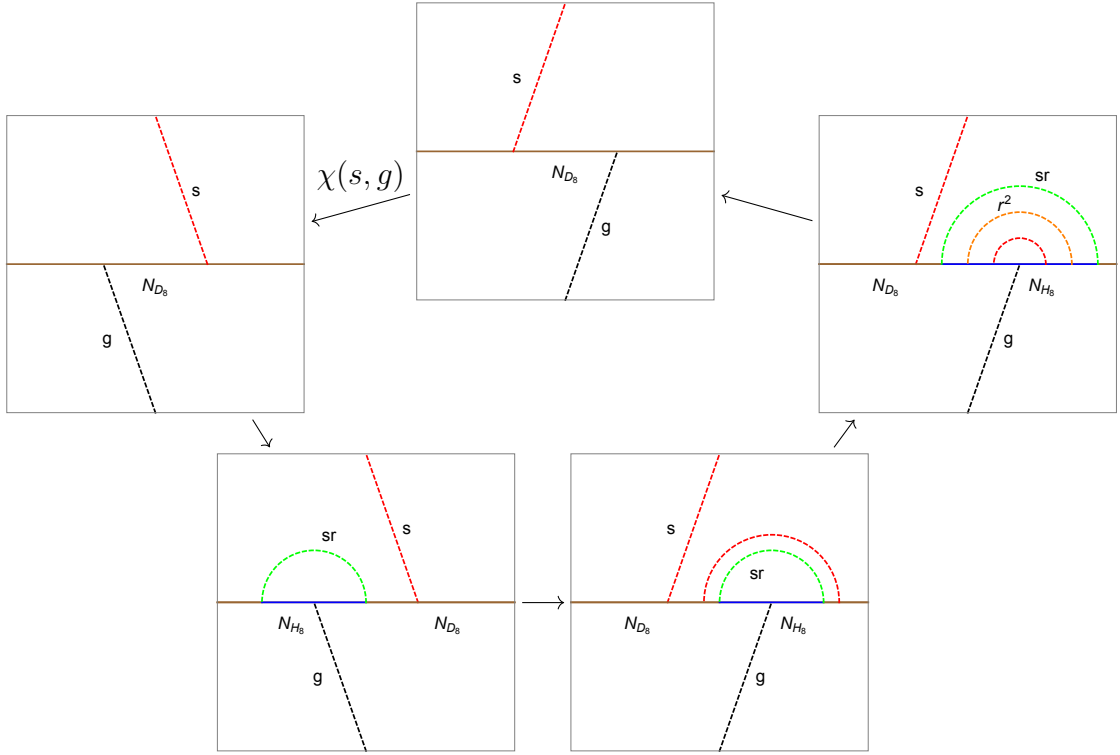
\begin{figure}[htb]
	\centering
	\begin{tikzpicture}
	\node[inner sep=0pt] (A) at (0,1.5)
	{\includegraphics[width=.25\textwidth]{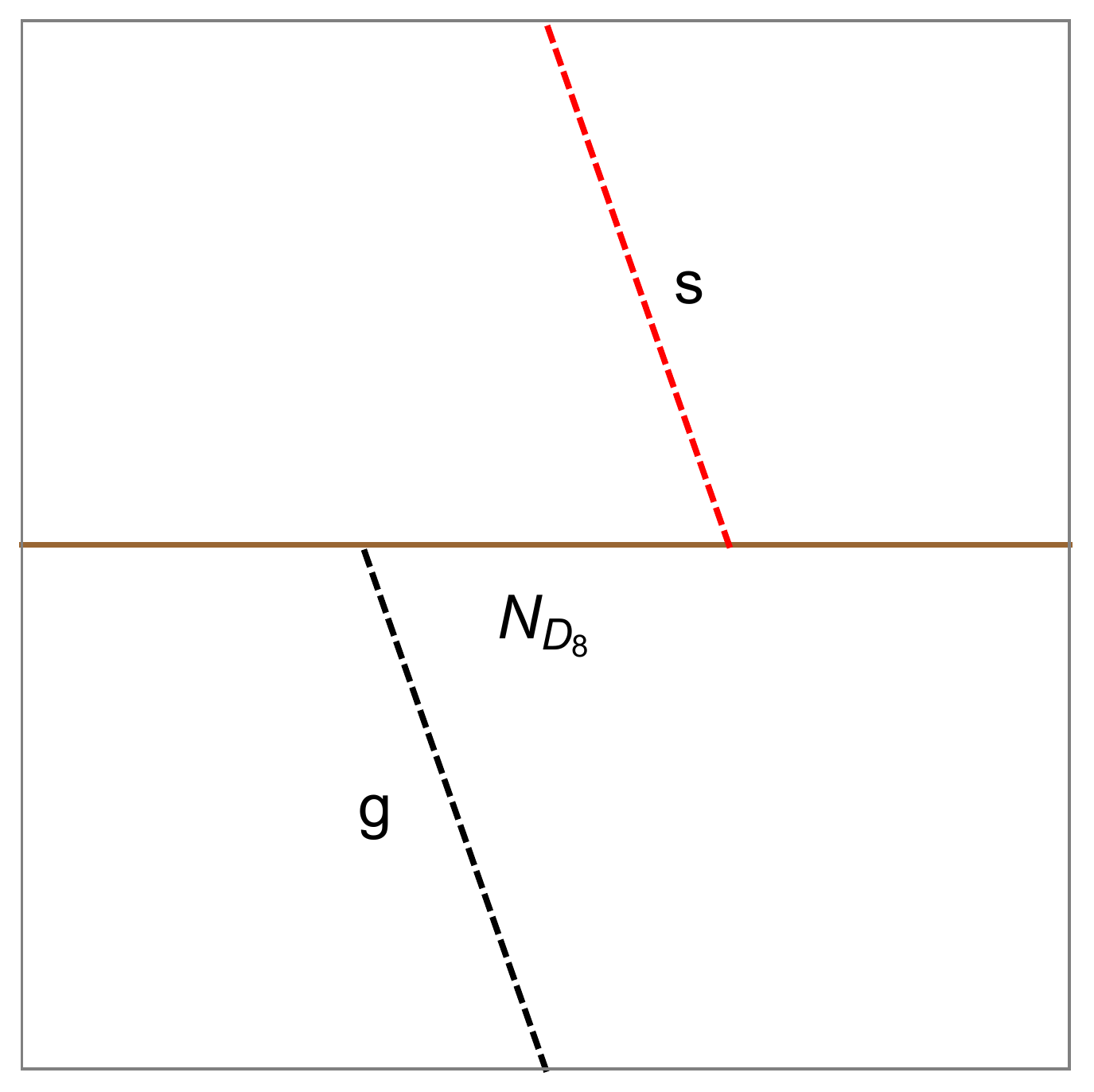}};
	\node[inner sep=0pt] (B) at (2.8,-3)
	{\includegraphics[width=.25\textwidth]{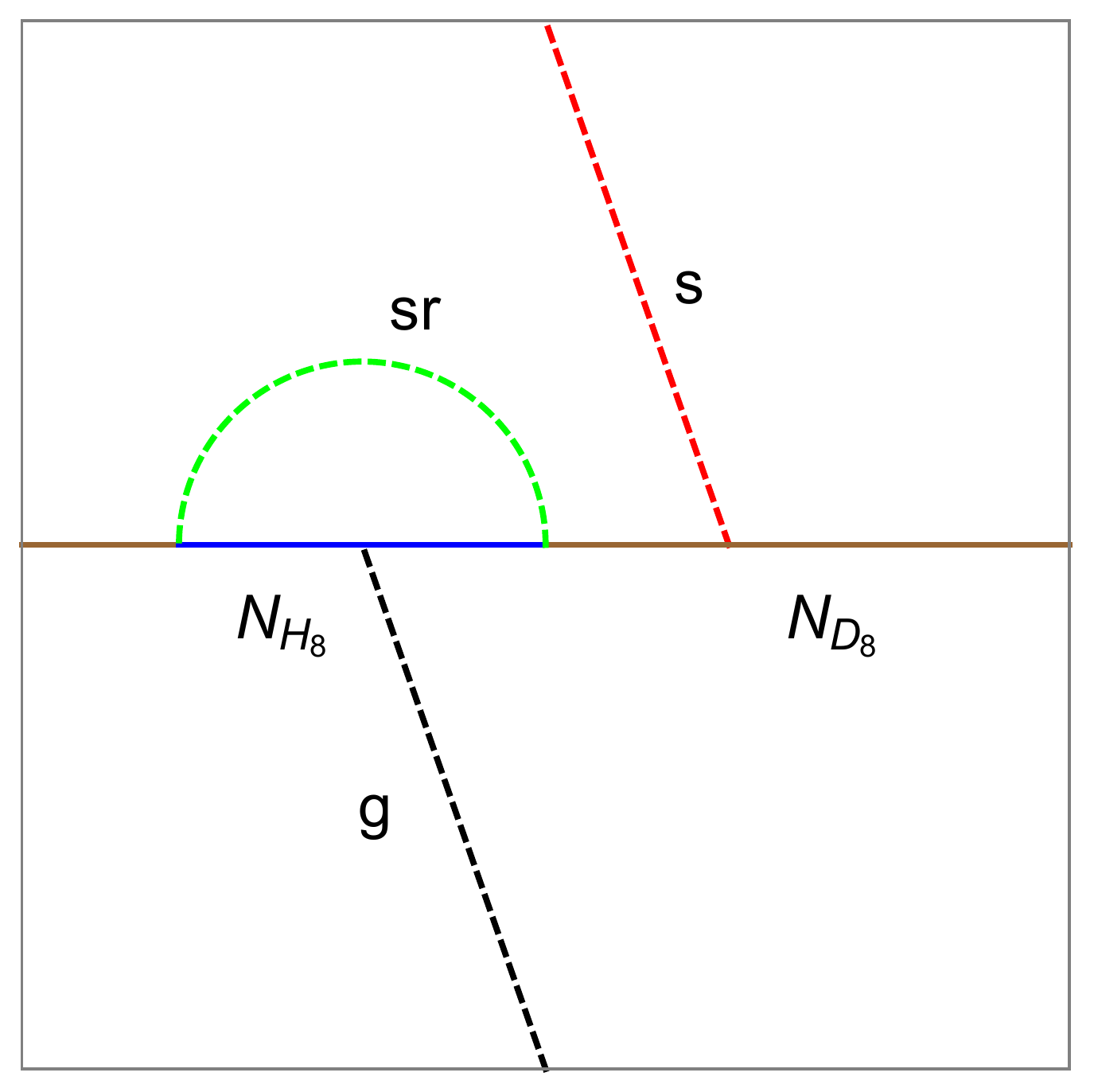}};
	\node[inner sep=0pt] (C) at (7.5,-3)
	{\includegraphics[width=.25\textwidth]{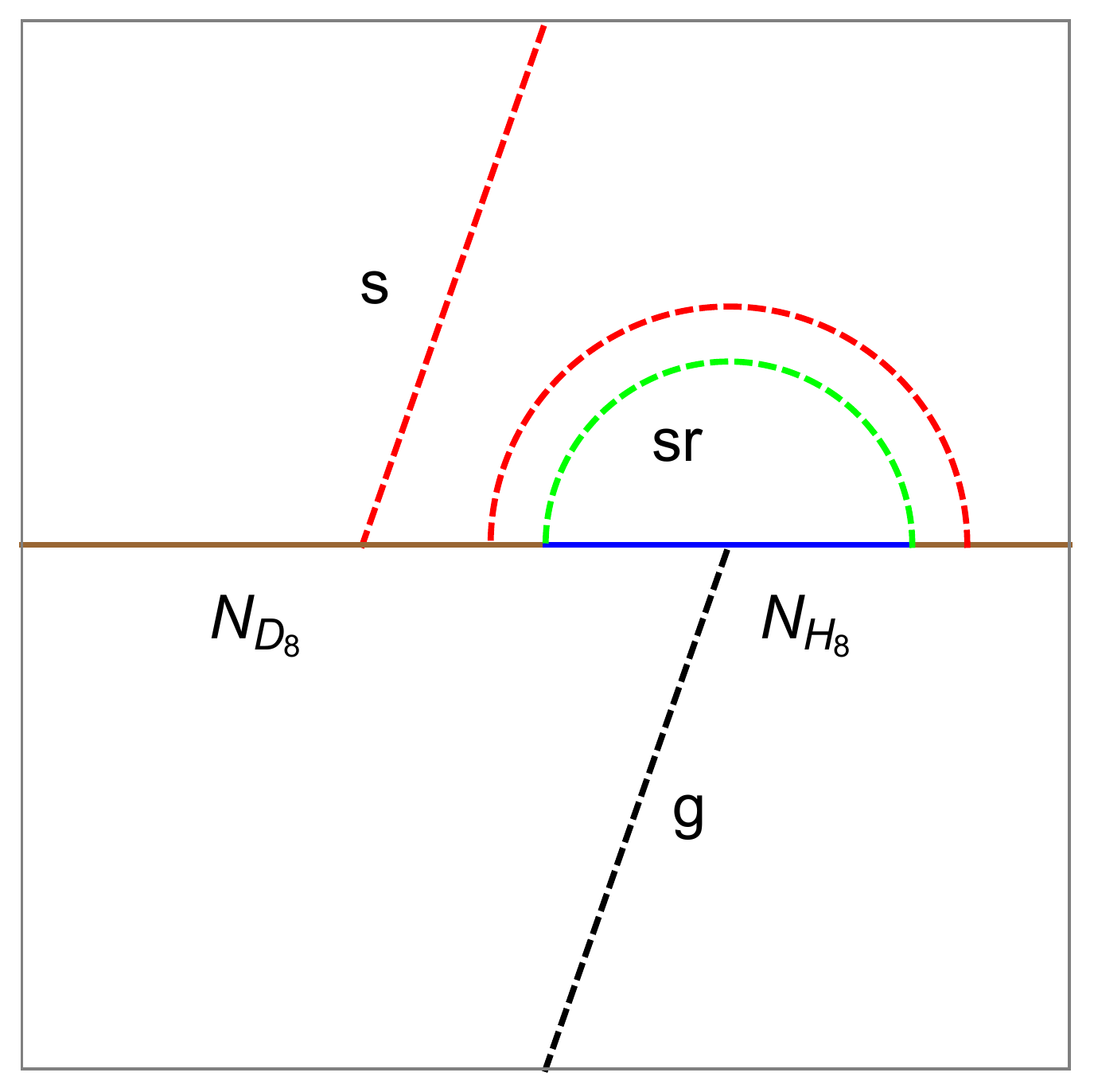}};
	\node[inner sep=0pt] (D) at (10.8,1.5)
	{\includegraphics[width=.25\textwidth]{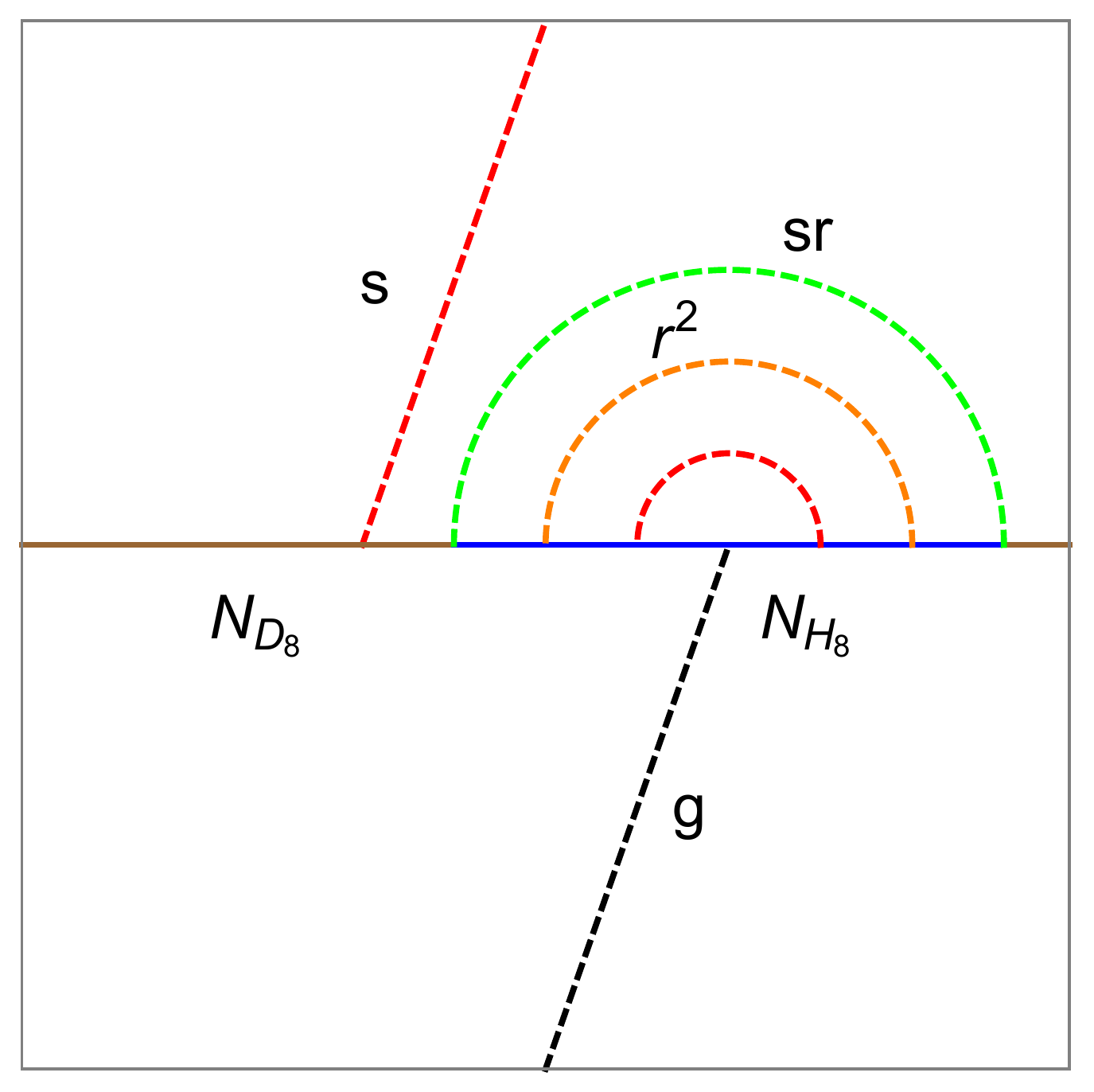}};
	\node[inner sep=0pt] (E) at (5.45,3)
	{\includegraphics[width=.25\textwidth]{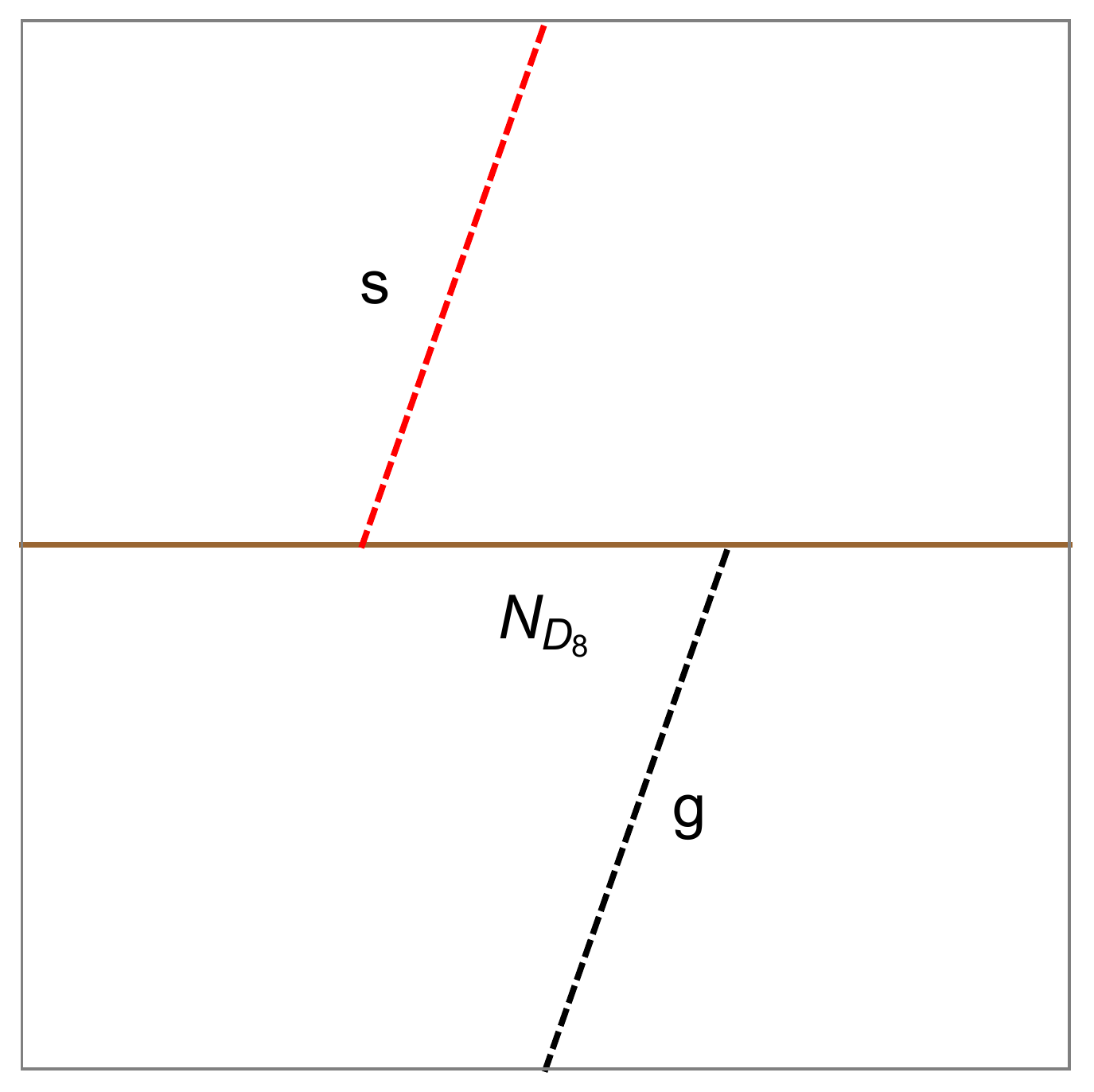}};
	\draw[->] (A) to [anchor=west] node [above right=-2] {} (B);
	\draw[->] (B) to [anchor=west] node [above=-2] {} (C);
	\draw[->] (C) to [anchor=west] node [above left=-2] {} (D);
	\draw[->] (D) to [anchor=west] node [above=2] {} (E);
	\draw[->] (E) to [anchor=west] node [above=2] {$\chi(s,g)$} (A);
	\end{tikzpicture}
	\caption{ 
		The bicharacter $\chi(s,g)$ for the duality defect $\cN_{D_8}$ with $g \in \mZ_2^2$ can be obtained from the F-symbols of the duality defect $\cN_{H_8}$ ($\cN_{diag}$ at the Ising$^2$ radius) as in the figure in the counter-clockwise order starting from the left-most figure and then applying fusion and crossing. Note that the first step involves the nucleation of an $sr$ bubble attached to an $\cN_{H_8}=\cN_{\rm diag}$ segment which may a priori introduce a sign but it's cancelled by the elimination of the some bubble in the fourth step. In addition, the fourth step involves pulling the symmetry TDL $g$ out of the inner two bubbles, and thus picks up a factor  $\chi_s(g,s)\chi_s(g,r^2)=\chi_s(g,sr^2)$ that involves the bicharacter of the duality defect $\cN_{H_8}$.
		}
	\label{fig:stackf}
\end{figure}

\subsubsection{$\bZ_4$ TY Categories}
 
 There are four $\bZ_4$ TY categories,
 \ie
 \TY(\mZ_4,\chi_\pm,\epsilon)
 \fe
labelled by the two bicharacters
 \ie \chi_\pm(a,b) = (\pm i)^{ab}, \qquad a,b \in \bZ_4\fe
 and the two Frobenius-Schur indicators $\epsilon = \pm 1$. None of the four TY categories admit fiber functors.  We determine which of these acts on the Ising$^2$ theory by studying the twisted partition functions. Note that since $r$ does not commute with the full chiral algebra of the Ising$^2$ CFT, we cannot use the $c = 1/2$ characters. Instead we use the description of the theory as an orbifold.

In fact, as for the $\mZ_2^2$ case discussed in the previous section, we find the $c=1$ orbifold branch realizes both $\mZ_4$ TY categories with $\epsilon=1$ at general $R$. The two duality TDLs $\cN_{\mZ_4^\pm}$ are related by stacking with the $\mZ_2$ symmetry defect associated to $s\in D_8$:
\ie
\TY(\mZ_4, \chi_{+},+)
\xleftrightarrow{~~\cN_{\mZ_4^\pm}\ \mapsto\ s \cN_{\mZ_4^\pm}~~}
\TY(\mZ_4, \chi_-,+).
\fe 
 
More explicitly, 
the corresponding duality defects $\cN_{\mZ_4^+}$ and $\cN_{\mZ_4^-}=s\cN_{\mZ_4^+} $ define the following twisted partition functions at general $R\geq \sqrt{2}$
\ie
Z_{1\cN_{\mZ_4^+}}(q,\bar q)=&
{1\over  |\eta|^2} \left(
\sum_{m,n\in \mZ,m-n\in 2\mZ}  
i^{n-m}q^{{1\over 2}\left({n\over R}+{mR\over 2}\right)^2}\bar q^{{1\over 2}\left({n\over R}-{mR\over 2}\right)^2}
+
\sum_{m,n\in \mZ}  (-1)^{m+n}q^{ m^2}\bar q^{ n^2}
\right)
\\
Z_{1\cN_{\mZ_4^-}}(q,\bar q)=&
{1\over  |\eta|^2} \left(
\sum_{m,n\in \mZ,m-n\in 2\mZ}  
i^{n+m}q^{{1\over 2}\left({n\over R}+{mR\over 2}\right)^2}\bar q^{{1\over 2}\left({n\over R}-{mR\over 2}\right)^2}
+
\sum_{m,n\in \mZ}  (-1)^{m+n}q^{ m^2}\bar q^{ n^2}
\right).
\label{Z4duality}
\fe
These twisted partitions have the following desired features. First, the duality defects annihilates operators that are charged under $\mZ_4$ and on the $\mZ_4$ invariant operators with eigenvalue $\pm 2$ as required by the fusion ring $\cN^2=1+r+r^2 +r^3$:
\ie
\widehat\cN_{\cN_{\mZ_4^+}}: \begin{cases}
	V^+_{n,w}\to 2 i^{n-w} V_{n,w}^+ \quad {~\rm for~} n-w \in 2\mZ
	\\
	j_{n^2} \bar j_{m^2} \to 2 j_{n^2} \bar j_{m^2}
	\\
	\text{all other primaries} \to 0
\end{cases}\quad 
\widehat\cN_{\mZ_4^-}: \begin{cases}
	V^+_{n,w}\to 2 i^{n+w} V_{n,w}^+ \quad {~\rm for~} n-w\in 2\mZ
	\\
	j_{n^2} \bar j_{m^2} \to 2 j_{n^2} \bar j_{m^2}
	\\
	\text{all other primaries} \to 0.
\end{cases}
\fe
They also solve the modular bootstrap equation for torus partition function with a single duality twist around the space or time cycle. 
In particular, the modular $S$-transform (using Poisson resummation) yields the spectrum of the duality-twisted sectors:
\ie
Z_{\cN_{\mZ_4^+}1}(q,\bar q)=&
{1\over  2|\eta|^2} \left(
\sum_{m,n\in \mZ,m-n\in 2\mZ}    
q^{{1\over 8}\left({n-1/2 \over R}+{(m+1/2) R\over 2}\right)^2}\bar q^{{1\over 8}\left({n-1/2\over R}-{(m+ 1/2) R\over 2}\right)^2}
+
\sum_{m,n\in \mZ}
q^{{1\over 4}(m+1/2)^2}\bar q^{{1\over4 }(n+1/2)^2}
\right)
\\
Z_{\cN_{\mZ_4^-}1}(q,\bar q)=&
{1\over  2|\eta|^2} \left(
\sum_{m,n\in \mZ,m-n\in 2\mZ}    
q^{{1\over 8}\left({n+1/2 \over R}+{(m+1/2) R\over 2}\right)^2}\bar q^{{1\over 8}\left({n+1/2\over R}-{(m+ 1/2) R\over 2}\right)^2}
+
\sum_{m,n\in \mZ}
q^{{1\over 4}(m+1/2)^2}\bar q^{{1\over4 }(n+1/2)^2}
\right),
\label{Z4dualityHS}
\fe
which decompose into Virasoro characters with positive integer degeneracies, as required, thanks to the shifts in the exponents of $q$ and $\bar q$. The derivations of these twisted partition functions will be given in \cite{toappear}.

 \subsubsection{Ising$^{2\star}$ and Triality}\label{secisingstar}
 
When considering $\bZ_2^2$ gauging of Ising$^2$, it is critical to understand not only how the symmetry acts on local operators, but also on certain non-local operators in the symmetry-twisted sectors. Recall $s$ and $sr^2$ are the spin-flip operators for each Ising chain. In the tensor product theory, all $s$-twisted sector operators are $sr^2$-even and vice versa. However, by further tensoring with a $\bZ_2^2$ SPT and letting $s$ and $sr^2$ act simultaneously as the Ising spin flip and as one of the symmetry generators in this SPT, we obtain a variation of Ising$^2$ which we call Ising$^{2\star}$ after \cite{verresen2019gapless}, where this theory was studied as a possible topologically enriched critical point. This theory may also be obtained by applying the $\bZ_2^2$ SPT entangler to the Ising$^2$ Hamiltonian, similar in spirit to the decorated domain walls of \cite{Scaffidi_2017}.

In this theory, all operators in the $s$-twisted sector are $sr^2$-odd and vice versa. Therefore, the theory we obtain by gauging the $\bZ_2^2$ symmetry, Ising$^{2\star}/\bZ_2^2$, looks very different from Ising$^2$, which was self-dual under this transformation. While the lightest operators in the $s$ and $sr^2$-twisted sectors of that theory contributed the two order parameters $\sigma_{1,2}$ of that theory, we see that in Ising$^{2\star}/\bZ_2^2$, they are projected out. In fact, by studying the scaling spectrum and the nearby gapped phases (e.g. $\epsilon_1 + \epsilon_2$ tunes Ising$^{2\star}/\bZ_2^2$ to the SPT or trivial phase depending on the sign) we see Ising$^{2\star}/\bZ_2^2$ is actually the free-fermion point of the compact boson branch!\footnote{We also check directly the orbifold partition function of Ising$^{2\star}$ and match to that of the Dirac point of $c=1$ CFT.} Note that this transformation is involutive, so gauging the magnetic $\bZ_2^2$ symmetry of the compact boson we again obtain Ising$^{2\star}$.

Tensoring with the $\bZ_2^2$ SPT (which fixes the compact boson branch \cite{verresen2019gapless}) and $\bZ_2^2$ gauging together generate an $S_3$ action on these three theories. The order 3 element (the composition of the two order 2 generators) is the triality we studied in Section \ref{secnality}. In \cite{verresen2019gapless}, the authors (including one of us) found a phase diagram (see also \cite{bridgeman}) which includes all three of these CFTs in the $c=1$ moduli space where the three CFTs merge from the KT point. At this point, the triality becomes a fusion category symmetry which we analyze in our follow-up \cite{toappear}.
  
  \section{Discussion}
  
  We have shown how the fusion category anomalies of \cite{theOGs} fit into a broader context of classifying gapped phases with fusion category symmetry by studying boundary conditions of Turaev-Viro theory. However, our anomaly in-flow structure is unsatisfying in that the fusion category symmetry is always ``gauged" in the bulk. It is very interesting to ask whether there are non-degenerate gapped phases in 2+1D with fusion category symmetry such that the associated anomalous theories occur on their boundary, as in the usual theory of SPT phases and their anomalous boundaries.
  
  There is at least one obstruction to doing this. Indeed, suppose a fusion category with a non-integer quantum dimension were to act by topological surface operators in 2+1D. If this 2+1D theory is ``modular-invariant'' on a 3-torus, then we can adapt the arguments of \cite{theOGs} to show that it must have degenerate ground-states---the symmetry is anomalous in 2+1D as well! At present, we don't see a way around this difficulty.
  
  We note also that there is no apparent group law for these anomalies, as a typical fusion ring $R$ lacks a diagonal map $R \to R \times R$, meaning a tensor product of theories with $R$ symmetry doesn't have a canonical $R$ symmetry itself (we expect the $F$-symbols to change under such embeddings, so we speak in terms of the fusion ring only). Thus it is not clear what it would mean for the anomaly theory to be invertible, if it existed. On the other hand, if $R$ has a special generator, such as the duality line in the Ising$_+$ category, then we can sometimes assign an order to the anomaly. In the case of Ising$_+$, the diagonal duality line in Ising$_+\times$Ising$_+$ generates an anomaly-free $\bZ_2 \times \bZ_2$-TY category ${\rm Rep}(H_8)$. In a sense we can say this anomaly is a $\bZ_2$ anomaly. On the other hand, by results of \cite{Tambara2000}, if we take a $\bZ_m$-TY, $m$ odd, all $\bZ_m^n$-TY categories we obtain by this procedure are anomalous, so it is like a $\bZ$ anomaly. Likewise there is no obvious algebraic structure for stacking 1+1D gapped phases.
  
  Actually, simply constructing a 2+1D or 3+1D theory with a (higher) fusion category symmetry would be quite interesting. Non-invertible topological surface operators in Chern-Simons theory were constructed in \cite{Kapustin2}, but these surface operators are not \emph{dualizable}, meaning that they cannot fuse to the vacuum. Technically, such objects do not form a fusion category, but they may still be interesting for constraining RG flows.
  
  One could also search for other dualities to use to define fusion category symmetries in higher dimensions. Most familiar dualities like electric-magnetic duality however define ordinary invertible surface operators at their self-dual points \cite{Gaiotto_2009,Kapustin_2009,Vidal_2009}. It is more fruitful to look for dualities which hold only after gauging a symmetry, as in Kramers-Wannier. Note also that while TQFTs seem to be highly constrained in 3+1D and beyond, in particular that higher dimensional Turaev-Viro might be a kind of finite gauge theory \cite{Lan_2018}, it still may be that there are many interesting higher fusion categories, although many of them will be Morita equivalent.
  
  Although we have chosen to focus on field theory, it is also possible to study fusion category symmetry on the lattice. For instance, in the anyon chain models \cite{Feiguin_2007,BuicanGromov} where the fusion category defining the Hilbert space automatically acts as a symmetry. For ordinary global symmetries $G$, one normally says that an anomaly-free symmetry is one which can be realized on-site, that is by tensor product unitary operators. However, a fusion category symmetry like Kramers-Wannier duality sends local operators (such as the order parameter) to nonlocal operators (the disorder operator), so it cannot be realized in such a way, anomalous or not. This is related to how there is no ``trivial" $\cA$-SPT since there is no group law for fiber functors. In fact, we have seen that for the Tambara-Yamagami categories based on $G$, only a ``maximally nontrivial" $G$-SPT can realize a nondegenerate gapped symmetric phase (cf. $\alpha$'s nondegeneracy in Section \ref{secselfdualspt}), so all these ground-states are entangled. Fusion category symmetries are realized by matrix product operator (MPO) symmetries on the lattice \cite{Williamson_2016,Bultinck_2017}. It would be very interesting to understand what distinguishes an anomalous MPO from an anomaly-free one.

\bibliographystyle{plain}
\bibliography{refs.bib}

\end{document}